\documentclass[12pt]{article}
\usepackage{graphicx}
\usepackage[textwidth=8em,textsize=small]{todonotes}
\usepackage{natbib}
\usepackage{amssymb}
\usepackage[utf8]{inputenc}
\usepackage{mathtools}
\usepackage{amsmath,amsfonts,graphicx}
\usepackage[colorlinks, citecolor={blue}]{hyperref}
\usepackage{float}
\usepackage{bbm}
\usepackage{subfigure}
\usepackage{algorithm}
\usepackage{algpseudocode}
\usepackage{url} 

\newtheorem{assumption}{Assumption}
\newtheorem{theorem}{Theorem}[section]
\newtheorem{lemma}[theorem]{Lemma}
\newtheorem{corollary}[theorem]{Corollary}

\newcommand{\blind}{1}

\begin{document}

\def\spacingset#1{\renewcommand{\baselinestretch}%
{#1}\small\normalsize} \spacingset{1}


\if1\blind
{
  \title{\bf Functional Mixed Membership Models}
  \author{Nicholas Marco\\
    Department of Biostatistics, University of California,
Los Angeles, USA.\\
    and \\
    Damla \c{S}ent\"{u}rk \\
    Department of Biostatistics, University of California,
Los Angeles, USA. \\
    and \\
    Shafali Jeste \\
    Division of Neurology and Neurological Institute,\\ Children’s Hospital Los Angeles, Los Angeles, USA.\\
    and \\
    Charlotte DiStefano \\
    Division of Psychiatry, Children’s Hospital Los Angeles, Los Angeles, USA.\\
    and \\
    Abigail Dickinson \\
    Department of Psychiatry and Biobehavioral Sciences,\\ University of California, Los Angeles, USA.\\
    and \\
    Donatello Telesca \thanks{
    The authors gratefully acknowledge \textit{funding from the NIH/NIMH R01MH122428-01 (DS,DT)}}\hspace{.2cm}\\
    Department of Biostatistics, University of California,
Los Angeles, USA.\\
    }
  \maketitle
} \fi

\if0\blind
{
  \bigskip
  \bigskip
  \bigskip
  \begin{center}
    {\LARGE\bf Functional Mixed Membership Models}
\end{center}
  \medskip
} \fi

\bigskip

\newpage
\begin{abstract}
    Mixed membership models, or partial membership models, are a flexible unsupervised learning method that allows each observation to belong to multiple clusters. In this paper, we propose a Bayesian mixed membership model for functional data. By using the multivariate Karhunen-Loève theorem, we are able to derive a scalable representation of Gaussian processes that maintains data-driven learning of the covariance structure. Within this framework, we establish conditional posterior consistency given a known feature allocation matrix. Compared to previous work on mixed membership models, our proposal allows for increased modeling flexibility, with the benefit of a directly interpretable mean and covariance structure. Our work is motivated by studies in functional brain imaging through electroencephalography (EEG) of children with autism spectrum disorder (ASD). In this context, our work formalizes the clinical notion of ``spectrum'' in terms of feature membership proportions. Supplementary materials, including proofs, are available online. The R package \textsc{BayesFMMM} is available to fit functional mixed membership models.
\end{abstract}
\noindent%
\vfill

\newpage
\spacingset{1.5} 
\section{Introduction}

Cluster analysis often aims to identify homogeneous subgroups of statistical units within a data set \citep{clusteringHandbook}. A typical underlying assumption of both heuristic and model-based procedures posits the existence of a finite number of sub-populations, from which each sample is extracted with some probability, akin to the idea of {\it uncertain membership}. A natural extension of this framework allows each observation to belong to multiple clusters simultaneously; leading to the concept of mixed membership models, or partial membership models \citep{BleiJordan03, Erosheva04}, akin to the idea of {\it mixed membership}. This manuscript aims to extend the class of mixed membership models to functional data.

Functional data analysis (FDA) is concerned with the statistical analysis of the realizations of a random function. In the functional clustering framework, the random function is often conceptualized as a mixture of stochastic processes, so that each realization is assumed to be sampled from one of $K<\infty$ cluster-specific sub-processes. The literature on functional clustering is well established, and several analytical strategies have been proposed to handle different sampling designs and to ensure increasing levels of model flexibility \citep{james2003clustering, chiou2007functional, petrone2009hybrid, jacques2014model}. 
 
\begin{figure}
    \centering
     \includegraphics[width=.9\textwidth]{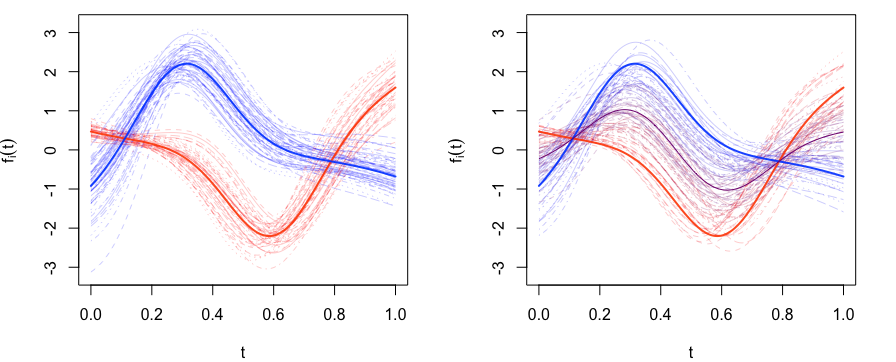}
    \caption{Generative model illustration. (Left panel) Data generated under the functional clustering framework. (Right panel) Data generated under a mixed membership framework.}
    \label{fig:fmmm}
\end{figure}

Beyond clustering, the ability to model an observation's membership on a spectrum is particularly advantageous when we consider applications to biomedical data. For example, in diagnostic settings, the severity of symptoms may vary from person to person. Therefore, it is important to ask which symptomatic features characterize the sample and whether subjects exhibit one or more of these traits. Within this general context, our work is motivated by functional brain imaging studies of neurodevelopmental disorders, such as autism spectrum disorder (ASD). In this setting, typical patterns of neuronal activity are examined through the use of brain imaging technologies (e.g. electroencephalography (EEG)), and result in observations which are naturally characterized as random functions on a specific evaluation domain. We are primarily interested in the discovery of distinct functional features that characterize the sample and the explanation of how these features determine subject-level functional heterogeneity. To our knowledge, the literature on FDA is still silent on the concept of mixed membership, defining both the methodological and scientific need for a rigorous statistical formulation of functional mixed membership models. Figure \ref{fig:fmmm} shows a conceptual illustration of the difference between data generated under the {\it uncertain membership} framework versus data generated under the {\it mixed membership} framework. In the mixed membership framework, components of the pure mixture can be continuously mixed to generate higher levels of realized heterogeneity in an observed sample.  

One of the first uses of mixed membership models appeared in the field of population genetics. \citet{pritchard2000inference} used the idea of mixed membership to group individuals into subpopulations, while accounting for the presence of admixed cases. 
\cite{BleiJordan03} introduced the concept of latent Dirichlet allocation to model mixed membership in the context of information retrieval associated with text corpora and other discrete data. These ideas have been extended to sampling models in the exponential family of distributions by \cite{heller2008statistical}, and a general discussion of mixed versus partial membership models for continuous distributions is presented in \cite{Gruhl2014}.
Prior distributions generalizing the concept of mixed membership to an arbitrary number of features have been discussed by \cite{heller2008statistical}, who suggested leveraging the Indian buffet process \citep{Griffiths2011} to create a non-parametric mixed membership model. A comprehensive review and probabilistic generalizations of Bayesian clustering have been discussed by \cite{Broderick2013}.

In this manuscript, we introduce the concept of mixed membership functions to the broader field of functional data analysis. Assuming a known number of latent functional features, we show how a functional mixed membership process is naturally defined through a simple extension of finite Gaussian process (GP) mixture models, by allowing mixing proportions to be indexed at the level of individual sample paths.  Some sophistication is introduced through the multivariate treatment of the underlying functional features to ensure sampling models of adequate expressivity. Specifically, we represent a family of multivariate GPs through the multivariate Karhunen-Loève construction of \citet{happ2018multivariate}. Using this idea, we jointly model the mean and covariance structures of the underlying functional features and derive a sampling model through the ensuing finite-dimensional marginal distributions.  Since na\"{i}ve GP mixture models assume that observations can only belong to one cluster, they only require the univariate treatment of the underlying GPs. However, when sample paths are allowed partial membership in multiple clusters, we need to assume independence between clusters or estimate the cross-covariance functions. Since the assumption of independence does not often hold in practice, we propose a model that allows us to jointly model the covariance and cross-covariance functions through the eigenfunctions of the multivariate covariance operator. Typically, this representation would require sampling on a constrained space to ensure that the eigenfunctions are mutually orthogonal. However, in this context, we are able to relax these constraints and maintain many of the desirable theoretical properties of our model by extending the multiplicative gamma process shrinkage prior proposed by \citet{bhattacharya2011sparse}. 

The remainder of the paper is organized as follows. Section \ref{sec: FPM} formalizes the concept of functional mixed membership as a natural extension of functional clustering. Section \ref{sec: Inference} discusses Bayesian inference through posterior simulation and establishes conditions for weak posterior consistency. Section \ref{sec: SimulatioCase} contains two simulation studies to assess operating characteristics on engineered data, and illustrates the application of the proposed model to a case study involving functional brain imaging through electroencephalography (EEG). Finally, we conclude our exposition with a critical discussion of the methodological challenges and related literature in Section \ref{discussion}.

\section{Functional Mixed Membership}
\label{sec: FPM}
Functional data analysis (FDA) focuses on methods to analyze sample paths of an underlying continuous stochastic process $f: \mathcal{T} \rightarrow \mathbb{R}$, where $\mathcal{T}$ is a compact subset of $\mathbb{R}^d$. In the i.i.d. setting, a typical characterization of these stochastic processes relies on the definition of the mean function, $\mu(t) = \mathbb{E}\left(f(t)\right)$, and the covariance function, $C(s,t) = \text{Cov}\left(f(s), f(t)\right)$, where $t,s \in \mathcal{T}$. In this paper, we focus on jointly modeling $K$ underlying stochastic processes, where each stochastic process represents one cluster. In this section, we will proceed under the assumption that the number of clusters, $K$, is known \textit{a priori}. While $K$ is fixed for the technical developments of the paper, Section \ref{sim_study_2} discusses the use of various information criteria for the selection of the optimal number of features.

Let $f^{(k)}_{clust}: \mathcal{T} \rightarrow \mathbb{R}$ denote the underlying stochastic process associated with the $k^{th}$ cluster. Let $\mu^{(k)}_{clust}$ and $C^{(k)}_{clust}$ be, respectively, the mean and covariance function corresponding to the $k^{th}$ stochastic process.  To ensure desirable properties, such as decompositions, FDA often makes the assumption that the random functions are elements in a Hilbert space. Specifically, we assume $f^{(k)}_{clust} \in L^2(\mathcal{T})$ $\left(\int_{\mathcal{T}} |f^{(k)}(t)|^2 \text{d}t < \infty \right)$. Within this context, a typical model-based representation of functional clustering assumes that each underlying process is a latent Gaussian process, $f^{(k)}_{clust} \sim \mathcal{GP}\left(\mu^{(k)}_{clust}, C^{(k)}_{clust}\right)$, with sample paths $f_i$ $(i=1,2,\ldots,N)$, assumed to be independently drawn from a finite GP mixture
$$p\left(f_i \mid \rho^{(1:K)}, \mu^{(1:K)}_{clust}, C^{(1:K)}_{clust}\right) = \sum_{k=1}^K \rho^{(k)}\;\mathcal{GP}\left(f_i\mid \mu^{(k)}_{clust}, C^{(k)}_{clust}\right);$$
where $\rho^{(k)}$ is the mixing proportion (fraction of sample paths) for component $k$, s.t. $\sum_{k=1}^K \rho^{(k)} = 1$. Equivalently, by introducing latent variables $\tilde{\boldsymbol{\pi}}_i = \left(\tilde{\pi}_{i1}, \ldots, \tilde{\pi}_{iK}\right)$, s.t. $ \tilde{\boldsymbol{\pi}}_i \sim_{iid} \mbox{Multinomial}(1;\, \rho^{(1)}, \ldots, \rho^{(K)})$, 
it is easy to show that, 
\begin{equation}
    \label{eq: GPFMM}
    f_i \mid \tilde{\boldsymbol{\pi}}_1,\dots, \tilde{\boldsymbol{\pi}}_N, \mu^{(1:K)}_{clust}, C^{(1:K)}_{clust} \sim \mathcal{GP}\left(\sum_{k=1}^K \tilde{\pi}_{ik}\mu^{(k)}_{clust}, \sum_{k=1}^K \tilde{\pi}_{ik}C^{(k)}_{clust}\right).
\end{equation}
Mixed membership is naturally defined by generalizing the finite mixture model to allow each sample path to belong to a finite positive number of mixture components, which we call {\it functional features} \citep{heller2008statistical, Broderick2013}. Let $f^{(k)}$ denote the underlying Gaussian process associated with the $k^{th}$ functional feature where $\mu^{(k)}$ and $C^{(k)}$ denote the corresponding mean and covariance functions (note that $f^{(k)}$ need not be equal in distribution to $f^{(k)}_{clust}$). By introducing continuous latent variables $\mathbf{z}_i = (Z_{i1},\ldots, Z_{iK})'$, where $Z_{ik} \in [0,1]$ and $\sum_{k=1}^K Z_{ik} = 1$,  the general form of the proposed mixed membership model follows the following convex combination:
\begin{equation}
    f_i \mid \mathbf{z}_1,\dots, \mathbf{z}_N =_d \sum_{k=1}^K Z_{ik} f^{(k)}.
    \label{eq: GPPMM_f}
\end{equation}
In Equation \ref{eq: GPFMM}, since each observation only belongs to one cluster, the Gaussian processes, $f^{(k)}_{clust}$, are most commonly assumed to be mutually independent. 
However, the same assumption is too restrictive for the model in Equation \ref{eq: GPPMM_f}. Thus, we will let $C^{(k,k')}$ denote the cross-covariance function between $f^{(k)}$ and $f^{(k')}$, and denote with $\boldsymbol{\mathcal{C}}$ the collection of covariance and cross-covariance functions. Therefore, the sampling model for the proposed mixed membership scheme can be written as
\begin{equation}
    \label{eq: GPPMM}
    f_i \mid \mathbf{z}_1,\dots, \mathbf{z}_N, \mu^{(1:K)}, \boldsymbol{\mathcal{C}} \sim \mathcal{GP}\left(\sum_{k=1}^K Z_{ik}\mu^{(k)}, \sum_{k=1}^K Z_{ik}^2 C^{(k)} + \sum_{k=1}^K \sum_{k'\ne k} Z_{ik}Z_{ik'} C^{(k,k')}\right).
\end{equation}
 Given a finite evaluation grid ${\bf t}_i = (t_{i1}, \ldots ,t_{in_i})'$, the process' finite-dimensional marginal distribution can be characterized s.t.
\begin{equation}
    \label{eq: NPMM}
    f_i({\bf t}_i) \mid  \mathbf{z}_1,\dots, \mathbf{z}_N, \mu^{(1:K)}, \boldsymbol{\mathcal{C}} \sim \mathcal{N}\left(\sum_{k=1}^K Z_{ik}\mu^{(k)}({\bf t}_i), C_i({\bf t}_i, {\bf t}_i)\right),
\end{equation}
where $C_i({\bf t}_i, {\bf t}_i) = \sum_{k=1}^K Z_{ik}^2C^{(k)}({\bf t}_i, {\bf t}_i) + \sum_{k=1}^K \sum_{k'\ne k} Z_{ik}Z_{ik'} C^{(k,k')}({\bf t}_i, {\bf t}_i)$. A detailed comparison between our model and other similar models such as functional factor analysis models and FPCA can be found in Section B.3 of the Supplementary Materials. Since we do not assume that the functional features are independent, a concise characterization of the $K$ stochastic processes is needed in order to ensure that our model is scalable and computationally tractable. In Section \ref{section: MKL}, we review the multivariate Karhunen-Loève theorem \citep{happ2018multivariate}, which will provide a joint characterization of the latent functional features, $\left(f^{(1)}, f^{(2)},\ldots,f^{(K)}\right)$. 
Using this joint characterization, we are able to specify a scalable sampling model for the finite-dimensional marginal distribution found in Equation \ref{eq: NPMM}, which is described in full detail in Section \ref{section: FPMP}.

\subsection{Multivariate Karhunen-Loève Characterization}
\label{section: MKL}

To aid in computation, we assume that $f^{(k)}$ is a smooth function and is in the $P$-dimensional subspace $\boldsymbol{\mathcal{S}} \subset L^2(\mathcal{T})$, spanned by a set of linearly independent square-integrable basis functions $\{b_1, \dots, b_P\}$ ($b_p: \mathcal{T} \rightarrow \mathbb{R}$). While the choice of basis functions is user-defined, in practice B-splines (or the tensor product of B-splines for $\mathcal{T} \subset \mathbb{R}^d$ for $d \ge 2$) are a common choice due to their flexibility. \citet{de1978practical} has shown that for any smooth function, the distance between the true function and the approximation constructed using B-splines with uniform knots goes to zero as the number of knots goes to infinity. In practice, the number of basis functions used is often proportional to the average number of time points observed for each functional observation (Section E of the web-based appendix contains a detailed discussion on the choice of $P$). Alternative basis systems can be selected in relation to application-specific considerations.

The multivariate Karhunen-Loève (KL) theorem, proposed by \citet{ramsay2005principal}, can be used to jointly decompose our $K$ stochastic processes. To allow our model to handle higher-dimensional functional data, such as images, we will use the extension of the multivariate KL theorem for different dimensional domains proposed by \citet{happ2018multivariate}. Although they state the theorem in full generality, we will only consider the case where $f^{(k)} \in \boldsymbol{\mathcal{S}}$. In this section, we will show that the multivariate KL theorem for different dimensional domains still holds under our assumption that $f^{(k)} \in \boldsymbol{\mathcal{S}}$, given that the conditions in Lemma \ref{lemma_compact} are satisfied.

We will start by defining the multivariate function $\mathbf{f}(\mathbf{t})$ in the following way:
\begin{equation}
    \mathbf{f}(\mathbf{t}) := \left(f^{(1)}\left(t^{(1)}\right), f^{(2)}\left(t^{(2)}\right), \dots, f^{(K)}\left(t^{(K)}\right)\right),
\end{equation}
where $\mathbf{t}$ is a $K$-tuple such that $\mathbf{t} = \left(t^{(1)}, t^{(2)}, \dots, t^{(K)}\right)$, where $t^{(1)}, t^{(2)}, \dots, t^{(K)} \in \mathcal{T}$. This construction allows for the joint representation of $K$ stochastic processes, each at different points in their domain. Since $f^{(k)} \in \boldsymbol{\mathcal{S}}$, we have $\mathbf{f}\in \boldsymbol{\mathcal{H}}:= \boldsymbol{\mathcal{S}} \times \boldsymbol{\mathcal{S}} \times \dots \times \boldsymbol{\mathcal{S}} := \boldsymbol{\mathcal{S}}^K$. We define the corresponding mean of $\mathbf{f}(\mathbf{t})$, such that
$\boldsymbol{\mu}(\mathbf{t}) := \left(\mu^{(1)}\left(t^{(1)}\right), \mu^{(2)}\left(t^{(2)}\right), \dots, \mu^{(K)}\left(t^{(K)}\right) \right)$,
where $\boldsymbol{\mu}(\mathbf{t}) \in \boldsymbol{\mathcal{H}}$, and the mean-centered cross-covariance functions as $C^{(k,k')}(s,t) := \text{Cov}\left(f^{(k)}(s) - \mu^{(k)}(s), f^{(k')}(t) - \mu^{(k')}(t)\right)$, where $s,t \in \mathcal{T}$.
\begin{lemma}
$\boldsymbol{\mathcal{S}}$ is a closed linear subspace of $L^2(\mathcal{T})$.
\label{lemma_comp_sub}
\end{lemma} 
Proofs for all results are given in the Supplementary Materials. From Lemma \ref{lemma_comp_sub}, since $\boldsymbol{\mathcal{S}}$ is a closed linear subspace of $L^2(\mathcal{T})$, we have that $\boldsymbol{\mathcal{S}}$ is a Hilbert space with respect to the inner product $\langle f, g \rangle_{\boldsymbol{\mathcal{S}}}= \int_{\mathcal{T}} f(t)g(t) \text{d}t$ where $f,g \in \boldsymbol{\mathcal{S}}$. By defining the inner product on $\boldsymbol{\mathcal{H}}$ as 
\begin{equation}
\label{inner_product}
    \langle\mathbf{f}, \mathbf{g} \rangle_{\boldsymbol{\mathcal{H}}}  := \sum_{k=1}^K \int_{\mathcal{T}} f^{(k)}\left(t^{(k)}\right) g^{(k)}\left(t^{(k)}\right)\text{d}t^{(k)},
\end{equation}
where $\mathbf{f}, \mathbf{g} \in \boldsymbol{\mathcal{H}}$, we have that $\boldsymbol{\mathcal{H}}$ is the direct sum of Hilbert spaces. Since $\boldsymbol{\mathcal{H}}$ is the direct sum of Hilbert spaces, $\boldsymbol{\mathcal{H}}$ is a Hilbert space \citep{reed1972methods}. Thus, we have constructed a subspace $\boldsymbol{\mathcal{H}}$ using our assumption that $f^{(k)}\in \boldsymbol{\mathcal{S}}$, which satisfies Proposition 1 in \cite{happ2018multivariate}. Letting $\mathbf{f} \in \boldsymbol{\mathcal{H}}$, we can define the covariance operator $\boldsymbol{\mathcal{K}}: \boldsymbol{\mathcal{H}} \rightarrow \boldsymbol{\mathcal{H}}$ element-wise in the following way:
\begin{equation}
    \left(\boldsymbol{\mathcal{K}}\mathbf{f}\right)^{(k)}(\mathbf{t}) := \left\langle C^{(\cdot,k)}\left(\cdot,t^{(k)}\right),\mathbf{f} \right\rangle_{\boldsymbol{\mathcal{H}}} = \sum_{k'=1}^K \int_{\mathcal{T}} C^{(k',k)}\left(s,t^{(k)}\right) f^{(k')}(s) \text{d}s.
    \label{Cov_op}
\end{equation}
Since we made the assumption $f^{(k)} \in \boldsymbol{\mathcal{S}}$, we can simplify Equation \ref{Cov_op}. Since $\boldsymbol{\mathcal{S}}$ is the span of the basis functions, we have $f^{(k)}(t) - \mu^{(k)}(t) = \sum_{p=1}^P \theta_{(k,p)}b_p(t) = \mathbf{B}'(t)\boldsymbol{\theta}_k,$
for some $\boldsymbol{\theta}_k \in \mathbb{R}^P$ and $\mathbf{B}'(t) = [b_1(t) \cdots b_P(t)]$. Thus, we can see that 
\begin{equation}
    C^{(k,k')}(s,t) = \text{Cov}\left(\mathbf{B}'(s)\boldsymbol{\theta}_k,  \mathbf{B}'(t)\boldsymbol{\theta}_{k'}\right) = \mathbf{B}'(s)\text{Cov}\left(\boldsymbol{\theta}_k, \boldsymbol{\theta}_{k'}\right)\mathbf{B}(t),
    \label{Cov_simp}
\end{equation}
 and we can rewrite Equation \ref{Cov_op} as $\left(\boldsymbol{\mathcal{K}}\mathbf{f}\right)^{(k)}(\mathbf{t}) = \sum_{k'=1}^K \int_{\mathcal{T}} \mathbf{B}'(s)\text{Cov}\left(\boldsymbol{\theta}_{k'}, \boldsymbol{\theta}_k\right)\mathbf{B}\left(t^{(k)}\right) f^{(k')}(s) \text{d}s,$
 for some $\mathbf{f} \in \mathcal{H}$. 
 The following lemma establishes conditions under which $\boldsymbol{\mathcal{K}}$ is a bounded and compact operator, which is a necessary condition for the multivariate KL decomposition to exist.
 
 \begin{lemma}
 $\boldsymbol{\mathcal{K}}$ is a bounded and compact operator if we have that (1) the basis functions, $b_1, \dots, b_P$, are uniformly continuous and (2) there exists $C_{max} \in \mathbb{R}$ such that $\left|\text{Cov}\left(\theta_{(k,p)}, \theta_{(k',p')}\right)\right| \le C_{max}$.
 \label{lemma_compact}
 \end{lemma}
 Assuming the conditions specified in lemma \ref{lemma_compact} hold, by the Hilbert-Schmidt theorem, since $\boldsymbol{\mathcal{K}}$ is a bounded, compact, and self-adjoint operator, we know that there exists real eigenvalues $\lambda_1, \dots, \lambda_{KP}$ and a complete set of eigenfunctions $\boldsymbol{\Psi}_1, \dots, \boldsymbol{\Psi}_{KP} \in \boldsymbol{\mathcal{H}}$ such that $\boldsymbol{\mathcal{K}}\boldsymbol{\Psi}_p = \lambda_p \boldsymbol{\Psi}_p,$ for $p = 1, \dots, KP.$ Since $\boldsymbol{\mathcal{K}}$ is a non-negative operator, we know that $\lambda_p \ge 0$ for $p = 1, \dots, KP$ (\cite{happ2018multivariate}, Proposition 2). From Theorem VI.17 of \citet{reed1972methods}, we have that the positive, bounded, self-adjoint, and compact operator $\boldsymbol{\mathcal{K}}$ can be written as $\boldsymbol{\mathcal{K}}\mathbf{f} = \sum_{p=1}^{KP} \lambda_p \left\langle  \boldsymbol{\Psi}_p, \mathbf{f}\right\rangle_{\boldsymbol{\mathcal{H}}} \boldsymbol{\Psi}_p.$
Thus, from Equation \ref{Cov_op}, we have that 
\begin{align}
    \nonumber
     \sum_{k'=1}^K \int_{\mathcal{T}} C^{(k',k)}\left(s,t^{(k)}\right) f^{(k')}(s) \text{d}s & = \sum_{k'=1}^K\int_{\mathcal{T}}\left(\sum_{p=1}^{KP} \lambda_p \Psi^{(k')}_p(s)\Psi_p^{(k)}\left(t^{(k)}\right)\right)f^{(k')}(s) \text{d}s,
\end{align}
where $\Psi_p^{(k)}(t^{(k)})$ is the $k^{th}$ element of $\boldsymbol{\Psi}_p(\mathbf{t})$. Thus, we can see that the covariance kernel can be written as the finite sum of eigenfunctions and eigenvalues,
\begin{equation}
    C^{(k,k')}(s,t) = \sum_{p=1}^{KP} \lambda_p \Psi^{(k)}_p(s)\Psi_p^{(k')}(t).
    \label{Cov_expansion}
\end{equation}
Since we are working under the assumption that the random function,  $\mathbf{f} \in \boldsymbol{\mathcal{H}}$, we can use a modified version of the Multivariate KL theorem. 
Considering that $\{\boldsymbol{\Psi}_1, \dots, \boldsymbol{\Psi}_{KP}\}$ form a complete basis for $\boldsymbol{\mathcal{H}}$ and $\mathbf{f}(\mathbf{t}) - \boldsymbol{\mu}(\mathbf{t}) \in \boldsymbol{\mathcal{H}}$, we have
\begin{equation}
    \nonumber
     \mathbf{f}(\mathbf{t}) - \boldsymbol{\mu}(\mathbf{t})\; =\; \mathbf{P}_{\boldsymbol{\mathcal{H}}} \circ \left(\mathbf{f} - \boldsymbol{\mu}\right)(\mathbf{t})   \;=\; \sum_{p=1}^{KP} \left\langle\boldsymbol{\Psi}_p, \mathbf{f} - \boldsymbol{\mu} \right\rangle_{\boldsymbol{\mathcal{H}}} \boldsymbol{\Psi}_p(\mathbf{t}),
\end{equation}
where $\mathbf{P}_{\boldsymbol{\mathcal{H}}}$ is the projection operator onto $\boldsymbol{\mathcal{H}}$. Letting $\rho_p = \left\langle\boldsymbol{\Psi}_p, \mathbf{f} - \boldsymbol{\mu} \right\rangle_{\boldsymbol{\mathcal{H}}}$, we have $\mathbf{f}(\mathbf{t}) - \boldsymbol{\mu}(\mathbf{t}) = \sum_{p =1}^{KP}\rho_p \boldsymbol{\Psi}_p(\mathbf{t})$, where $\mathbb{E}(\rho_p) = 0$ and $\text{Cov}(\rho_p, \rho_{p'}) = \lambda_p \delta_{pp'}$ \citep{happ2018multivariate}. 

Since $\boldsymbol{\Psi}_p \in \boldsymbol{\mathcal{H}}$ and $\boldsymbol{\mu} \in \boldsymbol{\mathcal{H}}$, there exist $\boldsymbol{\nu}_k \in \mathbb{R}^P$ and $\boldsymbol{\phi}_{kp} \in \mathbb{R}^P$ such that $\boldsymbol{\mu}^{(k)}(\mathbf{t}) =  \boldsymbol{\nu}_k'\mathbf{B}\left(t^{(k)}\right)$ and $ \sqrt{\lambda_p}\boldsymbol{\Psi}^{(k)}_p(\mathbf{t}) =  \boldsymbol{\phi}_{kp}'\mathbf{B}\left(t^{(k)}\right):= \boldsymbol{\Phi}_p^{(k)}(\mathbf{t})$. 
These scaled eigenfunctions, $\boldsymbol{\Phi}_p^{(k)}(\mathbf{t})$, are used over $\boldsymbol{\Psi}^{(k)}_p(\mathbf{t})$ because they fully specify the covariance structure of the latent features, as described in Section \ref{section: FPMP}. From a modeling prospective, the scaled eigenfunction parameterization is advantageous, as it admits a prior model based on the multiplicative gamma process shrinkage prior proposed by \citet{bhattacharya2011sparse}, allowing for adaptive regularized estimation \citep{Shamshoian2020}.
Thus we have
\begin{equation}
    \mathbf{f}^{(k)}(\mathbf{t}) = \boldsymbol{\mu}^{(k)}(\mathbf{t}) + \sum_{p=1}^{KP}\chi_p \boldsymbol{\Phi}_p^{(k)}(\mathbf{t})=  \boldsymbol{\nu}_k'\mathbf{B}\left(t^{(k)}\right) + \sum_{p=1}^{KP}\chi_p \boldsymbol{\phi}_{kp}'\mathbf{B}\left(t^{(k)}\right),
    \label{f_model}
\end{equation}
where $\chi_p = \left\langle(\lambda_p)^{-1/2}\boldsymbol{\Psi}_p, \mathbf{f} - \boldsymbol{\mu} \right\rangle_{\boldsymbol{\mathcal{H}}}$, $\mathbb{E}(\chi_p) = 0$, and $\text{Cov}(\chi_p, \chi_{p'}) = \delta_{pp'}$. Equation \ref{f_model} gives us a way to jointly decompose any realization of the $K$ stochastic processes, $\mathbf{f} \in \boldsymbol{\mathcal{H}}$, as a finite weighted sum of basis functions. 
\begin{corollary}
 If $\chi_{p}\sim_{iid} \mathcal{N}(0,1)$, then the random function ${\bf f}(\mathbf{t})$ follows a multivariate $\mathcal{GP}$ with means $\boldsymbol{\mu}^{(k)}(\mathbf{t})=\boldsymbol{\nu}'_k{\bf B}\left(t^{(k)}\right)$, and cross-covariance functions $C^{(k,k')}(s,t) = {\bf B}'(s)\sum_{p=1}^{KP} (\boldsymbol{\phi}_{kp}\boldsymbol{\phi}'_{k'p}){\bf B}(t)$. 
 \label{corollary_GP}
 \end{corollary}

\subsection{Functional Mixed Membership Process}
\label{section: FPMP}

In this section, we will describe a Bayesian additive model that allows for the constructive representation of mixed memberships for functional data. Our model allows for direct inference on the mean and covariance structures of the $K$ stochastic processes, which are often of scientific interest. To aid computational tractability, we will use the joint decomposition described in Section \ref{section: MKL}. In this section, we will assume that $f^{(k)} \in \boldsymbol{\mathcal{S}}$, and that the conditions of Lemma \ref{lemma_compact} hold. 

We aim to model the mixed membership of $N$ observed sample paths, $\{\mathbf{Y}_i(.)\}_{i=1}^N$, to $K$ latent functional features ${\bf f} = \left(f^{(1)}, f^{(2)},\ldots, f^{(K)}\right)$. Assuming that each path is observed at $n_i$ evaluation points ${\bf t}_i = [t_{i1} \cdots t_{in_i}]'$, without loss of generality, we define a sampling model for the finite-dimensional marginals of ${\bf Y}_i({\bf t}_i)$. To allow for path-specific partial membership, we extend the finite mixture model in Equation \ref{eq: GPFMM} and introduce path-specific mixing proportions  $Z_{ik}\in (0,1)$, s.t. $\sum_{k=1}^K Z_{ik} = 1$, for $i=1,2, \ldots,N$.  

Let $\mathbf{S}(\mathbf{t}_i) = [\mathbf{B}(t_1) \cdots \mathbf{B}(t_{n_i})] \in \mathbb{R}^{P \times n_i}$, $\chi_{im}\sim \mathcal{N}(0,1)$, and $\boldsymbol{\Theta}$ denote the collection of model parameters. Let $M \le KP$ be the number of eigenfunctions used to approximate the covariance structure of the $K$ stochastic processes. Using the decomposition of $f^{(k)}(t)$ in Equation \ref{f_model} and assuming a normal distribution on ${\bf Y}_i({\bf t}_i)$, we obtain: 
\begin{equation}
    {\bf Y}_i({\bf t}_i)|\boldsymbol{\Theta}  \sim  \mathcal{N}\left\{\sum_{k=1}^K
    Z_{ik}\left(\mathbf{S}'(\mathbf{t}_i) \boldsymbol{\nu}_k
   + \sum_{m=1}^M\chi_{im}\mathbf{S}'({\bf t}_i) \boldsymbol{\phi}_{km}\right),\; \sigma^2 \mathbf{I}_{n_i}\right\}.
    \label{BFPM_model}
\end{equation}
If we integrate out the latent $\chi_{im}$ variables, we obtain a more transparent form for the proposed functional mixed membership process. Specifically, we have  

\begin{equation}
    \mathbf{Y}_i(\mathbf{t}_i)| \boldsymbol{\Theta}_{-\chi} \sim \mathcal{N}\left\{\sum_{k=1}^KZ_{ik}\mathbf{S}'(\mathbf{t}_i) \boldsymbol{\nu}_k,\; \mathbf{V}_i + \sigma^2\mathbf{I}_{n_i}\right\},
    \label{BFPM_model_int_chi}
\end{equation}
where $\boldsymbol{\Theta}_{-\chi}$ is the collection of our model parameters excluding the $\chi_{im}$ variables, and $\mathbf{V}_i =  \sum_{k=1}^K\sum_{k'=1}^K Z_{ik}Z_{ik'}\left\{\mathbf{S}'(\mathbf{t}_i)\sum_{m=1}^{M}\left(\boldsymbol{\phi}_{km}\boldsymbol{\phi}'_{k'm}\right)\mathbf{S}(\mathbf{t}_i)\right\}.$
Thus, for a sample path, we have that the mixed membership mean is a convex combination of the functional feature means, and the mixed membership covariance, is a weighted sum of the covariance and  cross-covariance functions between different functional features, following from the multivariate KL characterization in Section \ref{section: MKL}. Furthermore, from Equation \ref{Cov_expansion} it is easy to show that,  for large enough $M$, we have 
$\mathbf{S}'(\mathbf{t}_i)\sum_{m=1}^{M}\left(\boldsymbol{\phi}_{kp}\boldsymbol{\phi}'_{k'p}\right)\mathbf{S}(\mathbf{t}_i) \approx C^{(k,k')}(\mathbf{t}_i,\mathbf{t}_i),$with equality when $M = KP$.

Mixed membership models can be thought of as a generalization of clustering. As such, these stochastic schemes are characterized by an inherent lack of likelihood identifiability. A typical source of non-indentifiability is the common \textit{label switching} problem. To deal with the label switching problem, a relabelling algorithm can be derived for this model directly from the work of \citet{stephens2000dealing}. A second source of non-identifiabilty stems from allowing $Z_{ik}$ to be continuous random variables. 
Specifically, consider a model with 2 features and let $\boldsymbol{\Theta}_0$ be the set of ``true'' parameters. Let $Z_{i1}^* = 0.5(Z_{i1})_0$ and $Z_{i2}^* =  (Z_{i2})_0 + 0.5(Z_{i1})_0$. If we let $\boldsymbol{\nu}_1^* = 2(\boldsymbol{\nu}_1)_0 - (\boldsymbol{\nu}_2)_0$, $\boldsymbol{\nu}_2^* = (\boldsymbol{\nu}_2)_0$, $\boldsymbol{\phi}_{1m}^* = 2(\boldsymbol{\phi}_{1m})_0 - (\boldsymbol{\phi}_{2m})_0$,  $\boldsymbol{\phi}_{2m}^* = (\boldsymbol{\phi}_{2m})_0 $, $\chi_{im}^* = (\chi_{im})_0$, and $(\sigma^2)^* = \sigma^2_0$, we have $P(Y_i(t)|\boldsymbol{\Theta}_0) = P(Y_i(t)|\boldsymbol{\Theta}^*)$ (Equation \ref{BFPM_model}). Thus, we can see that our model is not identifiable, and we will refer to this problem as the \textit{rescaling problem}. 
To address the rescaling problem, additional assumptions such as the \textit{separability} condition \citep{papadimitriou1998latent, mcsherry2001spectral, azar2001spectral, chen2022learning} or the \textit{sufficiently scattered} condition \citep{huang2016anchor, jang2019minimum, chen2022learning} are often made on the allocation parameters. The separability condition assumes that at least one observation belongs entirely to each feature, making it relatively easy to incorporate into a sampling model. To ensure that the separability condition holds in a two-feature model, we use the Membership Rescale Algorithm (Algorithm 2 in the Supplementary Materials). Compared to the separability condition, the sufficiently scattered condition makes weaker geometric assumptions on the sampling model when $K \ge 3$, but it is more difficult to implement in a sampling scheme. Section C.4 in the Supplementary Materials has an in-depth review of the two common assumptions made to address the rescaling problem.


A third non-identifiability problem may arise numerically as a form of concurvity, 
i.e. when $\boldsymbol{\nu}_{k'} \propto \boldsymbol{\phi}_{km}$ in Equation \ref{BFPM_model}. 
Typically, an overestimation of the magnitude of $\boldsymbol{\phi}_{km}$, may result in small variance estimates for the allocation parameters (smaller credible intervals). 
To address this problem, we can assume the separability condition; however, under weaker geometric assumptions this problem may persist, but is typically of little practical relevance.

\subsection{Prior Distributions and Model Specification}
\label{section: Priors}
The sampling model in Section \ref{section: FPMP}, allows a practitioner to select how many eigenfunctions are to be used in the approximation of the covariance function. In the case of $M = KP$, we have a fully saturated model and can represent any realization $\mathbf{f}\in \boldsymbol{\mathcal{H}}$. 
 In Equation \ref{f_model}, $\boldsymbol{\Phi}$ parameters are mutually orthogonal (where orthogonality is defined by the inner product defined in Equation \ref{inner_product}), and have a magnitude proportional to the square root of the corresponding eigenvalue, $\lambda_p$. Thus, a modified version of the multiplicative gamma process shrinkage prior proposed by \citet{bhattacharya2011sparse} will be used as our prior for $\boldsymbol{\phi}_{km}$. By using this prior, we promote shrinkage across the $\boldsymbol{\phi}_{km}$ coefficient vectors, with increasing prior shrinkage towards zero as $m$ increases. 
 
 To facilitate MCMC sampling from the posterior target 
 we will remove the assumption that $\boldsymbol{\Phi}$ parameters are mutually orthogonal. Even though $\boldsymbol{\Phi}$ parameters can no longer be thought of as scaled eigenfunctions, posterior inference can still be conducted on the eigenfunctions by post-processing posterior samples of $\boldsymbol{\phi}_{km}$ (given that we can recover the true covariance operator). Specifically, given posterior samples from $\boldsymbol{\phi}_{km}$, we obtain posterior realizations for the covariance function and then calculate the eigenpairs of the posterior samples of the covariance operator using the method described in \citet{happ2018multivariate}. Thus, letting $\phi_{kpm}$ be the $p^{th}$ element of $\boldsymbol{\phi}_{km}$, we have
 
$$\phi_{kpm}|\gamma_{kpm}, \tilde{\tau}_{mk} \sim \mathcal{N}\left(0, \gamma_{kpm}^{-1}\tilde{\tau}_{mk}^{-1}\right), \;\;\; \gamma_{kpm} \sim \Gamma\left(\nu_\gamma /2 , \nu_\gamma /2\right), \;\;\; \tilde{\tau}_{mk} = \prod_{n=1}^m \delta_{nk},$$
$$ \delta_{1k}|a_{1k} \sim \Gamma(a_{1k}, 1), \;\;\; \delta_{jk}|a_{2k} \sim \Gamma(a_{2k}, 1), \;\;\; a_{1k} \sim \Gamma(\alpha_1, \beta_1), \;\;\; a_{2k} \sim \Gamma(\alpha_2, \beta_2),$$
where $1 \le k \le K$, $1 \le p \le P$, $1 \le m \le M$, and $2 \le j \le M$. To promote shrinkage across the $M$ matrices, we need $\delta_{jk} > 1$. Thus, letting $\alpha_2 > \beta_2$, we have $\mathbb{E}(\delta_{jk}) > 1$, which will promote shrinkage. In this construction, we allow for different rates of shrinkage across different functional features, which is particularly important in cases where the covariance functions of different features have different magnitudes. In cases where the magnitudes of the covariance functions are different, we would expect $\delta_{mk}$ to be relatively smaller in the $k$ associated with the functional feature with a large covariance function.

To promote adaptive smoothing in the mean function, we will use a first-order random walk penalty proposed by \citet{lang2004bayesian}. The first-order random walk penalty penalizes differences in adjacent B-spline coefficients. In the case where $\mathcal{T} \subset \mathbb{R}$, we have that
$P(\boldsymbol{\nu}_k|\tau_k) \propto exp\left(-\frac{\tau_k}{2}\sum_{p =1}^{P - 1}\left(\nu_{pk}'- {\nu}_{(p+1)k}\right)^2\right),$
 for $k = 1, \dots, K$, where $\tau_k \sim \Gamma(\alpha, \beta)$ and $\nu_{pk}$ is the $p^{th}$ element of $\boldsymbol{\nu}_k$. Since we have $Z_{ik} \in (0,1)$ and $\sum_{k=1}^K Z_{ik} = 1$, it is natural to consider prior Dirichlet sampling for $\mathbf{z}_i = [Z_{i1} \cdots Z_{iK}]$. Therefore, following \citet{heller2008statistical}, we have
$$\mathbf{z}_i\mid \boldsymbol{\pi}, \alpha_3 \sim_{iid} Dir(\alpha_3\boldsymbol{\pi}), \;\;\; \boldsymbol{\pi} \sim Dir(\mathbf{c}), \;\;\;
\alpha_3 \sim Exp(b)$$
for $i = 1,\dots, N$. Lastly, we will use a conjugate prior for our random error component of the model, such that $\sigma^2 \sim IG(\alpha_0,  \beta_0)$. Although we relax the assumption of orthogonality, in Section C.1 of the Supplementary Materials, we outline an alternative sampling scheme where we impose the condition that the $\boldsymbol{\Phi}$ parameters form orthogonal eigenfunctions using the work of \citet{kowal2017bayesian}.
\section{Posterior Inference}
\label{sec: Inference}
Statistical inference is based on Markov chain Monte Carlo samples from the posterior distribution, by using the Metropolis-within-Gibbs algorithm. While the na\"{i}ve sampling scheme is relatively simple, ensuring good exploration of the posterior target can be challenging due to the potentially multimodal nature of the posterior distribution. Specifically, some sensitivity of the results to the starting values of the chain can be observed for some data. Section C.2 of the Supplementary Materials outlines an algorithm for the selection of informed starting values. Furthermore, to mitigate sensitivity to chain initialization, we also implemented a tempered transition scheme, which improves the mixing of the Markov chain by allowing for transitions between modal configuration of the target.  Implementation details for the proposed tempered transition scheme are reported in Section C.3 of the Supplementary Materials. Additional information on sampling schemes, as well as how to construct simultaneous confidence intervals, can be found in Section D of the Supplementary Materials.

\subsection{Weak Posterior Consistency}
In the previous section, we saw that the $\boldsymbol{\Phi}$ parameters are not assumed to be mutually orthogonal. By removing this constraint, we facilitate MCMC sampling from the posterior target and 
can perform inference on the eigenpairs of the covariance operator, as long as we can recover the covariance structure. Due to the complex identifiability issues mentioned in Section \ref{section: FPMP}, establishing weak posterior consistency with unknown allocation parameters unattainable, even if we include the orthogonality constraint on the $\boldsymbol{\Phi}$ parameters. However, the model becomes identifiable when we condition on the allocation parameters. In this section, we show that we can achieve conditional weak posterior consistency without enforcing the orthogonality constraint of the $\boldsymbol{\Phi}$ parameter. Therefore, we will show that conditional on the allocation parameters, relaxing the orthogonality constraint does not affect our ability to recover the mean and covariance structure of the $K$  underlying stochastic processes from Equation \ref{BFPM_model_int_chi}.
Let $\boldsymbol{\Pi}$ be the prior distribution on $\boldsymbol{\omega} := \left\{\boldsymbol{\nu}_1, \dots, \boldsymbol{\nu}_K, \boldsymbol{\Sigma}_{11}, \dots, \boldsymbol{\Sigma}_{1K},\dots, \boldsymbol{\Sigma}_{KK}, \sigma^2 \right\}$, where $\boldsymbol{\Sigma}_{kk'} := \sum_{p=1}^{KP}\left(\boldsymbol{\phi}_{kp}\boldsymbol{\phi}'_{k'p}\right)$. We will be proving weak posterior consistency with respect to the parameters $\boldsymbol{\Sigma}_{kk'}$ because the parameters $\boldsymbol{\phi}_{kp}$ are non-identifiable. Since the covariance and cross-covariance structure are completely specified by the $\boldsymbol{\Sigma}_{kk'}$ parameters, 
the lack of identifiability of the $\boldsymbol{\phi}_{kp}$ parameters bears no importance on inferential considerations.
We will denote the true set of parameter values as $\boldsymbol{\omega}_0 =\left\{(\boldsymbol{\nu}_1)_0, \dots, (\boldsymbol{\nu}_K)_0, (\boldsymbol{\Sigma}_{11})_0, \dots, (\boldsymbol{\Sigma}_{1K})_0,\dots, (\boldsymbol{\Sigma}_{KK})_0, \sigma^2_0 \right\}.$
To prove weak posterior consistency, we will make the following assumptions:
\begin{assumption}
Each sample path , $\mathbf{Y}_i$ ($i = 1, \dots, N$), is observed at time points $\{t_{i1}, \dots, t_{in_i}\}$, where $KP < n_i \le n_{max}$ for some $n_{max} \in \mathbb{Z}^+$.
\label{uniform_obs}
\end{assumption}
\begin{assumption}
The variables $Z_{ik}$ are known a-priori for $i =1, \dots, N$ and $k = 1,\dots, K$.
\label{known_Z}
\end{assumption}
\begin{assumption}
The true parameter that models random noise is positive ($\sigma^2_0 > 0$).
\label{positive_var}
\end{assumption}
In order to prove weak posterior consistency, we will first specify the following quantities related to the Kullback–Leibler (KL) divergence. Following the notation of \citet{choi2005posterior}, we define the following quantities:
\begin{equation}
    \nonumber
    \Lambda_i(\boldsymbol{\omega}_0, \boldsymbol{\omega}) = \text{log}\left(\frac{f_i(\mathbf{Y}_i;\boldsymbol{\omega}_0)}{f_i(\mathbf{Y}_i;\boldsymbol{\omega})} \right),\;\;
    K_i(\boldsymbol{\omega}_0, \boldsymbol{\omega})  = \mathbb{E}_{\boldsymbol{\omega}_0}(\Lambda_i(\boldsymbol{\omega}_0, \boldsymbol{\omega})),\;\; 
    V_i(\boldsymbol{\omega}_0, \boldsymbol{\omega})  = \text{Var}_{\boldsymbol{\omega}_0}(\Lambda_i(\boldsymbol{\omega}_0, \boldsymbol{\omega})),
\end{equation}
 where $f_i(\mathbf{Y}_i;\boldsymbol{\omega}_0)$ is the likelihood under $\boldsymbol{\omega}_0$. To simplify the notation, we define the following two quantities:
 \begin{align}
     \nonumber
     \boldsymbol{\mu}_i &= \sum_{k=1}^KZ_{ik}\mathbf{S}'(\mathbf{t}) \boldsymbol{\nu}_k,\\
     \nonumber 
     \boldsymbol{\Sigma}_i &= \sum_{k=1}^K\sum_{k'=1}^K Z_{ik}Z_{ik'}\left(\mathbf{S}'(\mathbf{t})\sum_{p=1}^{KP}\left(\boldsymbol{\phi}_{kp}\boldsymbol{\phi}'_{k'p}\right)\mathbf{S}(\mathbf{t})\right) + \sigma^2\mathbf{I}_{n_i} = \mathbf{U}_i'\mathbf{D}_i\mathbf{U}_i + \sigma^2\mathbf{I}_{n_i},
 \end{align}
 where $\mathbf{U}_i'\mathbf{D}_i\mathbf{U}_i$ is the corresponding spectral decomposition. Let $d_{il}$ be the $l^{th}$ diagonal element of $\mathbf{D}_i$. Let $\boldsymbol{\Omega}_\epsilon(\boldsymbol{\omega}_0)$ be the set of parameters such that the KL divergence is less than some $\epsilon > 0$ ($\boldsymbol{\Omega}_\epsilon(\boldsymbol{\omega}_0):= \{\boldsymbol{\omega}: K_i(\boldsymbol{\omega}_0, \boldsymbol{\omega}) < \epsilon \text{ for all i} \}$). Let $a, b \in \mathbb{R}$ be such that $a > 1$ and $b > 0$, and define $\mathcal{B}(\boldsymbol{\omega}_0) := \left\{\boldsymbol{\omega}: \frac{1}{a}\left((d_{il})_0 + \sigma^2_0\right) \le d_{il} + \sigma^2 \le a\left((d_{il})_0 + \sigma^2_0\right), \|\left(\boldsymbol{\mu}_i\right)_0 - \boldsymbol{\mu}_i\| \le b\right\}.$
 \begin{lemma}
 Let $\mathcal{C}(\boldsymbol{\omega}_0, \epsilon) :=\mathcal{B}(\boldsymbol{\omega}_0) \cap \boldsymbol{\Omega}_\epsilon(\boldsymbol{\omega}_0)$. Thus, for $\boldsymbol{\omega}_0 \in \boldsymbol{\Omega}$ and $\epsilon > 0$, there exist $a > 1$ and $b > 0$ such that (1) $\sum_{i=1}^\infty \frac{V_i(\boldsymbol{\omega}_0, \boldsymbol{\omega})}{i^2} < \infty$, for any $\boldsymbol{\omega} \in \mathcal{C}(\boldsymbol{\omega}_0, \epsilon)$ and (2) $\boldsymbol{\Pi}\left(\boldsymbol{\omega} \in  \mathcal{C}(\boldsymbol{\omega}_0, \epsilon)\right) > 0$.
 \label{pos_prior_prob}
 \end{lemma}
Lemma \ref{pos_prior_prob} shows that the prior probability of our parameters being arbitrarily close (defined by KL divergence) to the true parameters is positive. Since $\mathbf{Y}_1, \dots, \mathbf{Y}_N$ are not identically distributed, condition (1) in Lemma \ref{pos_prior_prob} is necessary to prove Lemma \ref{weak_consistency}. 

\begin{lemma}
Under assumptions \ref{uniform_obs}-\ref{positive_var}, the posterior distribution, $\boldsymbol{\Pi}_N(. | \mathbf{Y}_1, \dots, \mathbf{Y}_N)$, is weakly consistent at $\boldsymbol{\omega}_0 \in \boldsymbol{\Omega}$.
\label{weak_consistency}
\end{lemma}
All proofs are provided in the Supplementary Materials. Lemma \ref{weak_consistency} shows us that, conditional on the allocation parameters, we are able to recover the covariance structure of the $K$ stochastic processes. Thus, relaxation of the orthogonality constraint on the $\boldsymbol{\phi}_{km}$ parameters does not affect our ability to perform posterior inference on the main functions of scientific interest. Inference on the eigenstructure can still be performed by calculating the eigenvalues and eigenfunctions of the covariance operator using the MCMC samples of the $\boldsymbol{\phi}_{km}$ parameter. Finally, we point out that, in most cases, the parameters $Z_{ik}$ are unknown. Although theoretical guarantees for consistent estimation of the latent mixed allocation parameters are still elusive, we provide some empirical evidence of convergence in Section \ref{sim_study_1}.

\section{Case Studies and Experiments on Simulated Data}
\label{sec: SimulatioCase}

\subsection{Simulation Study 1}
\label{sim_study_1}
In this simulation study, we examine how well our model can recover the mean and covariance functions when we vary the number of observed functions. The model used in this section is a mixed membership model with $2$ functional features ($K = 2$), which can be represented by a basis constructed of $8$ B-splines ($P = 8$), and uses $3$ eigenfunctions ($M = 3$). We consider the case where we have $40$, $80$, and $160$ observed functional observations, where each observation is uniformly observed at $100$ time points. We then simulated 50 data sets for each of the three cases ($N = 40, 80, 160$). To measure how well we recovered the the mean, covariance, and cross-covariance functions, we estimate the integrated relative mean square error (R-MISE), where
$\text{R-MISE} = \frac{\int \{f(t) - \hat{f}(t)\}^2 \text{d}t}{\int f(t)^2 \text{d}t} \times 100\%$ and $\hat{f}$ is the estimated posterior median of the function $f$.
 To measure how well we recovered the allocation parameters, $Z_{ik}$, we calculated the root-mean-square error (RMSE). Section B.1 of the Supplementary Materials goes into more detail of how the model parameters were drawn, how estimates of all quantities of interest were calculated in this simulation, and includes visualizations of some of the recovered covariance structures.
\begin{figure}[htb]
    \centering
    \includegraphics[width=.9\textwidth]{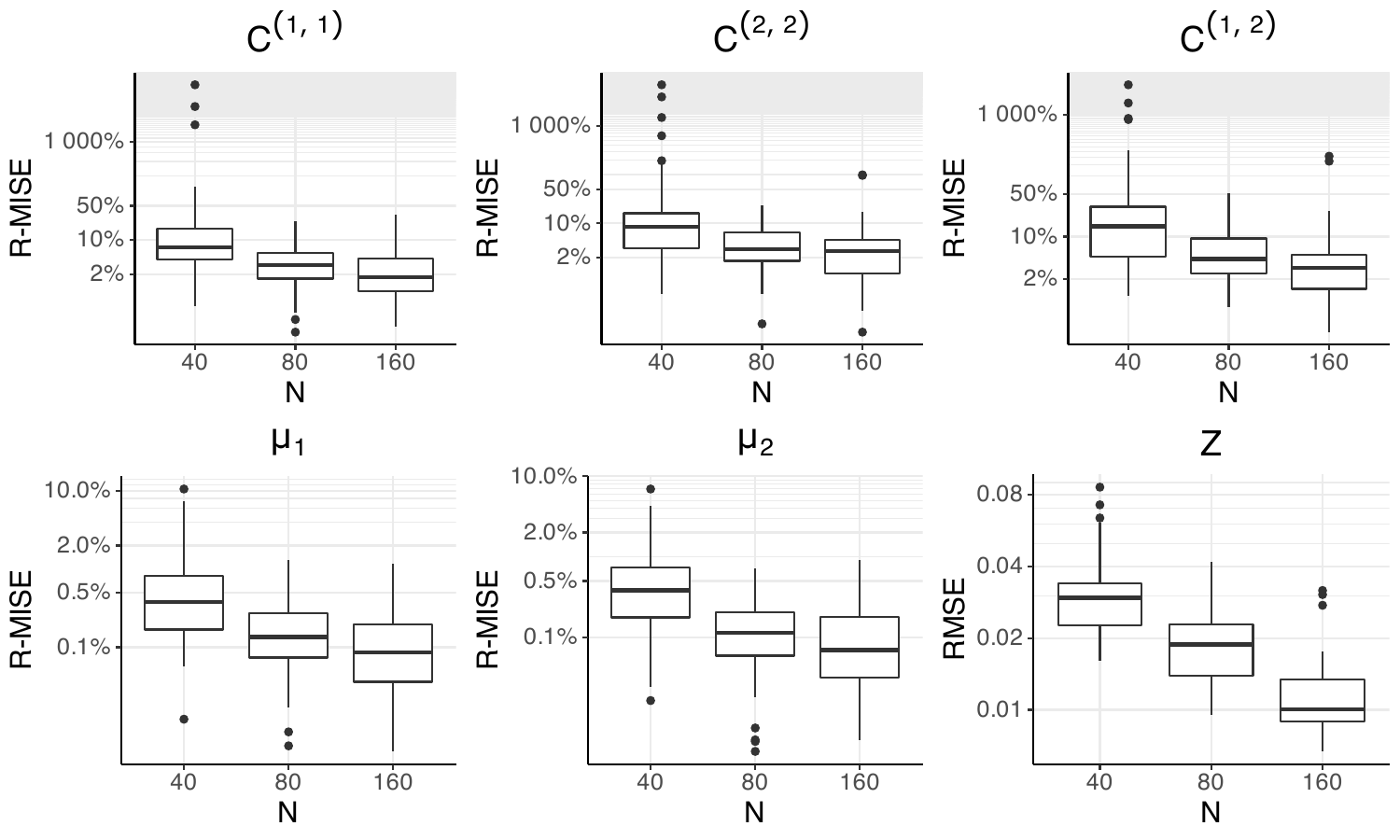}
    \caption{R-MISE values for the latent feature means and cross-covariances, as well as RMSE values for the allocation parameters, evaluated as sample size increases.}
    \label{fig:sim_1_RMISE}
\end{figure}

 From Figure \ref{fig:sim_1_RMISE}, we can see that we have a good recovery of the mean structure with as few as $40$ functional observations. While the R-MISE of the mean functions improves as we increase the number of functional observations ($N$), this improvement will likely have little practical impact. However, when looking at the recovery of the covariance and cross-covariance functions, we can see that the R-MISE noticeably decreases as more functional observations are added. As the recovery of the mean and covariance structures improves, the recovery of the allocation structure $(\mathbf{Z})$ improves. A supplementary simulation study examining how well our model can recover the mean, covariance, and allocation structures in a three-feature model ($K=3$) can be found in Section B.1 of the Supplementary Materials.

\subsection{Simulation Study 2}
\label{sim_study_2}

Choosing the number of latent features can be challenging, especially when no prior knowledge is available for this quantity. Information criteria, such as the AIC, BIC, or DIC, are often used to aid practitioners in the selection of a data-supported value for $K$. 
In this simulation, we evaluate how various types of information criteria, as well as simple heuristics such as the ``elbow-method'', perform in recovering the true number of latent features. To do this, we simulate 10 different data sets, each with 200 functional observations, from a mixed membership model with three functional features. We then calculate these information criteria on the 10 data sets for mixed membership models where $K = 2,3,4,5$. Additional information on how the simulation study was conducted, as well as definitions of the information criterion, can be found in Section B.2 of the Supplementary Materials. As specified, the optimal model should have the largest BIC, the smallest AIC, and the smallest DIC. 

\begin{figure}
    \centering
    \includegraphics[width=.9\textwidth]{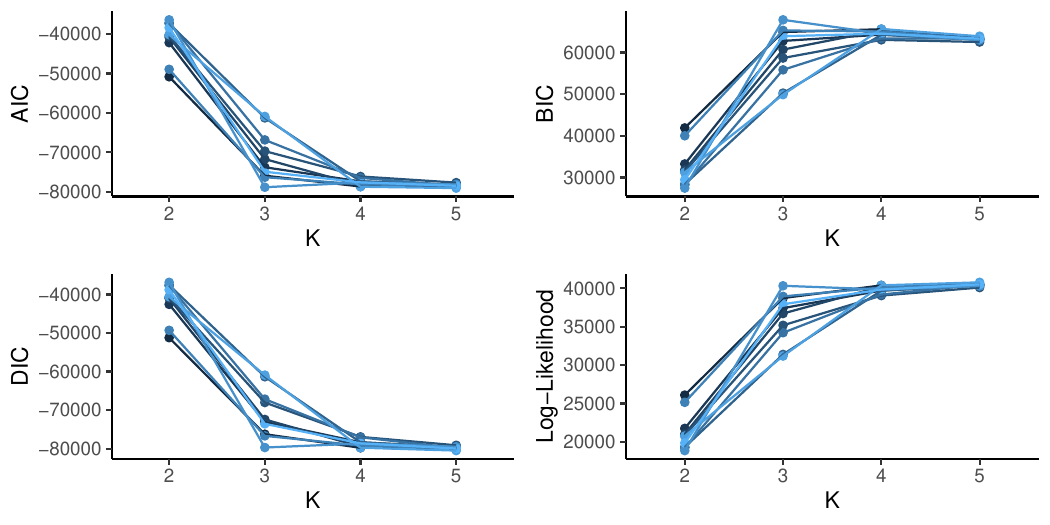}
    \caption{AIC, BIC, DIC, and the average log-likelihood evaluated for each of the 10 simulated data sets.}
    \label{fig:information_criterion}
\end{figure}

From the simulation results presented in Figure \ref{fig:information_criterion}, we can see that on average each information criterion overestimated the number of functional features in the model. While the three information criteria seem to greatly penalize models that do not have enough features, they do not seem to punish overfit models to a great enough extent.  As expected, the log-likelihood increases as we add more features; however, we can see that there is often an elbow at $K=3$ (Figure \ref{fig:information_criterion}). Using the ``elbow-method'' led to selecting the correct number of latent functional features 8 times out of 10, while BIC picked the correct number of latent functional features twice. DIC and AIC were found to be the least reliable information criteria, only choosing the correct number of functional features once. Thus, through empirical consideration, we suggest using the ``elbow-method'' along with the BIC to aid in the final selection of the number of latent features to be interpreted in analyses. Although formal considerations of model-selection consistency are out of scope for the current contribution, we maintain that some of these techniques are best interpreted in the context of data exploration, with a potential for great improvement in interpretation if semi-supervised considerations allow for an {\it a-priori} informed choice of $K$.

\begin{figure}
    \centering
    \includegraphics[width=.9\textwidth]{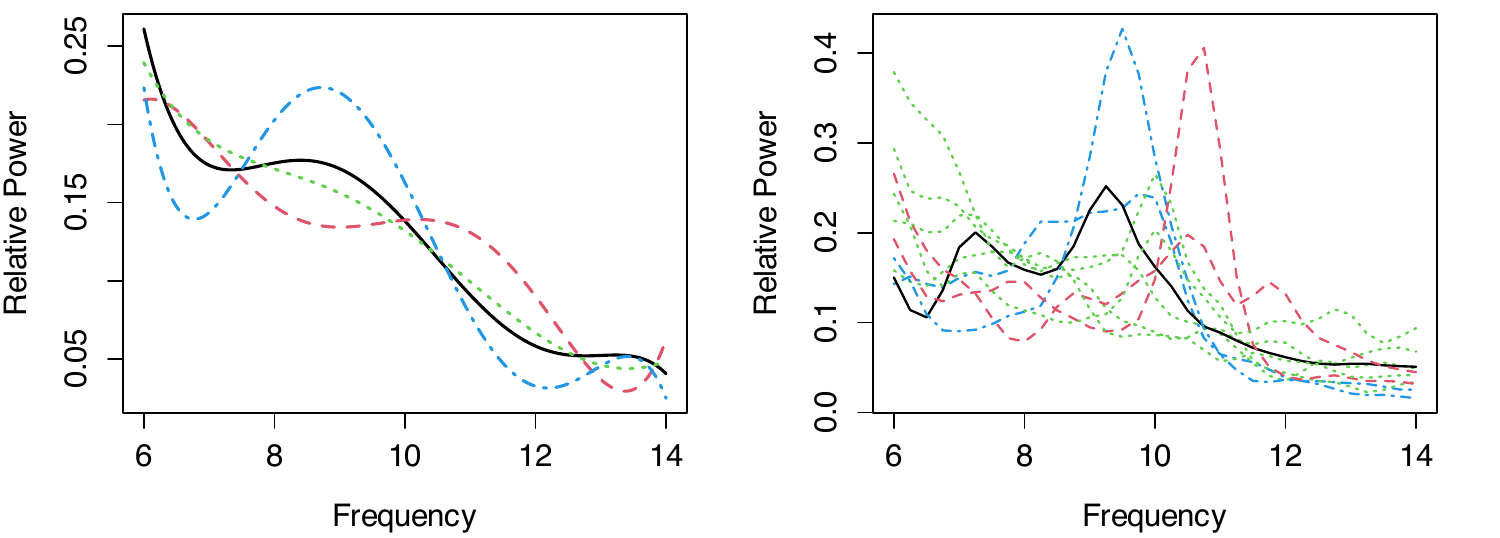}
    \caption{(Left Panel)
    Recovered means of model-based functional clustering with 4 clusters. (Right Panel) Alpha frequency patterns for a sample of EEG recordings from the T8 electrode of children (TD and ASD), styled by estimated cluster membership.}
    \label{fig:td_asd}
\end{figure}

\subsection{A Case Study of EEG in ASD}
\label{EEG_Case_Study}

Autism spectrum disorder (ASD) is a term used to describe individuals with a collection of social communication deficits and restricted or repetitive sensory-motor behaviors \citep{lord2018autism}. While once considered a very rare disorder with very specific symptoms, today the definition is more broad and is now thought of as a spectrum. Some individuals with ASD may have minor symptoms, while others may have very severe symptoms and require lifelong support. 
In this case study, we will use electroencephalogram (EEG) data that was previously analyzed by \citet{scheffler2019covariate} in the context of regression. The data set consists of EEG recordings of 39 typically developing (TD) children and 58 children with ASD between the ages of 2 and 12 years of age, which were analyzed in the frequency domain. 


\citet{scheffler2019covariate} observed that the data from the T8 electrode, corresponding to the right temporal region, were correlated with ASD diagnosis, so we will specifically use the data from the T8 electrode in our mixed membership model. We focus our analysis to the alpha band of frequencies (6 to 14 Hz), whose patterns at rest are thought to play a role in neural coordination and communication between distributed brain regions. As clinicians examine the sample paths for the two cohorts, shown in Figure \ref{fig:td_asd} (right panel), they are often interested in the location of a single prominent peak in the spectral density located within the alpha frequency band called the peak alpha frequency (PAF). This quantity has previously been associated with neural development in TD children \citep{Rodriguez2017}. \citet{scheffler2019covariate} found that as TD children age, the peak becomes more prominent and the PAF shifts to a higher frequency. Conversely, a discernible PAF pattern is attenuated for most children with ASD compared to their TD counterpart.  

A visual examination of the data in Figure \ref{fig:td_asd} (right panel) anticipates the potential inadequacy of cluster analysis in this application, as path-specific heterogeneity does not seem to define well-separated subpopulations. In fact, if we cluster our data using the model in \cite{pya2021fdamocca} with $K = 4$ (BIC selection), we find cluster means of dubious interpretability and poor separation of sample paths between clusters.

\begin{figure}[h]
    \centering
    \includegraphics[width=.9\textwidth]{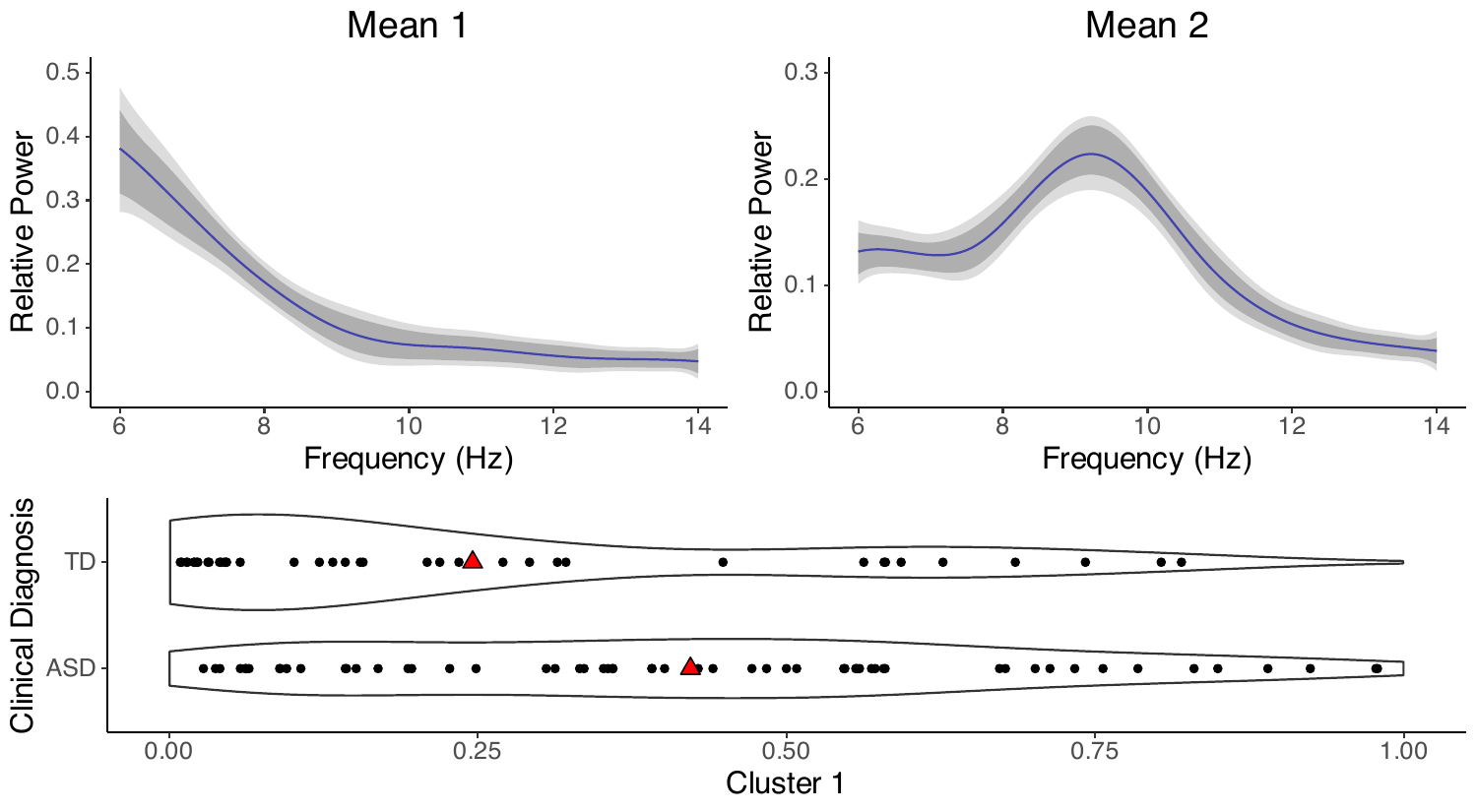}
    \caption{(Top Panel) Posterior median and 95\% credible intervals (pointwise CI in dark gray and simultaneous CI in light gray) of the mean function for each functional feature. (Bottom Panel) Posterior median of each individual's (feature-1)-membership, stratified by clinical cohort (cohort average (feature-1)-membership denoted by triangles).}
    \label{fig:mean}
\end{figure}

In contrast to classical clustering, we use a mixed membership model with only two functional features ($K = 2$), (AIC-BIC selection). 
We note that the enhanced flexibility of mixed membership models induces parsimony in the number of pure mixture components supported by the data. In particular, the mean function of the first feature, depicted in Figure \ref{fig:mean}, can be interpreted as $1/f$ noise or \textit{ pink noise}. This component noise is expected to be found in every individual to some extent, but we can see that the first feature has no discernible alpha peak. The mean function of the second feature captures a distinct alpha peak, which is typically observed in EEGs of older neurotypical individuals. These two features help to differentiate between periodic (alpha waves) and aperiodic ($1/f$ trend) neural activity patterns, which coexist in the EEG spectrum. In this context, a model of {\it uncertain membership} would be necessarily inadequate to describe the heterogeneity of the observed sample path, as we would not naturally think of subjects in our sample as belonging to one or the other cluster. Instead, assuming {\it mixed membership} between the two-feature processes is likely to represent a more realistic data generating mechanism, as we conceptualize periodic and aperiodic neural activity patterns to mix continuously.

Figure \ref{fig:mean} shows that TD children are highly likely to heavily load feature 2 (well-defined PAF), whereas ASD children exhibit a higher level of heterogeneity. The cohort average (feature-1) membership, indicated by the triangles in Figure \ref{fig:mean}, shows that the average (feature-1)-membership for children with ASD is 0.42 compared to 0.25 for TD children. It is important to note that the allocation parameters constitute compositional data, meaning that using a Euclidean metric to measure mean membership may be inappropriate in cases where we have more than two features. Therefore, the use of Fréchet means \citep{frechet1948elements} with a suitable metric for compositional data may be more appropriate to measure the centroids of the allocation parameters when we have more than two functional features.
In general, these loadings suggest that clear alpha peaks take longer to emerge in children with ASD, compared to their typically developing counterparts.


In general, our findings confirm related evidence in the scientific literature on developmental neuroimaging but offer a completely novel point of view in quantifying group membership as part of a spectrum. Additional details on how the study was conducted, a posterior predictive check for a subset of individuals, and an in-depth comparison between our method and alternative methods such as FPCA and functional clustering are reported in Section B.3 of the Supplementary Materials. An extended analysis of multi-channel EEG for the same case study is reported in Section B.4 of the Supplementary Materials, investigating how spatial patterns vary across the scalp.

\section{Discussion}
\label{discussion}
This manuscript introduces the concept of functional mixed membership to the broader field of functional data analysis. Mixed membership is defined as a natural generalization of the concept of \textit{uncertain membership}, which underlies the many approaches to functional clustering discussed in the literature.  Our discussion is carried out within the context of Bayesian analysis. In this paper, a coherent and flexible sampling model is introduced by defining a mixed membership Gaussian process through projections on the linear subspace spanned by a suitable set of basis functions. Within this context, we leverage the multivariate KL formulation \citep{happ2018multivariate}, to define a model ensuring weak conditional posterior consistency. 

Our work is closely related to the approach introduced by \citet{heller2008statistical}, who extended the theory of mixed membership models in the Bayesian framework for 
multivariate exponential family data. For Gaussian marginal distributions, their representation implies multivariate normal observations, with the natural parameters modeled as convex combinations of the individual cluster's natural parameters. Although intuitively appealing, this idea has important drawbacks. 
Crucially, the differential entropy 
of observations in multiple clusters is constrained to be smaller than the minimum differential entropy across clusters, which may not be realistic in many sampling scenarios.

Information criteria are often used to aid the choice of $K$ in the context of mixture models.  However, in Section \ref{sim_study_2}, we saw that they often overestimated the number of features in our model. In simulations, we observed that using the ``elbow-method'' could lead to the selection of the correct number of features with good frequency. The literature has discussed non-parametric approaches to feature allocation models by using, for example, the Indian Buffet Processes \citep{Griffiths2011}, but little is known about operating characteristics of these proposed procedures and little has been discussed about the ensuing need to carry out statistical inference across changing dimensions. \citet{rousseau2011asymptotic}, as well as \citet{nguyen2013convergence}, proved that under certain conditions, overfitted mixture models have a posterior distribution that converges to a set of parameters where the only components with positive weight will be the ``true'' parameters. 
 However, by allowing the membership parameters to be a continuous random variable, we introduce a stronger type of nonidentifiability, which led to the \textit{rescaling problem} discussed in Section \ref{section: FPMP}. Therefore, more work is needed on the posterior convergence of overfitted models for mixed membership models. 


\section{Supplementary Materials}

\begin{description}
    \item[Supplementary Materials:] Contains proofs for all lemmas in the manuscript, as well as additional details for the case studies and simulation studies (BFMMM\_Supplement.pdf).
    \item[\textsc{BayesFMMM}:] The R package to fit functional mixed membership models can be found on GitHub at \url{https://github.com/ndmarco/BayesFMMM}.
    \item[Simulation Studies and Case Studies:] The R scripts for the case studies and simulation studies can be found on GitHub at \url{https://github.com/ndmarco/BFMMM_Functional_Sims}. Unfortunately, the data for the EEG case study cannot be made publicly available.
\end{description}

\bibliographystyle{rss}
\bibliography{BFPMM}

\end{document}


\def\spacingset#1{\renewcommand{\baselinestretch}%
{#1}\small\normalsize} \spacingset{1}


\if1\blind
{
  \title{\bf Web-based supporting materials for Functional Mixed Membership Models}
  \author{Nicholas Marco\thanks{
    The authors gratefully acknowledge \textit{please remember to list all relevant funding sources in the unblinded version}}\hspace{.2cm}\\
    Department of Biostatistics, University of California,
Los Angeles, USA.\\
    and \\
    Damla \c{S}ent\"{u}rk \\
    Department of Biostatistics, University of California,
Los Angeles, USA. \\
    and \\
    Shafali Jeste \\
    Division of Neurology and Neurological Institute, Children’s Hospital Los Angeles, Los Angeles, USA.\\
    and \\
    Charlotte DiStefano \\
    Division of Psychiatry, Children’s Hospital Los Angeles, Los Angeles, USA.\\
    and \\
    Abigail Dickinson \\
    Department of Psychiatry and Biobehavioral Sciences, University of California, Los Angeles, USA.\\
    and \\
    Donatello Telesca \\
    Department of Biostatistics, University of California,
Los Angeles, USA.\\
    }
  \maketitle
} \fi

\if0\blind
{
  \bigskip
  \bigskip
  \bigskip
  \begin{center}
    {\LARGE\bf Supplemental Materials for the Article:\\Functional Mixed Membership Models}
\end{center}
  \medskip
} \fi

\bigskip

\begin{abstract}
    Section A of the web-based supporting materials contains proofs of all of the lemmas found in the paper. Section B contains details on how the two simulation studies were conducted, as well as supplementary information on the two case studies. Section C contains details on the posterior distribution, computational strategies, and algorithms. Section D contains details on simulation-based posterior inference. Section E contains details on how to pick a set of basis functions when fitting a functional mixed membership model. Section F discusses how cross-covariance functions can affect the model flexibility as well as interpretation.
\end{abstract}
\noindent%
\vfill

\newpage
\spacingset{1.5} 

\appendix
\section{Proofs}
\subsection{Proof of Lemma 2.1}
\begin{proof}
We will first show that $\boldsymbol{\mathcal{S}}$ is a linear subspace of $L^2(\mathcal{T})$. Let $w_1, w_2 \in \boldsymbol{\mathcal{S}}$, and let $\alpha_1, \alpha_2 \in \mathbb{R}$. Since $\boldsymbol{\mathcal{S}}$ is the space spanned by the square-integrable basis functions $b_1, \dots b_P$ $\left(\boldsymbol{\mathcal{S}} = \left\{\sum_{p=1}^P a_pb_p: a_i \in \mathbb{R} \right\}\right)$, we can write $w_1 = \sum_{p=1}^P\delta_p b_p$ and $w_2 = \sum_{p=1}^P\beta_pb_p$ for some $\delta_p, \beta_p \in \mathbb{R}$. Therefore we have that 
$$\alpha_1 w_1 + \alpha_2 w_2 = \alpha_1\left(\sum_{p=1}^P\delta_p b_p\right) + \alpha_2 \left(\sum_{p=1}^P\beta_pb_p \right).$$
Letting $\gamma_p = \alpha_1 \delta_p + \alpha_2 \beta_p$, we have that
$$\alpha_1 w_1 + \alpha_2 w_2 = \sum_{p=1}^P \gamma_p b_p \in \boldsymbol{\mathcal{S}}.$$
Therefore, by definition, we know that $\boldsymbol{\mathcal{S}}$ is a linear subspace of $L^2(\mathcal{T})$. Next, we will show that $\boldsymbol{\mathcal{S}}$ is a closed linear subspace. Let $f_n$ be a Cauchy sequence in $\boldsymbol{\mathcal{S}}$. Thus by definition, for some $\epsilon > 0$, there exists a $m \in \mathbb{N}$ such that for $i,j > m$ we have 
\begin{equation}
    \|f_i - f_j\|_{\boldsymbol{\mathcal{S}}} < \epsilon.
    \label{def_cauchy}
\end{equation}
Since $f_i, f_j \in \boldsymbol{\mathcal{S}}$, we know that $f_i, f_j \in \text{span}\{b_1, \dots, b_P\}$. Thus, using the Gram-Schmidt process, we know that there exists an orthonormal set of functions such that $\text{span}\left\{b_1, \dots, b_P\right\}$ $= \text{span}\left\{\tilde{b}_1, \dots, \tilde{b}_P\right\}$. Thus, we can expand $f_i$ and $f_j$ so that $f_i = \sum_{p=1}^P \alpha_{ip}\tilde{b}_p$ and $f_j = \sum_{p=1}^P\alpha_{jp}\tilde{b}_p$. Thus we can rewrite Equation \ref{def_cauchy} as
\begin{align}
    \nonumber
    \|f_i - f_j\|_{\boldsymbol{\mathcal{S}}} & = \left(\left\langle \sum_{p=1}^P (\alpha_{ip} - \alpha_{jp})\tilde{b}_p, \sum_{p=1}^P (\alpha_{ip} - \alpha_{jp})\tilde{b}_p \right\rangle \right)^{1/2} \\
    \nonumber & = \left(\sum_{p=1}^P\left\langle (\alpha_{ip} - \alpha_{jp})\tilde{b}_p, (\alpha_{ip} - \alpha_{jp})\tilde{b}_p \right\rangle \right)^{1/2} \\
    \nonumber & = \left( \sum_{p=1}^P\left|\left|(\alpha_{ip} - \alpha_{jp})\tilde{b}_p\right|\right|_{\boldsymbol{\mathcal{S}}} \right)^{1/2}\\
    \nonumber & = \left( \sum_{p=1}^P\int_{\mathcal{T}}\left((\alpha_{ip} - \alpha_{jp})\tilde{b}_p(t)\right)^2\text{d}t \right)^{1/2} \\
     & = \left( \sum_{p=1}^P(\alpha_{ip} - \alpha_{jp})^2\int_{\mathcal{T}}\tilde{b}_p(t)^2\text{d}t \right)^{1/2}.
     \label{norm_fifj}
\end{align}
Since $\tilde{b}_p(t)$ are orthonormal, we know that $\int_{\mathcal{T}}\tilde{b}_p(t)^2\text{d}t =1 $. Thus from Equations \ref{def_cauchy} and \ref{norm_fifj}, for $i, j > m$, we have that
\begin{align}
    \nonumber
    \epsilon & > \|f_i - f_j\|_{\boldsymbol{\mathcal{S}}}  \\
     \nonumber &= \left( \sum_{p=1}^P(\alpha_{ip} - \alpha_{jp})^2\int_{\mathcal{T}}\tilde{b}_p(t)^2\text{d}t \right)^{1/2}\\
      \nonumber & = \left(\sum_{p=1}^P(\alpha_{ip} - \alpha_{jp})^2 \right)^{1/2}.\\
\end{align}
Thus we can see that the sequence $\alpha_{ip}$ is a Cauchy sequence. Since the Euclidean space is a complete metric space, there exists $\alpha_p \in \mathbb{R}$ such that $\alpha_{ip} \rightarrow \alpha_p$. Letting $f = \sum_{p=1}^P \alpha_p \tilde{b}_p(t)$, we have 
\begin{align}
    \nonumber
    \|f_i - f\| & = \left(\sum_{p=1}^P(\alpha_{ip} - \alpha)^2\int_{\mathcal{T}}\tilde{b}_p(t)^2\text{d}t \right)^{1/2}\\
    \nonumber
    & = \left(\sum_{p=1}^P(\alpha_{ip} - \alpha_{p})^2 \right)^{1/2}\\
    &  = \sum_{p=1}^P\left| \left|\alpha_{ip} - \alpha_{p} \right| \right|,
    \label{convergent_f}
\end{align}
for all $i,j > m$. By the definition of $\alpha_{ip} \rightarrow \alpha_{p}$, we know that for $\epsilon_2 = \frac{\epsilon_1}{P}$, there exists a $m_1 \in \mathbb{N}$, such that for all $i > m_1$ we have $\left| \left|\alpha_{ip} - \alpha_{p} \right| \right| < \epsilon_1$ for $p = 1, \dots, P$. Thus, from Equation \ref{convergent_f}, we see that for all $i > m_1$, we have
$$\|f_i - f\| < \epsilon_2.$$
Thus by definition, we have that the Cauchy sequence is convergent and that $\boldsymbol{\mathcal{S}}$ is a closed linear subspace.
\end{proof}

\subsection{Proof of Lemma 2.2}
 \begin{proof}
 We will start by fixing $\epsilon > 0$. Notice that since $b_j$ are uniformly continuous functions and $\mathcal{T}$ is a closed and bounded domain, we know that $b_j$ is bounded. Thus, let $R$ be such that $|b_j(s)| < R$ for $j = 1, \dots, P$ and any $s \in \mathcal{T}$. Let $\tilde{\epsilon} := \frac{\epsilon}{P^2 RC_{max}}$, where $C_{max}$ is defined in (2) of Lemma 2.2. Since $b_1, \dots,  b_k$ are uniformly continuous we have that there exists $\delta_i >0$ such that for all $t, t_* \in \mathcal{T}$, we have
 \begin{equation}
     \|t - t_*\| < \delta_i \implies |b_i(t) - b_i(t_*)| < \tilde{\epsilon},
     \label{def_unif_cont}
 \end{equation}
 for $i = 1, \dots, P$. Define $\tilde{\delta} = \min_{i}\delta_i$. Thus from Equation 9 in the main text, if $\|t - t_*\| < \tilde{\delta}$, then we have
 \begin{align}
    \nonumber
     \left|C^{(i,j)}(s,t) - C^{(i,j)}(s,t_*)\right| & = \left|\mathbf{B}'(s)\text{Cov}\left(\boldsymbol{\theta}_i, \boldsymbol{\theta}_j\right)\left[\mathbf{B}(t)- \mathbf{B}(t_*)\right]\right|\\
     \nonumber
     & = \left|\sum_{k=1}^P\sum_{l=1}^P b_k(s) \text{Cov}\left(\theta_{(i,k)}, \theta_{(j,l)} \right)\left[b_l(t) - b_l(t_*)\right]\right|\\
     \nonumber 
     & \le \left|\sum_{k=1}^P\sum_{l=1}^P b_k(s) C_{max}\left[b_l(t) - b_l(t_*)\right]\right|\\
     \nonumber 
     & \le \sum_{k=1}^P\sum_{l=1}^P \left|b_k(s) C_{max} \left[b_l(t) - b_l(t_*)\right]\right|\\ 
     \nonumber
     & = \sum_{k=1}^P\sum_{l=1}^P \left|b_k(s) C_{max} \right|\left|b_l(t) - b_l(t_*)\right| .
 \end{align}
 From Equation \ref{def_unif_cont}, we have that
 \begin{align}
    \nonumber
     \left|C^{(i,j)}(s,t) - C^{(i,j)}(s,t_*)\right| & < \sum_{k=1}^P\sum_{l=1}^P \left|b_k(s) C_{max} \right|\tilde{\epsilon} \\
     \nonumber
     &  \le \sum_{k=1}^P\sum_{l=1}^P R C_{max} \tilde{\epsilon}\\
     \nonumber 
     & = \epsilon.
 \end{align}
Thus we have that for any $\epsilon > 0$, there exists a $\tilde{\delta} >0$, such that for any $t,t_*,s \in \mathcal{T}$ and $1 \le i \le j \le K$, we have
\begin{equation}\|t - t_*\| < \tilde{\delta} \implies \left|C^{(i,j)}(s,t) - C^{(i,j)}(s,t_*)\right| < \epsilon.
\label{uniform_cont}
\end{equation}
Consider $\boldsymbol{\mathcal{B}}_Z := \{\mathbf{f} \in \boldsymbol{\mathcal{H}} : \|\mathbf{f}\| < Z\}$ for some $Z \in \mathbb{R}^+$. We will show that the family of functions $\boldsymbol{\mathcal{K}}\mathbf{f}_{\boldsymbol{\mathcal{B}}_Z} := \left\{\boldsymbol{\mathcal{K}}\mathbf{f}: \mathbf{f} \in \boldsymbol{\mathcal{B}}_Z \right\}$ is an equicontinuous set of functions. We will fix $\epsilon_1 > 0$.
Letting $\mathbf{f} \in \boldsymbol{\mathcal{B}}_Z$ and $t^{(i)}, t_*^{(i)} \in \mathcal{T}$ such that $\|t^{(i)} - t_*^{(i)}\| < \tilde{\delta}$, we have from Equation 7 of the main text that
\begin{align}
    \nonumber
    \left|\left(\boldsymbol{\mathcal{K}}\mathbf{f}\right)^{(i)}(\mathbf{t}) - \left(\boldsymbol{\mathcal{K}}\mathbf{f}\right)^{(i)}(\mathbf{t}_*)\right| & = \left| \sum_{k=1}^K \int_{\mathcal{T}} C^{(k,i)}\left(s,t^{(i)}\right)f^{(k)}(s)  - C^{(k,i))}\left(s,t_*^{(i)}\right)f^{(k)}(s) \text{d}s \right|\\
    \nonumber
    & \le  \sum_{k=1}^K \int_{\mathcal{T}} \left|C^{(k,i)}\left(s,t^{(i)}\right)f^{(k)}(s)  - C^{(k,i)}\left(s,t_*^{(i)}\right)f^{(k)}(s)\right| \text{d}s \\
    & =  \sum_{k=1}^K \int_{\mathcal{T}} \left|C^{(k,i)}\left(s,t^{(i)}\right)  - C^{(k,i)}\left(s,t_*^{(i)}\right)\right|\left|f^{(k)}(s)\right| \text{d}s.
    \label{diff_cov_op}
\end{align}
Thus from Equation \ref{uniform_cont} we have that $\left|\left(C^{(k,i)}\left(s,t^{(i)}\right) - C^{(k,i)}\left(s,t_*^{(i)}\right)\right)\right| < \epsilon$. Notice that since $\mathbf{f} \in \mathcal{H}$, we know that $f^{(k)}(s)$ can be written as $f^{(k)}(s) = \sum_{i=1}^P a_ib_i(s)$. Since the sum of uniformly continuous functions is also a uniformly continuous function, we know that $f^{(k)}$ is uniformly continuous. Therefore, since $\mathcal{T}$ is a closed and bounded domain, we know that $f^{(k)}$ is bounded. Let $M_1$ be such that $|f^{(k)}| < M_1$. Thus we can write Equation \ref{diff_cov_op} as
\begin{align}
    \nonumber
    \left|\left(\boldsymbol{\mathcal{K}}\mathbf{f}\right)^{(i)}(\mathbf{t}) - \left(\boldsymbol{\mathcal{K}}\mathbf{f}\right)^{(i)}(\mathbf{t}_*)\right| & < \sum_{k=1}^K \int_{\mathcal{T}} \epsilon M_1 \text{d}s \\
    \nonumber
    & = \epsilon M_1 \sum_{k=1}^K \int_{\mathcal{T}} 1\text{d}s.
\end{align}
Since $\mathcal{T}$ is a compact subset of $\mathbb{R}^d$, by the Bolzano-Weierstrass theorem we know that $\mathcal{T}$ is closed and bounded. Therefore, let $B$ be such that $\int_{\mathcal{T}} 1 \text{d}t = B$. Thus, for $i = 1, \dots, K$, we have
\begin{align}
    \nonumber
    \left|\left(\boldsymbol{\mathcal{K}}\mathbf{f}\right)^{(i)}(\mathbf{t}) - \left(\boldsymbol{\mathcal{K}}\mathbf{f}\right)^{(i)}(\mathbf{t}_*)\right| & <  \epsilon M_1 KB.
\end{align}
Since $$\left|\left|\left(\boldsymbol{\mathcal{K}}\mathbf{f}\right)(\mathbf{t}), \left(\boldsymbol{\mathcal{K}}\mathbf{f}\right)(\mathbf{t}_*)\right|\right| = \left(\sum_{i=1}^K \left|\left(\boldsymbol{\mathcal{K}}\mathbf{f}\right)^{(i)}(\mathbf{t}) - \left(\boldsymbol{\mathcal{K}}\mathbf{f}\right)^{(i)}(\mathbf{t}_*)\right|^2\right)^{1/2},$$
we know that
$$\left|\left|\left(\boldsymbol{\mathcal{K}}\mathbf{f}\right)(\mathbf{t}), \left(\boldsymbol{\mathcal{K}}\mathbf{f}\right)(\mathbf{t}_*)\right|\right| < \epsilon M_1K^{3/2}B.$$
If we let $\epsilon = \frac{\epsilon_1}{M_1K^{3/2}B}$ $\left(\tilde{\epsilon} = \frac{\epsilon_1}{M_1 K^{3/2} BP^2 RC_{max}}\right)$, then we have that $\left|\left|\left(\boldsymbol{\mathcal{K}}\mathbf{f}\right)(\mathbf{t}), \left(\boldsymbol{\mathcal{K}}\mathbf{f}\right)(\mathbf{t}_*)\right|\right| < \epsilon_1$.Thus, from the assumption that $b_j$ are uniformly continuous (Equation \ref{def_unif_cont}), we know there exists a $\tilde{\delta}$ such that for $j = 1, \dots, K$, we have that 
\begin{equation}
    \|t - t_*\| < \tilde{\delta} \implies |b_j(t) - b_j(t_*)| < \frac{\epsilon_1}{M_1 K^{3/2} BP^2 RC_{max}}
    \label{pick_delta} \implies \left|\left|\left(\boldsymbol{\mathcal{K}}\mathbf{f}\right)(\mathbf{t}), \left(\boldsymbol{\mathcal{K}}\mathbf{f}\right)(\mathbf{t}_*)\right|\right| < \epsilon_1.
\end{equation}
Thus, by definition, we have proved that $\boldsymbol{\mathcal{K}}\mathbf{f}_{\boldsymbol{\mathcal{B}}_Z}$ is an equicontinuous set of functions. Next, we show that $\boldsymbol{\mathcal{K}}\mathbf{f}_{\boldsymbol{\mathcal{B}}_Z}$ is a family of uniformly bounded functions. If $t \in \mathcal{T}$, then we have
\begin{align}
    \nonumber
    |\left(\boldsymbol{\mathcal{K}}\mathbf{f}\right)^{(i)}(\mathbf{t})| & = \left|\sum_{k=1}^K \int_{\mathcal{T}} \mathbf{B}'(s)\text{Cov}\left(\boldsymbol{\theta}_k, \boldsymbol{\theta}_i\right)\mathbf{B}\left(t^{(i)}\right) f^{(k)}(s) \text{d}s\right| \\
    \nonumber
    & = \left|\sum_{k=1}^K \int_{\mathcal{T}} \sum_{p=1}^P\sum_{l=1}^P b_p(s) \text{Cov}\left(\theta_{(k,l)}, \theta_{(i,p)} \right)b_l\left(t^{(i)}\right) f^{(k)}(s) \text{d}s\right|\\
    \nonumber
    & = \left|\sum_{k=1}^K  \sum_{p=1}^P\sum_{l=1}^P  b_l\left(t^{(i)}\right) \text{Cov}\left(\theta_{(k,l)}, \theta_{(i,p)} \right)\int_{\mathcal{T}}b_p(s) f^{(k)}(s)\text{d}s\right| \\
    \nonumber
    & \le \left(\sum_{k=1}^K  \sum_{p=1}^P\sum_{l=1}^P  \left|b_l\left(t^{(i)}\right) \text{Cov}\left(\theta_{(k,l)}, \theta_{(i,p)} \right)\int_{\mathcal{T}}b_p(s) f^{(k)}(s)\text{d}s\right|\right)\\
    \nonumber
    & = \left(\sum_{k=1}^K  \sum_{p=1}^P\sum_{l=1}^P  \left|b_l\left(t^{(i)}\right) \right|\left|\text{Cov}\left(\theta_{(k,l)}, \theta_{(i,p)} \right)\right|\left|\int_{\mathcal{T}}b_p(s) f^{(k)}(s)\text{d}s\right|\right).
\end{align}
Using the $R$ defined such that $|b_j(s)| < R$ for all $s \in \mathcal{T}$, and condition (b), we have that
\begin{align}
    \nonumber
    |\left(\boldsymbol{\mathcal{K}}\mathbf{f}\right)^{(i)}(\mathbf{t})| & < \left(\sum_{k=1}^K  \sum_{p=1}^P\sum_{l=1}^P  RC_{max}\left|\int_{\mathcal{T}}b_p(s) f^{(k)}(s)\text{d}s\right|\right).
\end{align}
Using Hölder's Inequality, we have
\begin{align}
    \nonumber
    |\left(\boldsymbol{\mathcal{K}}\mathbf{f}\right)^{(i)}(\mathbf{t})|& < \left(\sum_{k=1}^K  \sum_{p=1}^P\sum_{l=1}^P  RC_{max}\left(\int_{\mathcal{T}}\left|b_p(s)\right|^2\text{d}s\right)^{1/2} \left(\int_{\mathcal{T}}\left|f^{(k)}(s)\right|^2\text{d}s\right)^{1/2}\right)\\
    \nonumber
    & < \left(\sum_{k=1}^K  \sum_{p=1}^P\sum_{l=1}^P  RC_{max}\left(\int_{\mathcal{T}}R^2\text{d}s\right)^{1/2} \left(\int_{\mathcal{T}}\left|f^{(k)}(s)\right|^2\text{d}s\right)^{1/2}\right).
\end{align}
Since $\boldsymbol{\mathcal{K}}$ is the direct sum of Hilbert spaces, we know that if $\mathbf{f} \in \boldsymbol{\mathcal{B}}_Z$, then $\|f^{(j)}\| < Z$ for all $j$, since $\|\mathbf{f}\| = \sum_{j=1}^K \|f^{(j)}\|$. Since $\|f^{(j)}\| = \int_{\mathcal{T}} f^{(j)}(s)^2\text{d}s = \int_{\mathcal{T}} |f^{(j)}(s)|^2\text{d}s$, we know that $\int_{\mathcal{T}} |f^{(k)}(s)|^2\text{d}s < Z$. Thus, we have
\begin{align}
    \nonumber
    |\left(\boldsymbol{\mathcal{K}}\mathbf{f}\right)^{(i)}(\mathbf{t})| & < \left(\sum_{k=1}^K  \sum_{p=1}^P\sum_{l=1}^P  RC_{max}\left(R^2B\right)^{1/2} \left(Z\right)^{1/2}\right)\\
    & =KP^2R^2C_{max}B^{1/2}Z^{1/2} < \infty.
    \label{bound_family_cov}
\end{align}
Since $\|\boldsymbol{\mathcal{K}}\mathbf{f}\|_{\boldsymbol{\mathcal{H}}}^2 = \sum_{i=1}^K |\left(\boldsymbol{\mathcal{K}}\mathbf{f}\right)^{(i)}(\mathbf{t})|^2$, we have that
$$\|\boldsymbol{\mathcal{K}}\mathbf{f}\|_{\boldsymbol{\mathcal{H}}}^2 < K^{3/2}P^2R^2C_{max}B^{1/2}Z^{1/2} < \infty.$$
Thus we know that $\boldsymbol{\mathcal{K}}\mathbf{f}_{\boldsymbol{\mathcal{B}}_Z}$ is a bounded equicontinuous set of functions.  Therefore, using Ascoli's Theorem (\cite{reed1972methods}, page 30), we know that for every sequence $\mathbf{f}_n \in \boldsymbol{\mathcal{B}}_Z$, the set $\boldsymbol{\mathcal{K}}\mathbf{f}_{\boldsymbol{\mathcal{B}}_Z}$ has a subsequence that converges (\cite{reed1972methods}, page 199). Therefore, $\boldsymbol{\mathcal{K}}\mathbf{f}_{\boldsymbol{\mathcal{B}}_Z}$ is precompact, which implies that $\boldsymbol{\mathcal{K}}$ is compact.

We will now show that $\boldsymbol{\mathcal{K}}$ is a bounded operator. Let $\mathbf{f} \in \boldsymbol{\mathcal{B}}_Z$. Thus, we have
$$\|\boldsymbol{\mathcal{K}}\mathbf{f}\|_{\boldsymbol{\mathcal{H}}}^2 = \sum_{i=1}^K\int_{\mathcal{T}}|\left(\boldsymbol{\mathcal{K}}\mathbf{f}\right)^{(i)}(\mathbf{t})|^2 \text{d}t.$$
From Equation \ref{bound_family_cov}, we have that 
\begin{align}
    \nonumber
    \|\boldsymbol{\mathcal{K}}\mathbf{f}\|_{\boldsymbol{\mathcal{H}}}^2 & < \sum_{i=1}^K\int_{\mathcal{T}}\left(KP^2R^2C_{max}B^{1/2}Z^{1/2}\right)^2 \text{d}{t} \\
    \nonumber
    & = K^3P^4R^4C_{max}^2BZ\int_{\mathcal{T}}\text{d}{t}.
\end{align}
 Using the $B$ defined above ($\int_{\mathcal{T}} 1 \text{d}t < B$), we have that
\begin{align}
    \nonumber
    \|\boldsymbol{\mathcal{K}}\mathbf{f}\|_{\boldsymbol{\mathcal{H}}}^2  &= K^3P^4R^4C_{max}^2BZ \int_\mathcal{T}1\text{d}t \\
    \nonumber
    & < K^3P^4R^4C_{max}^2B^2Z < \infty
\end{align}
Therefore we have that $\boldsymbol{\mathcal{K}}$ is a bounded linear operator. Therefore, if conditions (a) and (b) are met, then $\boldsymbol{\mathcal{K}}$ is a compact and bounded linear operator.
\end{proof}

\subsection{Proof of Lemma 3.1}
We will start be explicitly defining the functions $\Lambda_i(\boldsymbol{\omega}_0 ,\boldsymbol{\omega})$, $K_i(\boldsymbol{\omega}_0, \boldsymbol{\omega})$, and $V_i(\boldsymbol{\omega}_0, \boldsymbol{\omega})$. Thus we have
 \begin{align}
     \nonumber
    \Lambda_i(\boldsymbol{\omega}_0 ,\boldsymbol{\omega}) = & \text{log}\left(\frac{\left|\left(\boldsymbol{\Sigma}_i\right)_0\right|^{-1/2} \text{exp}\left\{-\frac{1}{2}\left(\mathbf{Y}_i - \left(\boldsymbol{\mu}_i\right)_0\right)'\left(\boldsymbol{\Sigma}_i\right)^{-1}_0\left(\mathbf{Y}_i - \left(\boldsymbol{\mu}_i\right)_0\right)\right\}}{\left|\boldsymbol{\Sigma}_i\right|^{-1/2} \text{exp}\left\{-\frac{1}{2}\left(\mathbf{Y}_i - \boldsymbol{\mu}_i\right)'\left(\boldsymbol{\Sigma}_i\right)^{-1}\left(\mathbf{Y}_i - \boldsymbol{\mu}_i\right)\right\}} \right)\\
    \nonumber
    =& -\frac{1}{2}\left[\text{log}\left(\left|\left(\boldsymbol{\Sigma}_i\right)_0 \right|\right) - \text{log}\left(\left|\boldsymbol{\Sigma}_i \right|\right) \right]  \\
    \nonumber
    & -\frac{1}{2} \left[\left(\mathbf{Y}_i - \left(\boldsymbol{\mu}_i\right)_0\right)'\left(\boldsymbol{\Sigma}_i\right)^{-1}_0\left(\mathbf{Y}_i - \left(\boldsymbol{\mu}_i\right)_0\right) - \left(\mathbf{Y}_i - \boldsymbol{\mu}_i\right)'\left(\boldsymbol{\Sigma}_i\right)^{-1}\left(\mathbf{Y}_i - \boldsymbol{\mu}_i\right) \right]\\
    \nonumber
    =& -\frac{1}{2}\left[\sum_{l=1}^{n_i}\text{log}\left((d_{il})_0 + \sigma^2_0\right) - \text{log}\left(d_{il} + \sigma^2\right) \right]  \\
    & -\frac{1}{2} \left[\left(\mathbf{Y}_i - \left(\boldsymbol{\mu}_i\right)_0\right)'\left(\boldsymbol{\Sigma}_i\right)^{-1}_0\left(\mathbf{Y}_i - \left(\boldsymbol{\mu}_i\right)_0\right) - \left(\mathbf{Y}_i - \boldsymbol{\mu}_i\right)'\left(\boldsymbol{\Sigma}_i\right)^{-1}\left(\mathbf{Y}_i - \boldsymbol{\mu}_i\right) \right]
 \end{align}
 \begin{align}
     \nonumber
     K_i(\boldsymbol{\omega}_0, \boldsymbol{\omega}) =& -\frac{1}{2}\left[\sum_{l=1}^{n_i}\text{log}\left((d_{il})_0 + \sigma^2_0\right) - \text{log}\left(d_{il} + \sigma^2\right) \right] \\
     \nonumber
     & -\frac{1}{2} \mathbb{E}_{\boldsymbol{\omega}_0}\left[\left(\mathbf{Y}_i - \left(\boldsymbol{\mu}_i\right)_0\right)'\left(\boldsymbol{\Sigma}_i\right)^{-1}_0\left(\mathbf{Y}_i - \left(\boldsymbol{\mu}_i\right)_0\right) -\left(\mathbf{Y}_i - \boldsymbol{\mu}_i\right)'\left(\boldsymbol{\Sigma}_i\right)^{-1}\left(\mathbf{Y}_i - \boldsymbol{\mu}_i\right) \right]\\
     \nonumber
      =& -\frac{1}{2}\left[\sum_{l=1}^{n_i}\text{log}\left((d_{il})_0 + \sigma^2_0\right) - \text{log}\left(d_{il} + \sigma^2\right) \right] \\
     & -\frac{1}{2} \left[n_i - \left(\text{tr}\left(\boldsymbol{\Sigma}_i^{-1}\left(\boldsymbol{\Sigma}_i\right)_0\right) + \left(\left(\boldsymbol{\mu}_i\right)_0- \boldsymbol{\mu}_i\right)'\left(\boldsymbol{\Sigma}_i\right)^{-1}\left(\left(\boldsymbol{\mu}_i\right)_0 - \boldsymbol{\mu}_i\right) \right) \right]
 \end{align}
 \begin{align}
     \nonumber
     V_i(\boldsymbol{\omega}_0, \boldsymbol{\omega}) = & \frac{1}{4}\text{Var}_{\boldsymbol{\omega}_0}\left[\left(\mathbf{Y}_i - \left(\boldsymbol{\mu}_i\right)_0\right)'\left(\boldsymbol{\Sigma}_i\right)^{-1}_0\left(\mathbf{Y}_i - \left(\boldsymbol{\mu}_i\right)_0\right) - \left(\mathbf{Y}_i - \boldsymbol{\mu}_i\right)'\left(\boldsymbol{\Sigma}_i\right)^{-1}\left(\mathbf{Y}_i - \boldsymbol{\mu}_i\right) \right] \\
     \nonumber
     = & \frac{1}{4}\text{Var}_{\boldsymbol{\omega}_0}\left[ \mathbf{Y}_i'\left(\left(\boldsymbol{\Sigma}_i \right)_0^{-1} + \boldsymbol{\Sigma}_i^{-1} \right)\mathbf{Y}_i - 2\mathbf{Y}_i'\left(\left( \boldsymbol{\Sigma}_i \right)_0^{-1}\left(\boldsymbol{\mu}_i\right)_0 + \boldsymbol{\Sigma}_i^{-1}\boldsymbol{\mu}_i\right)\right]
 \end{align}
 Letting $\mathbf{M}_v = \left(\boldsymbol{\Sigma}_i \right)_0^{-1} + \boldsymbol{\Sigma}_i^{-1}$, and $\mathbf{m}_v = \left( \boldsymbol{\Sigma}_i \right)_0^{-1}\left(\boldsymbol{\mu}_i\right)_0 + \boldsymbol{\Sigma}_i^{-1}\boldsymbol{\mu}_i$, we have
 \begin{align}
     \nonumber
     V_i(\boldsymbol{\omega}_0, \boldsymbol{\omega}) = & \frac{1}{4}\text{Var}_{\boldsymbol{\omega}_0}\left[(\mathbf{Y}_i - \mathbf{M}^{-1}_v\mathbf{m}_v)'\mathbf{M}_v(\mathbf{Y}_i - \mathbf{M}^{-1}_v\mathbf{m}_v) \right]\\
     \nonumber
     = & \frac{1}{4}\left[2\text{tr}\left(\mathbf{M}_v\left(\boldsymbol{\Sigma}_i \right)_0\mathbf{M}_v\left(\boldsymbol{\Sigma}_i \right)_0\right) + 4\left(\left(\boldsymbol{\mu}_i\right)_0 - \mathbf{M}^{-1}_v\mathbf{m}_v\right)' \left(\boldsymbol{\Sigma}_i \right)_0\left(\left(\boldsymbol{\mu}_i\right)_0 - \mathbf{M}^{-1}_v\mathbf{m}_v\right) \right]\\
     \nonumber
     = & \frac{1}{2}\left[n_i + 2 \text{tr}\left( \boldsymbol{\Sigma}_i^{-1}\left(\boldsymbol{\Sigma}_i\right)_0\right) + \text{tr}\left( \boldsymbol{\Sigma}_i^{-1}\left(\boldsymbol{\Sigma}_i\right)_0\boldsymbol{\Sigma}_i^{-1}\left(\boldsymbol{\Sigma}_i\right)_0\right)\right]\\
     & + \left(\left(\boldsymbol{\mu}_i\right)_0- \boldsymbol{\mu}_i\right)'\left(\boldsymbol{\Sigma}_i^{-1}\left(\boldsymbol{\Sigma}_i\right)_0\boldsymbol{\Sigma}_i^{-1}\right)\left(\left(\boldsymbol{\mu}_i\right)_0 - \boldsymbol{\mu}_i\right).
     \label{Var_theta}
 \end{align}
 Let $\boldsymbol{\Omega}_\epsilon(\boldsymbol{\omega}_0) = \{\boldsymbol{\omega}: K_i(\boldsymbol{\omega}_0, \boldsymbol{\omega}) < \epsilon \; \text{for all i}\}$ for some $\epsilon > 0$. We will assume that $\sigma^2_0 > 0$. Consider the set $\mathcal{B}(\boldsymbol{\omega}_0) = \{\boldsymbol{\omega}: \frac{1}{a}((d_{il})_0 + \sigma^2_0) \le d_{il} + \sigma^2 \le a((d_{il})_0 + \sigma^2_0 ), \|\left(\boldsymbol{\mu}_i\right)_0 - \boldsymbol{\mu}_i\| \le b\}$ for some $a,b \in \mathbb{R}$ such that $a > 1$ and $b > 0$. Thus for a fixed $\boldsymbol{\omega}_0 \in \boldsymbol{\Omega}$ and any $\boldsymbol{\omega} \in \mathcal{C}(\boldsymbol{\omega}_0, \epsilon) :=\mathcal{B}(\boldsymbol{\omega}_0) \cap \boldsymbol{\Omega}_\epsilon(\boldsymbol{\omega}_0)$, we can bound $V_i(\boldsymbol{\omega}_0, \boldsymbol{\omega})$. We will let $\lambda_r(\mathbf{A})$ denote the $r^{th}$ eigenvalue of the matrix $\mathbf{A}$, and $\lambda_{max}(\mathbf{A})$ denote the largest eigenvalue of $\mathbf{A}$. Thus we have 
$$\text{tr}\left( \boldsymbol{\Sigma}_i^{-1}\left(\boldsymbol{\Sigma}_i\right)_0\right) \le n_i \lambda_{max}\left( \boldsymbol{\Sigma}_i^{-1}\left(\boldsymbol{\Sigma}_i\right)_0\right)\le \frac{n_ia}{\sigma_0^2}\left(\max_l((d_{il})_0 + \sigma_0^2)\right)$$
     $$\text{tr}\left( \boldsymbol{\Sigma}_i^{-1}\left(\boldsymbol{\Sigma}_i\right)_0\boldsymbol{\Sigma}_i^{-1}\left(\boldsymbol{\Sigma}_i\right)_0\right) \le \text{tr}\left( \boldsymbol{\Sigma}_i^{-1}\left(\boldsymbol{\Sigma}_i\right)_0\right)^2 \le \left(\frac{n_ia}{\sigma_0^2}\left(\max_l((d_{il})_0 + \sigma_0^2)\right)\right)^2$$
     \begin{align}
         \nonumber \left(\left(\boldsymbol{\mu}_i\right)_0- \boldsymbol{\mu}_i\right)'\left(\boldsymbol{\Sigma}_i^{-1}\left(\boldsymbol{\Sigma}_i\right)_0\boldsymbol{\Sigma}_i^{-1}\right)\left(\left(\boldsymbol{\mu}_i\right)_0 - \boldsymbol{\mu}_i\right) & \le b^2\lambda_{max}\left(\boldsymbol{\Sigma}_i^{-1}\left(\boldsymbol{\Sigma}_i\right)_0\boldsymbol{\Sigma}_i^{-1} \right)\\
         \nonumber & \le \frac{a^2b^2}{\sigma_0^4}\max_l((d_{il})_0 + \sigma_0^2)
         \end{align}
Thus we can see that for any $\boldsymbol{\omega} \in \mathcal{C}(\boldsymbol{\omega}_0, \epsilon)$, 
 \begin{align}
    \nonumber
     V_i(\boldsymbol{\omega}_0, \boldsymbol{\omega}) \le & \frac{1}{2}\left[ n_i + 2 \left(\frac{n_ia}{\sigma_0^2}\left(\max_l((d_{il})_0 + \sigma_0^2)\right)\right) + \left(\frac{n_ia}{\sigma_0^2}\left(\max_l((d_{il})_0 + \sigma_0^2)\right)\right)^2\right]\\
     \nonumber
     & +\frac{a^2b^2}{\sigma_0^4}\max_l((d_{il})_0 + \sigma_0^2)\\ 
     \nonumber = &  M_{V}.
 \end{align}
If we can bound $\max_l\left((d_{il})_0 + \sigma_0\right)$, then we have that $V_i(\boldsymbol{\omega}_0, \boldsymbol{\omega})$ is bounded (since $n_i \le n_{max}$). Let $\|\cdot\|_F$ be the Frobenius norm. Using the triangle inequality, we have
\begin{align}
    \nonumber \|(\boldsymbol{\Sigma}_i)_0\|_{F} & \le \sum_{k=1}^K\sum_{j=1}^K\sum_{p=1}^{KP} Z_{ij}Z_{ik}\|\mathbf{S}'(\mathbf{t}_i)(\boldsymbol{\phi}_{kp})_0(\boldsymbol{\phi}_{jp})_0'\mathbf{S}(\mathbf{t}_i)\|_{F} + \sigma^2_0\|\mathbf{I}_{n_i}\|_F\\
    \nonumber & \le \sum_{k=1}^K\sum_{j=1}^K\sum_{p=1}^{KP} \|\mathbf{S}'(\mathbf{t}_i)(\boldsymbol{\phi}_{kp})_0(\boldsymbol{\phi}_{jp})_0'\mathbf{S}(\mathbf{t}_i)\|_{F}  + \sigma^2_0\|\mathbf{I}_{n_i}\|_F\\
    \nonumber &  \le \sum_{k=1}^K\sum_{j=1}^K\sum_{p=1}^{KP}  \sqrt{\text{tr}\left(\mathbf{S}'(\mathbf{t}_i)(\boldsymbol{\phi}_{jp})_0(\boldsymbol{\phi}_{kp})_0'\mathbf{S}(\mathbf{t}_i)\mathbf{S}'(\mathbf{t}_i)(\boldsymbol{\phi}_{kp})_0(\boldsymbol{\phi}_{jp})_0'\mathbf{S}(\mathbf{t}_i)\right)} + \sqrt{n_{max}}\sigma^2_0\\
    \nonumber &  = \sum_{k=1}^K\sum_{j=1}^K\sum_{p=1}^{KP}  \sqrt{\text{tr}\left((\boldsymbol{\phi}_{kp})_0'\mathbf{S}(\mathbf{t}_i)\mathbf{S}'(\mathbf{t}_i)(\boldsymbol{\phi}_{kp})_0(\boldsymbol{\phi}_{jp})_0'\mathbf{S}(\mathbf{t}_i)\mathbf{S}'(\mathbf{t}_i)(\boldsymbol{\phi}_{jp})_0\right)} + \sqrt{n_{max}}\sigma^2_0 \\
    \nonumber &  = \sum_{k=1}^K\sum_{j=1}^K\sum_{p=1}^{KP} \|\mathbf{S}'(\mathbf{t}_i)(\boldsymbol{\phi}_{jp})_0\|_2\|\mathbf{S}'(\mathbf{t}_i)(\boldsymbol{\phi}_{kp})_0\|_2 + \sqrt{n_{max}}\sigma^2_0\\
    \nonumber & = M_{\boldsymbol{\Sigma}_0} < \infty,
\end{align}
for all $i \in \mathbb{N}$. Therefore, we know that $\max_l\left((d_{il})_0 + \sigma_0\right) \le M_{\boldsymbol{\Sigma}_0}$, as the Frobenius norm is the square root of the sum of the squared eigenvalues for a square matrix. Therefore, we have for all $i \in \mathbb{N}$ and $\boldsymbol{\omega} \in \mathcal{C}(\boldsymbol{\omega}_0, \epsilon)$, we have $$\frac{V_i(\boldsymbol{\omega}_0, \boldsymbol{\omega})}{i^2} \le \frac{M_{V}}{i^2}.$$
Since $\sum_{i=1}^\infty \frac{1}{i^2} = \frac{\pi^2}{6}$, we have $\sum_{i=1}^\infty \frac{M_{V}}{i^2} = \frac{M_{V}\pi^2}{6} < \infty$. Thus we have 
 \begin{equation}
 \sum_{i = 1}^\infty \frac{V_i(\boldsymbol{\omega}_0, \boldsymbol{\omega})}{i^2}  < \infty.
     \label{finite_variance}
 \end{equation}
 We will next show that for $\boldsymbol{\omega}_0 \in \boldsymbol{\Omega}$ and $\epsilon > 0$, $\boldsymbol{\Pi}(\mathcal{C}(\boldsymbol{\omega}_0),\epsilon) > 0$. Fix $\boldsymbol{\omega}_0 \in \boldsymbol{\Omega}$. While $(\boldsymbol{\phi}_{jp})_0$ may not be identifiable (for any orthogonal matrix $\mathbf{H}$, $(\boldsymbol{\phi}_{jp})_0 \mathbf{H}\mathbf{H}' (\boldsymbol{\phi}_{kp})_0 = (\boldsymbol{\phi}_{jp})_0 '  (\boldsymbol{\phi}_{kp})_0$), let $(\boldsymbol{\phi}_{jp})_0$ be such that $\sum_{p=1}^{KP} (\boldsymbol{\phi}_{jp})_0 ' (\boldsymbol{\phi}_{kp})_0 = (\boldsymbol{\Sigma}_{jk})_0$. Thus we can define the following sets:
 \begin{align}
    \nonumber
     \boldsymbol{\Omega}_{\boldsymbol{\phi}_{jp}}  &= \left\{\boldsymbol{\phi}_{jp} :(\boldsymbol{\phi}_{jp})_0 \le \boldsymbol{\phi}_{jp} \le (\boldsymbol{\phi}_{jp})_0 + \epsilon_1\mathbf{1}_P \right\}\\
     \nonumber
     \boldsymbol{\Omega}_{\boldsymbol{\nu}_k}  &= \left\{\boldsymbol{\nu}_k : (\boldsymbol{\nu}_k)_0 \le \boldsymbol{\nu}_k \le (\boldsymbol{\nu}_k)_0 + \epsilon_2\mathbf{1}_P\right\} \\
     \nonumber
     \boldsymbol{\Omega}_{\sigma^2} &= \left\{\sigma^2:  \sigma^2_0 \le \sigma^2 \le (1 + \epsilon_1)\sigma^2_0 \right\}.
 \end{align}
We define $\boldsymbol{\epsilon}_{1jp}$ and $\boldsymbol{\epsilon}_{2k}$ such that each element of $\boldsymbol{\epsilon}_{1jp}$ is between 0 and $\epsilon_1$, and each element of $\boldsymbol{\epsilon}_{2k}$ is between 0 and $\epsilon_2$. Therefore, $(\boldsymbol{\phi}_{jp})_0 + \boldsymbol{\epsilon}_{1jp} \in \boldsymbol{\Omega}_{\boldsymbol{\phi}_{jp}}$ and $(\boldsymbol{\nu}_k)_0 + \boldsymbol{\epsilon}_{2k} \in \boldsymbol{\Omega}_{\boldsymbol{\nu}_k}$. 
We will define $$\boldsymbol{\Omega}_{\boldsymbol{\Sigma}_{jk}}:= \left\{\left.\sum_{p=1}^{KP} \boldsymbol{\phi}_{jp}'\boldsymbol{\phi}_{kp}\right| \boldsymbol{\phi}_{jp} \in \boldsymbol{\Omega}_{\boldsymbol{\phi}_{jp}}, \boldsymbol{\phi}_{kp} \in \boldsymbol{\Omega}_{\boldsymbol{\phi}_{kp}} \right\}.$$
Thus for $\boldsymbol{\Sigma}_i$ such that $\boldsymbol{\phi}_{jp} \in \boldsymbol{\Omega}_{\boldsymbol{\phi}_{jp}}$ and $\sigma^2 \in \boldsymbol{\Omega}_{\sigma^2}$, we have that 
\begin{align}
    \nonumber
    \boldsymbol{\Sigma}_i = & \sum_{k=1}^K \sum_{j=1}^K Z_{ik}Z_{ij} \left(\mathbf{S}'(\mathbf{t}_i)\sum_{p=1}^{KP}\left(\left((\boldsymbol{\phi}_{kp})_0 + \boldsymbol{\epsilon}_{1kp}\right)\left((\boldsymbol{\phi}_{jp})_0+ \boldsymbol{\epsilon}_{1jp}\right)'\right)\mathbf{S}(\mathbf{t}_i)\right) + (1 + \epsilon_\sigma)\sigma_0^2\mathbf{I}_{n_i}\\
    \nonumber
    = & (\boldsymbol{\Sigma}_i)_0 + \sum_{k=1}^K \sum_{j=1}^K\sum_{p=1}^{KP} Z_{ik}Z_{ij} \left(\mathbf{S}'(\mathbf{t}_i)\left(\left(\boldsymbol{\epsilon}_{1kp}\right)(\boldsymbol{\phi}_{jp})_0'\right)\mathbf{S}(\mathbf{t}_i)\right) \\
    \nonumber
    & + \sum_{k=1}^K \sum_{j=1}^K\sum_{p=1}^{KP} Z_{ik}Z_{ij} \left(\mathbf{S}'(\mathbf{t}_i)\left((\boldsymbol{\phi}_{kp})_0 \left( \boldsymbol{\epsilon}_{1jp}\right)'\right)\mathbf{S}(\mathbf{t}_i)\right) \\
    \nonumber
    & +\sum_{k=1}^K \sum_{j=1}^K\sum_{p=1}^{KP} Z_{ik}Z_{ij} \left(\mathbf{S}'(\mathbf{t}_i)\left(\left(\boldsymbol{\epsilon}_{1kp} \right)\left( \boldsymbol{\epsilon}_{1jp}\right)'\right)\mathbf{S}(\mathbf{t}_i)\right) + \epsilon_{\sigma}\sigma^2_0\mathbf{I}_{n_i}\\
    \nonumber
    = & (\boldsymbol{\Sigma}_i)_0 + \tilde{\boldsymbol{\Sigma}_i},
\end{align}
for some $\epsilon_{kp}$ and $\epsilon_\sigma$ such that $0 < \epsilon_\sigma \le \epsilon_1$. Thus, letting $\boldsymbol{\zeta}_{jkp} = \left(\mathbf{S}'(\mathbf{t}_i)\left(\left(\boldsymbol{\epsilon}_{1kp}\right)(\boldsymbol{\phi}_{jp})_0'\right)\mathbf{S}(\mathbf{t}_i)\right.$ $\left.+\mathbf{S}'(\mathbf{t}_i)\left((\boldsymbol{\phi}_{kp})_0 \left( \boldsymbol{\epsilon}_{1jp}\right)'\right)\mathbf{S}(\mathbf{t}_i) \right)$, we have
\begin{align}
    \nonumber
    \left|\left|Z_{ik}Z_{ij} \boldsymbol{\zeta}_{jkp}\right|\right|_F^2  \le & \left|\left| \boldsymbol{\zeta}_{jkp}\right|\right|_F^2\\
    \nonumber
    = & \text{tr}\left(\mathbf{S}'(\mathbf{t}_i)\left(\left(\boldsymbol{\epsilon}_{1kp}\right)(\boldsymbol{\phi}_{jp})_0'\right)\mathbf{S}(\mathbf{t}_i)\mathbf{S}'(\mathbf{t}_i)\left((\boldsymbol{\phi}_{jp})_0\left(\boldsymbol{\epsilon}_{1kp}\right)'\right)\mathbf{S}(\mathbf{t}_i)\right) \\
    \nonumber
    & + \text{tr}\left(\mathbf{S}'(\mathbf{t}_i)\left(\left(\boldsymbol{\epsilon}_{1kp}\right)(\boldsymbol{\phi}_{jp})_0'\right)\mathbf{S}(\mathbf{t}_i)\mathbf{S}'(\mathbf{t}_i)\left(\left( \boldsymbol{\epsilon}_{1jp}\right)(\boldsymbol{\phi}_{kp})_0 '\right)\mathbf{S}(\mathbf{t}_i)\right)\\
    \nonumber
    & + \text{tr}\left(\mathbf{S}'(\mathbf{t}_i)\left((\boldsymbol{\phi}_{kp})_0 \left( \boldsymbol{\epsilon}_{1jp}\right)'\right)\mathbf{S}(\mathbf{t}_i)\mathbf{S}'(\mathbf{t}_i)\left((\boldsymbol{\phi}_{jp})_0\left(\boldsymbol{\epsilon}_{1kp}\right)'\right)\mathbf{S}(\mathbf{t}_i)\right) \\
    \nonumber
    & + \text{tr}\left(\mathbf{S}'(\mathbf{t}_i)\left((\boldsymbol{\phi}_{kp})_0 \left( \boldsymbol{\epsilon}_{1jp}\right)'\right)\mathbf{S}(\mathbf{t}_i)\mathbf{S}'(\mathbf{t}_i)\left(\left( \boldsymbol{\epsilon}_{1jp}\right)(\boldsymbol{\phi}_{kp})_0 '\right)\mathbf{S}(\mathbf{t}_i)\right)\\
    \nonumber
     \le &  \epsilon_1^2\text{tr}\left((\boldsymbol{\phi}_{jp})_0'\mathbf{S}(\mathbf{t}_i)\mathbf{S}'(\mathbf{t}_i)(\boldsymbol{\phi}_{jp})_0\left(\mathbf{1}_{P}\right)'\mathbf{S}(\mathbf{t}_i)\mathbf{S}'(\mathbf{t}_i)\left(\mathbf{1}_{P}\right)\right) \\
    \label{simp_1} & + 2\text{tr}\left((\boldsymbol{\phi}_{jp})_0'\mathbf{S}(\mathbf{t}_i)\mathbf{S}'(\mathbf{t}_i)\left( \boldsymbol{\epsilon}_{1jp}\right)(\boldsymbol{\phi}_{kp})_0 '\mathbf{S}(\mathbf{t}_i)\mathbf{S}'(\mathbf{t}_i)\left(\boldsymbol{\epsilon}_{1kp}\right)\right)\\
    \nonumber
    & + \epsilon_1^2\text{tr}\left( \left( \mathbf{1}_{P}\right)'\mathbf{S}(\mathbf{t}_i)\mathbf{S}'(\mathbf{t}_i)\left(\mathbf{1}_{P}\right)(\boldsymbol{\phi}_{kp})_0 '\mathbf{S}(\mathbf{t}_i)\mathbf{S}'(\mathbf{t}_i)(\boldsymbol{\phi}_{kp})_0\right).
\end{align}
Using the Cauchy-Schwarz inequality, we can simplify Equation \ref{simp_1}, such that
\begin{align}
    \nonumber (\ref{simp_1}) & = 2\langle\mathbf{S}'(\mathbf{t}_i)(\boldsymbol{\phi}_{jp})_0, \mathbf{S}'(\mathbf{t}_i) \boldsymbol{\epsilon}_{1jp} \rangle \langle \mathbf{S}'(\mathbf{t}_i)(\boldsymbol{\phi}_{kp})_0, \mathbf{S}'(\mathbf{t}_i)\boldsymbol{\epsilon}_{1kp}  \rangle \\
    \nonumber & \le 2\|\mathbf{S}'(\mathbf{t}_i)(\boldsymbol{\phi}_{jp})_0\|_2 \|\mathbf{S}'(\mathbf{t}_i) \boldsymbol{\epsilon}_{1jp}\|_2 \| \mathbf{S}'(\mathbf{t}_i)(\boldsymbol{\phi}_{kp})_0\|_2 \|\mathbf{S}'(\mathbf{t}_i)\boldsymbol{\epsilon}_{1kp}\|_2\\
    \nonumber& \le 2\epsilon_1^2 \|\mathbf{S}'(\mathbf{t}_i)(\boldsymbol{\phi}_{jp})_0\|_2 \|\mathbf{S}'(\mathbf{t}_i) \mathbf{1}_{P}\|_2 \| \mathbf{S}'(\mathbf{t}_i)(\boldsymbol{\phi}_{kp})_0\|_2 \|\mathbf{S}'(\mathbf{t}_i)\mathbf{1}_{P}\|_2.
\end{align}
Let
\begin{align}
    \nonumber \tilde{M}_{ijkp} & = \|\mathbf{S}'(\mathbf{t}_i)\mathbf{1}_{P}\|_2^2\left[\|\mathbf{S}'(\mathbf{t}_i)(\boldsymbol{\phi}_{jp})_0\|_2^2 + \|\mathbf{S}'(\mathbf{t}_i)(\boldsymbol{\phi}_{kp})_0\|_2^2  \right] \\ 
 \nonumber & + 2\left(\|\mathbf{S}'(\mathbf{t}_i)(\boldsymbol{\phi}_{jp})_0\|_2 \|\mathbf{S}'(\mathbf{t}_i) \mathbf{1}_{P}\|_2 \| \mathbf{S}'(\mathbf{t}_i)(\boldsymbol{\phi}_{kp})_0\|_2 \|\mathbf{S}'(\mathbf{t}_i)\mathbf{1}_{P}\|_2\right).
\end{align} 
Since we know that the basis functions $b_1, \dots, b_P$ are uniformly continuous on the bounded interval $\mathcal{T}$, we know that the functions themselves are bounded. Therefore, we know that $\tilde{M}_{ijkp}$ is bounded above by some $\tilde{M}_{jkp}$ for $i = 1, \dots, N$. Therefore, we have that 
$$\left|\left|Z_{ik}Z_{ij} \boldsymbol{\zeta}_{jkp}\right|\right|_F^2  \le \epsilon_1^2 \tilde{M}_{jkp}.$$

In a similar fashion, we can show that 
 $$\|Z_{ik}Z_{ij} \left(\mathbf{S}'(\mathbf{t}_i)\left(\left(\boldsymbol{\epsilon}_{1kp} \right)\left(\boldsymbol{\epsilon}_{1jp}\right)'\right)\mathbf{S}(\mathbf{t}_i)\right)\|_F^2 \le \epsilon_1^2\|\mathbf{S}'(\mathbf{t}_i)\mathbf{1}_{P}\|_2^4 \le \epsilon_1^2 \tilde{S}_1$$
 where $\|\mathbf{S}'(\mathbf{t}_i)\mathbf{1}_{P}\|_2^4 \le \tilde{S}_1$ for $i = 1, \dots, N$, and
 $$\|\epsilon_{\sigma}\sigma^2_0\mathbf{I}_{n_i}\|^2_F \le \epsilon_1^2\sigma_0^4 n_i.$$
 By using the triangle inequality we have
 \begin{equation}
     \|\tilde{\boldsymbol{\Sigma}_i}\|_F \le \epsilon_1\left(\sum_{j=1}^K\sum_{k=1}^K\sum_{p=1}^{KP}\left(\sqrt{\tilde{M}_{jkp}}\right) + JK^2P\sqrt{\tilde{S}_1} + \sigma_0^2 \sqrt{n_{max}} \right) := \epsilon_1M_{\boldsymbol{\Sigma}}
     \label{Frobenius_Bound}
 \end{equation}
 for all $i \in \mathbb{N}$. By the Wielandt-Hoffman Theorem (\citet{golub2013matrix} Theorem 8.1.4), we have that
 $$\sum_{r=1}^R \left(\lambda_r\left((\boldsymbol{\Sigma}_i)_0 + \tilde{\boldsymbol{\Sigma}}_i\right) -\lambda_r\left((\boldsymbol{\Sigma}_i)_0\right)  \right)^2 \le \|\tilde{\boldsymbol{\Sigma}}_i\|^2_F,$$
 which implies that
 \begin{equation}
 \label{eq: bound_eigen}
     \max_{r}\left|\lambda_r\left((\boldsymbol{\Sigma}_i)_0 + \tilde{\boldsymbol{\Sigma}}_i\right) -\lambda_r\left((\boldsymbol{\Sigma}_i)_0\right) \right| \le \|\tilde{\boldsymbol{\Sigma}}_i\|_F
 \end{equation}
 where $\lambda_r(\mathbf{A})$ are the eigenvalues of the matrix $\mathbf{A}$.
 By using Equation \ref{Frobenius_Bound}, we can bound the log-determinant of the ratio of the two covariance matrices as follows
\begin{align}
     \nonumber
     \text{log}\left(\frac{|\boldsymbol{\Sigma}_i|}{|(\boldsymbol{\Sigma}_i)_0|} \right) & = \text{log}\left(\frac{\prod_{r=1}^{n_i}\lambda_r\left((\boldsymbol{\Sigma}_i)_0 + \tilde{\boldsymbol{\Sigma}}_i\right)}{\prod_{r=1}^{n_i}\lambda_r\left((\boldsymbol{\Sigma}_i)_0\right)} \right)\\
     \nonumber
     & \le \text{log}\left(\prod_{r=1}^{n_i}\frac{\left((d_{ir})_0 + \sigma_0^2\right) + \epsilon_1M_{\boldsymbol{\Sigma}}}{(d_{ir})_0 + \sigma_0^2}\right) \\
     & \le n_{max}\text{log}\left( 1 + \frac{\epsilon_1M_{\boldsymbol{\Sigma}}}{\sigma_0^2}\right).
     \label{Bound_log_det}
\end{align}
We can also bound $\text{tr}\left(\boldsymbol{\Sigma}_i^{-1}\left(\boldsymbol{\Sigma}_i\right)_0\right)$. To do this, we first consider the spectral norm, defined as $\|\mathbf{A}\|_2 = \sqrt{\mathbf{A}^*\mathbf{A}}$ for some matrix $\mathbf{A}$. In the case where $\mathbf{A}$ is symmetric, we have $\|\mathbf{A}\|_2 = \max_{r}|\lambda_r(\mathbf{A})|$. By the submultiplicative property of induced norms, we have that
\begin{equation}
\label{eq: submult}
    \max_{r}|\lambda_r(\mathbf{A}\mathbf{B})| = \|\mathbf{A}\mathbf{B}\|_2 \le \|\mathbf{A}\|_2 \|\mathbf{B}\|_2 = \max_{r}|\lambda_r(\mathbf{A})| \max_{r}|\lambda_r(\mathbf{B})|,
\end{equation}
for two symmetric matrices $\mathbf{A}$ and $\mathbf{B}$. By using the Sherman–Morrison–Woodbury formula, we can see that
\begin{align}
    \nonumber \boldsymbol{\Sigma}_i^{-1} &= \left((\boldsymbol{\Sigma}_i)_0 + \tilde{\boldsymbol{\Sigma}}_i \right)^{-1}\\
    \nonumber &= (\boldsymbol{\Sigma}_i)_0^{-1} - (\boldsymbol{\Sigma}_i)_0^{-1}\tilde{\boldsymbol{\Sigma}}_i\left((\boldsymbol{\Sigma}_i)_0 + \tilde{\boldsymbol{\Sigma}}_i \right)^{-1}.
\end{align}
Thus, we have that 
\begin{equation}
    \label{eq: SMW_formula}
    \boldsymbol{\Sigma}_i^{-1}\left(\boldsymbol{\Sigma}_i\right)_0 =  \mathbf{I}_{n_i} - (\boldsymbol{\Sigma}_i)_0^{-1}\tilde{\boldsymbol{\Sigma}}_i\left((\boldsymbol{\Sigma}_i)_0 + \tilde{\boldsymbol{\Sigma}}_i \right)^{-1}\left(\boldsymbol{\Sigma}_i\right)_0.
\end{equation}
Using Equation \ref{eq: submult}, we would like to bound the magnitude of the eigenvalues of \\
$(\boldsymbol{\Sigma}_i)_0^{-1}\tilde{\boldsymbol{\Sigma}}_i\left((\boldsymbol{\Sigma}_i)_0 + \tilde{\boldsymbol{\Sigma}}_i \right)^{-1}\left(\boldsymbol{\Sigma}_i\right)_0$. We know that $$\max_r\left|\lambda_r\left( \left((\boldsymbol{\Sigma}_i)_0 + \tilde{\boldsymbol{\Sigma}}_i \right)^{-1}\right)\right| \le \frac{1}{\sigma_0^2}$$
and
$$\max_r\left|\lambda_r(\tilde{\boldsymbol{\Sigma}}_i )\right| \le \epsilon_1 M_{\boldsymbol{\Sigma}},$$
with the second inequality coming from Equation \ref{Frobenius_Bound}. From Equation \ref{eq: SMW_formula} and basic properties of the trace, we have that
\begin{align}
    \nonumber \text{tr}\left(\boldsymbol{\Sigma}_i^{-1}\left(\boldsymbol{\Sigma}_i\right)_0 \right) & = \text{tr}\left(\mathbf{I}_{n_i} - (\boldsymbol{\Sigma}_i)_0^{-1}\tilde{\boldsymbol{\Sigma}}_i\left((\boldsymbol{\Sigma}_i)_0 + \tilde{\boldsymbol{\Sigma}}_i \right)^{-1}\left(\boldsymbol{\Sigma}_i\right)_0 \right) \\
    \nonumber & = \text{tr}\left(\mathbf{I}_{n_i} \right) - \text{tr}\left(  \tilde{\boldsymbol{\Sigma}}_i\left((\boldsymbol{\Sigma}_i)_0 + \tilde{\boldsymbol{\Sigma}}_i \right)^{-1}\left(\boldsymbol{\Sigma}_i\right)_0(\boldsymbol{\Sigma}_i)_0^{-1}\right) \\
    \nonumber & = \text{tr}\left(\mathbf{I}_{n_i} \right) - \text{tr}\left(  \tilde{\boldsymbol{\Sigma}}_i\left((\boldsymbol{\Sigma}_i)_0 + \tilde{\boldsymbol{\Sigma}}_i \right)^{-1}\right)
\end{align}
Thus, using the fact that the trace of a matrix is the sum of its eigenvalues, we have that 
    $$\text{tr}\left(\boldsymbol{\Sigma}_i^{-1}\left(\boldsymbol{\Sigma}_i\right)_0 \right) \le n_i + n_{max}\max_r\left|\lambda_r\left(\tilde{\boldsymbol{\Sigma}}_i\left((\boldsymbol{\Sigma}_i)_0 + \tilde{\boldsymbol{\Sigma}}_i \right)^{-1} \right)\right|.$$
Using the submultiplicative property stated in Equation \ref{eq: submult}, we have
\begin{equation}
\label{Bound_trace}
    \text{tr}\left(\boldsymbol{\Sigma}_i^{-1}\left(\boldsymbol{\Sigma}_i\right)_0 \right) \le n_i + \frac{n_{max}\epsilon_1M_{\boldsymbol{\Sigma}}}{\sigma_0^2}.
\end{equation}

Lastly, we can bound the quadratic term in $K_i(\boldsymbol{\omega}_0, \boldsymbol{\omega})$ in the following way:
\begin{align}
    \nonumber
    \left(\left(\boldsymbol{\mu}_i\right)_0- \boldsymbol{\mu}_i\right)'\left(\boldsymbol{\Sigma}_i\right)^{-1}\left(\left(\boldsymbol{\mu}_i\right)_0 - \boldsymbol{\mu}_i\right) &\le \left|\left|\left(\boldsymbol{\mu}_i\right)_0- \boldsymbol{\mu}_i \right| \right|_2^2 \max_{r} \lambda_r(\left(\boldsymbol{\Sigma}_i)^{-1}\right)\\
    \nonumber & \le \frac{1}{\sigma^2}\sum_{k=1}^K\|\mathbf{S}'(\mathbf{t}_i) (\boldsymbol{\nu}_k)_0 - \mathbf{S}'(\mathbf{t}_i) \boldsymbol{\nu}_k\|^2_2 \\
    \nonumber & = \frac{1}{\sigma^2}\sum_{k=1}^K\boldsymbol{\epsilon}_{2k}'\mathbf{S}(\mathbf{t}_i)\mathbf{S}'(\mathbf{t}_i)\boldsymbol{\epsilon}_{2k}  \\
    & \le \frac{Kn_{max}\epsilon_2^2}{\sigma^2_0} \lambda^{max}_{\mathbf{S}(\mathbf{t}_i)},
     \label{quadratic_bound}
\end{align}
where $\lambda^{max}_{\mathbf{S}(\mathbf{t}_i)}$ is the maximum eigenvalue of the matrix $\mathbf{S}(\mathbf{t}_i)\mathbf{S}'(\mathbf{t}_i)$. Since the maximum eigenvalue of $\mathbf{S}(\mathbf{t}_i)\mathbf{S}'(\mathbf{t}_i)$ (since the basis functions are bounded on $\mathcal{T}$), let $\lambda^{max}_{\mathbf{S}(\mathbf{t})}$ be an upper bound for $\lambda^{max}_{\mathbf{S}(\mathbf{t}_i)}$ (i.e. $\lambda^{max}_{\mathbf{S}(\mathbf{t}_i)} \le \lambda^{max}_{\mathbf{S}(\mathbf{t})}$ be an upper bound for $\lambda^{max}_{\mathbf{S}(\mathbf{t}_i)}$ for $i = 1, \dots, N$).
Thus letting 
\begin{equation}
\epsilon_1 < \min\left\{\frac{\sigma^2_0}{M_{\boldsymbol{\Sigma}}}\left(\text{exp}\left( \frac{2\epsilon}{3n_{max}} \right) - 1\right),\frac{2\epsilon\sigma^2_0}{3n_{max}M_{\boldsymbol{\Sigma}}}\right\}
\label{epsilon_1}
\end{equation}
and 
\begin{equation}
    \epsilon_2 < \sqrt{\frac{2\sigma_0^2\epsilon}{3Kn_{max}\lambda^{max}_{\mathbf{S}(\mathbf{t})}}},
\end{equation}
we have from equations \ref{Bound_log_det}, \ref{Bound_trace}, and \ref{quadratic_bound} that
$$K_i(\boldsymbol{\omega}_0,\boldsymbol{\omega}) < \epsilon \text{ for all } \boldsymbol{\omega} \in \boldsymbol{\Omega}_1$$
where $\boldsymbol{\Omega}_1 := \left(\bigtimes_{j=1}^K \bigtimes_{k = 1}^{K} \boldsymbol{\Omega}_{\boldsymbol{\Sigma}_{jk}} \right) \times \left(\bigtimes_{k=1}^K \boldsymbol{\Omega}_{\boldsymbol{\nu}_k} \right) \times \boldsymbol{\Omega}_{\sigma^2}$.
Letting $a > \max \left\{1 + \frac{\epsilon_1M_{\boldsymbol{\Sigma}}}{\sigma_0^2}, \left(1 - \frac{\epsilon_1M_{\boldsymbol{\Sigma}}}{\sigma_0^2} \right)^{-1} \right\}$ and $b > \sqrt{KR\epsilon_2^2\lambda^{max}_{\mathbf{S}(\mathbf{t}_i)}}$ in the definition of $\mathcal{C}(\boldsymbol{\omega}_0, \epsilon)$, we have $ \boldsymbol{\Omega}_1 \subset \mathcal{C}(\boldsymbol{\omega}_0, \epsilon)$. Let $H_{\boldsymbol{\phi}}$ be the set of hyperparameters corresponding to the $\boldsymbol{\phi}$ parameters, and let $\boldsymbol{\Pi}(\boldsymbol{\eta}_{\boldsymbol{\phi}})$ be the prior distribution on $\boldsymbol{\eta}_{\boldsymbol{\phi}}\in H_{\boldsymbol{\phi}}$. Thus we have that
\begin{align}
    \nonumber
    \boldsymbol{\Pi}\left(\boldsymbol{\omega} \in  \mathcal{C}(\boldsymbol{\omega}_0, \epsilon)\right) & \ge \int_{H_{\boldsymbol{\phi}}} \prod_{j=1}^K \prod_{p=1}^{KP} \prod_{r=1}^P \int_{(\phi_{jrp})_0}^{(\phi_{jrp})_0 + \epsilon_1} \sqrt{\frac{\gamma_{jrp}\tilde{\tau}_{pj}}{2\pi}} \text{exp}\left\{-\frac{\gamma_{jrp}\tilde{\tau}_{pj}}{2}\phi_{jrp}^2\right\} \text{d}\phi_{jrp} \text{d} \boldsymbol{\Pi}(\boldsymbol{\eta}_{\boldsymbol{\phi}})\\
    \nonumber & \times \prod_{k=1}^K \int_0^\infty \int_{(\boldsymbol{\nu}_k)_0}^{(\boldsymbol{\nu}_k)_0 + \epsilon_2\mathbf{1}}\left(\frac{\tau_k}{2\pi}\right)^{P/2} |\mathbf{P}|^{-1/2} \text{exp}\left\{\frac{\tau_k}{2}\boldsymbol{\nu}_k' \mathbf{P}\boldsymbol{\nu}_k \right\} \text{d}\boldsymbol{\nu}_k \text{d}\boldsymbol{\Pi}(\tau_k)\\
    \nonumber
    & \times \int_{\sigma_0^2}^{(1 + \epsilon_1) \sigma_0^2}\frac{\beta_0^{\alpha_0}}{\Gamma(\alpha_0)}(\sigma^2)^{-\alpha_0 - 1} \text{exp} \left\{- \frac{\beta_0}{\sigma^2} \right\} \text{d} \sigma^2.
\end{align}
Restricting the hyperparameters of $\boldsymbol{\phi}$ to only a subset of the support, say $\tilde{H}_{\boldsymbol{\phi}}$, where $$\tilde{H}_{\boldsymbol{\phi}} = \left\{\boldsymbol{\eta}_{\boldsymbol{\phi}}: \frac{1}{10} \le \gamma_{jrp} \le 10, 1 \le \delta_{pj} \le 2, 1 \le a_{1j} \le 10, 1 \le a_{2j} \le 10\right\},$$ we can see that there exists a $M_{\phi_{jrp}} > 0$ such that $$\sqrt{\frac{\gamma_{jrp}\tilde{\tau}_{pj}}{2\pi}} \text{exp}\left\{-\frac{\gamma_{jrp}\tilde{\tau}_{pj}}{2}\phi_{jrp}^2\right\} \ge M_{\phi_{jrp}},$$ for all $\phi_{jrp} \in [(\phi_{jrp})_0,(\phi_{jrp})_0 + \epsilon_1]$. Similarly, we can find a lower bound $M_{\tilde{H}_{\boldsymbol{\phi}}} > 0$, such that $$\int_{\tilde{H}_{\boldsymbol{\phi}}} \text{d}(\boldsymbol{\eta}_{\boldsymbol{\phi}}) \ge M_{\tilde{H}_{\boldsymbol{\phi}}}.$$ Similarly, if we bound $\tau_k$ such that $\frac{1}{10} \le \tau_k \le 10$, it is easy to see that there exists constants $M_{\boldsymbol{\nu}_k}, M_{\tau_k}, M_{\sigma^2} > 0$ such that
$$\left(\frac{\tau_k}{2\pi}\right)^{P/2} |\mathbf{P}|^{-1/2} \text{exp}\left\{\frac{\tau_k}{2}\boldsymbol{\nu}_k' \mathbf{P}\boldsymbol{\nu}_k \right\} \ge  M_{\boldsymbol{\nu}_k},$$
for all $\boldsymbol{\nu}_k \in [(\boldsymbol{\nu}_k)_0, (\boldsymbol{\nu}_k)_0 + \epsilon_2\mathbf{1}]$,
$$\int_{\frac{1}{10}}^{10} \boldsymbol{\Pi}(\tau_k) \ge M_{\tau_k},$$
and 
$$ \frac{\beta_0^{\alpha_0}}{\Gamma(\alpha_0)}(\sigma^2)^{-\alpha_0 - 1} \text{exp} \left\{- \frac{\beta_0}{\sigma^2} \right\} \ge M_{\sigma^2}$$
for all $\sigma^2 \in [\sigma^2_0, (1 + \epsilon_1)\sigma^2_0]$. Therefore we have that 
\begin{align}
    \nonumber
    \boldsymbol{\Pi}\left(\boldsymbol{\omega} \in  \mathcal{C}(\boldsymbol{\omega}_0, \epsilon)\right) & \ge M_{\tilde{H}_{\boldsymbol{\phi}}}  \prod_{j=1}^K \prod_{p=1}^{KP} \prod_{r=1}^{P} \epsilon_1 M_{\phi_{jrp}} \\
    \nonumber
    & \times \prod_{k=1}^K M_{\tau_k} \epsilon_2^PM_{\boldsymbol{\nu}_k} \\
    \nonumber
    & \times \epsilon_1\sigma^2_0 M_{\sigma^2_0}\\
    \nonumber
    & > 0.
\end{align}
Therefore, for $\epsilon > 0$, there exist $a$ and $b$ such that $\sum_{i=1}^\infty \frac{V_i(\boldsymbol{\omega}_0, \boldsymbol{\omega})}{i^2} < \infty$ for any $\boldsymbol{\omega} \in \mathcal{C}(\boldsymbol{\omega}_0, \epsilon)$ and $\boldsymbol{\Pi}\left(\boldsymbol{\omega} \in  \mathcal{C}(\boldsymbol{\omega}_0, \epsilon)\right) > 0$.

\subsection{Proof of Lemma 3.2}
 Following the notation of \citet{ghosal2017fundamentals}, we will let $P_{\boldsymbol{\omega}_0}^{(N)}$ denote the joint distribution of $\mathbf{Y}_1, \dots, \mathbf{Y}_N$ at $\boldsymbol{\omega}_0\in\boldsymbol{\Omega}$. To show the posterior distribution, $\boldsymbol{\Pi}_N(. | \mathbf{Y}_1, \dots, \mathbf{Y}_N)$, is weakly consistent at $\boldsymbol{\omega}_0 \in \boldsymbol{\Omega}$, we need to show that $\boldsymbol{\Pi}_N(\mathcal{U}^c| \mathbf{Y}_1, \dots, \mathbf{Y}_N) \rightarrow 0$  a.s. $[P_{\boldsymbol{\omega}_0}]$ for every weak neighborhood, $\mathcal{U}$ of $\boldsymbol{\omega}_0$. Following a similar notation to \citet{ghosal2017fundamentals}, let $\psi_N$ be measurable mappings, $\psi_N: \boldsymbol{\mathcal{S}}^N \times \boldsymbol{\mathcal{Z}}^N \rightarrow [0,1]$, where $\boldsymbol{\mathcal{Z}}$ is the sample space of $\{Z_{i1}, \dots, Z_{iK}\}$. Let $\psi_N(\mathbf{Y}_1, \dots, \mathbf{Y}_N, \mathbf{z}_1, \dots, \mathbf{z}_N)$ be the corresponding test function, and \linebreak $P_{\boldsymbol{\omega}}^N\psi_N = \mathbb{E}_{P_{\boldsymbol{\omega}}^N}\psi_N(\mathbf{Y}_1, \dots, \mathbf{Y}_N, \mathbf{z}_1,\dots, \mathbf{z}_N) = \int \psi_N \text{d}P_{\boldsymbol{\omega}}^N$, where $P_{\boldsymbol{\omega}}^N$ denotes the joint distribution on $\mathbf{Y}_1, \dots, \mathbf{Y}_N$ with parameters $\boldsymbol{\omega}$. Suppose there exist tests $\psi_N$ such that $P_{\boldsymbol{\omega}_0}^{N}\psi_N \rightarrow 0$, and $\text{sup}_{\boldsymbol{\omega} \in \mathcal{U}^c} P_{\boldsymbol{\omega}}^{N}(1-\psi_N) \rightarrow 0$. Since $\psi_N(\mathbf{Y}_1, \dots, \mathbf{Y}_N, \mathbf{z}_1,\dots, \mathbf{z}_N) \in [0,1]$, we have that 
 \begin{align}
     \nonumber
     \boldsymbol{\Pi}_n(U^c| \mathbf{Y}_1, \dots, \mathbf{Y}_N) & \le \boldsymbol{\Pi}_n(U^c| \mathbf{Y}_1, \dots, \mathbf{Y}_N) + \psi_N(\mathbf{Y}_1, \dots, \mathbf{Y}_N)\left( 1 - \boldsymbol{\Pi}_n(U^c| \mathbf{Y}_1, \dots, \mathbf{Y}_N)\right)\\
     & = \psi_N(\mathbf{Y}_1, \dots, \mathbf{Y}_N) + \frac{\left( 1- \psi_N(\mathbf{Y}_1, \dots, \mathbf{Y}_N)\right)\int_{U^c}\prod_{i=1}^N \frac{f_i(\mathbf{Y}_i; \boldsymbol{\omega})}{f_i(\mathbf{Y}_i; \boldsymbol{\omega}_0)}\text{d}\boldsymbol{\Pi}(\boldsymbol{\omega})}{\int_{\boldsymbol{\Omega}}\prod_{i=1}^N \frac{f_i(\mathbf{Y}_i; \boldsymbol{\omega})}{f_i(\mathbf{Y}_i; \boldsymbol{\omega}_0)}\text{d}\boldsymbol{\Pi}(\boldsymbol{\omega})}.
     \label{posterior}
 \end{align}
 To show that $\boldsymbol{\Pi}_n(U^c| \mathbf{Y}_1, \dots, \mathbf{Y}_N) \rightarrow 0$, it is sufficient to show the following three conditions:
 \begin{enumerate}
     \item $\psi_N(\mathbf{Y}_1, \dots, \mathbf{Y}_N,\mathbf{z}_1,\dots, \mathbf{z}_N) \rightarrow 0$ a.s. $[P_{\boldsymbol{\omega}_0}]$,
     \item $e^{\beta_1 N}\left( 1- \psi_N(\mathbf{Y}_1, \dots, \mathbf{Y}_N,\mathbf{z}_1,\dots, \mathbf{z}_N)\right)\int_{\mathcal{U}^c}\prod_{i=1}^N \frac{f_i(\mathbf{Y}_i; \boldsymbol{\omega})}{f_i(\mathbf{Y}_i; \boldsymbol{\omega}_0)}\text{d}\boldsymbol{\Pi}(\boldsymbol{\omega}) \rightarrow 0$ a.s. $[P_{\boldsymbol{\omega}_0}]$ for some $\beta_1 > 0$,
     \item $e^{\beta N}\left(\int_{\boldsymbol{\Omega}}\prod_{i=1}^N \frac{f_i(\mathbf{Y}_i; \boldsymbol{\omega})}{f_i(\mathbf{Y}_i; \boldsymbol{\omega}_0)}\text{d}\boldsymbol{\Pi}(\boldsymbol{\omega}) \right) \rightarrow \infty$ a.s. $[P_{\boldsymbol{\omega}_0}]$ for all $\beta > 0$.
 \end{enumerate}
 
 We will start by proving (c). Fix $\beta > 0$. Thus we have
 $$e^{\beta N}\left(\int_{\boldsymbol{\Omega}}\prod_{i=1}^N \frac{f_i(\mathbf{Y}_i; \boldsymbol{\omega})}{f_i(\mathbf{Y}_i; \boldsymbol{\omega}_0)}\text{d}\boldsymbol{\Pi}(\boldsymbol{\omega}) \right)= e^{\beta N}\left(\int_{\boldsymbol{\Omega}}\text{exp}\left[-\sum_{i=1}^N \text{log}\left(\frac{f_i(\mathbf{Y}_i; \boldsymbol{\omega}_0)}{f_i(\mathbf{Y}_i; \boldsymbol{\omega})}\right)\right]\text{d}\boldsymbol{\Pi}(\boldsymbol{\omega})\right).$$
 By Fatou's lemma, we have
 \begin{align}
    \nonumber & 
     \liminf_{N \rightarrow \infty} \int_{\boldsymbol{\Omega}}\text{exp}\left[\beta N -\sum_{i=1}^N \text{log}\left(\frac{f_i(\mathbf{Y}_i; \boldsymbol{\omega}_0)}{f_i(\mathbf{Y}_i; \boldsymbol{\omega})}\right)\right]\text{d}\boldsymbol{\Pi}(\boldsymbol{\omega}) \\
     \nonumber \ge & \int_{\boldsymbol{\Omega}}\liminf_{N \rightarrow \infty}\text{exp}\left[\beta N -\sum_{i=1}^N \text{log}\left(\frac{f_i(\mathbf{Y}_i; \boldsymbol{\omega}_0)}{f_i(\mathbf{Y}_i; \boldsymbol{\omega})}\right)\right]\text{d}\boldsymbol{\Pi}(\boldsymbol{\omega})
 \end{align} 
 Let $\beta >\epsilon > 0$ and $a,b > 0$ be defined such that lemma 3.1 holds. Since $\mathcal{C}(\boldsymbol{\omega}_0, \epsilon) \subset \boldsymbol{\Omega}$, we have that
 \begin{align}
     \nonumber & 
     \int_{\boldsymbol{\Omega}}\liminf_{N \rightarrow \infty}\text{exp}\left[\beta N -\sum_{i=1}^N \text{log}\left(\frac{f_i(\mathbf{Y}_i; \boldsymbol{\omega}_0)}{f_i(\mathbf{Y}_i; \boldsymbol{\omega})}\right)\right]\text{d}\boldsymbol{\Pi}(\boldsymbol{\omega})\\
     \nonumber \ge & \int_{\mathcal{C}(\boldsymbol{\omega}_0, \epsilon)}\liminf_{N \rightarrow \infty}\text{exp}\left[\beta N -\sum_{i=1}^N \text{log}\left(\frac{f_i(\mathbf{Y}_i; \boldsymbol{\omega}_0)}{f_i(\mathbf{Y}_i; \boldsymbol{\omega})}\right)\right]\text{d}\boldsymbol{\Pi}(\boldsymbol{\omega})
 \end{align}
 By Kolmogorov's strong law of large numbers for non-identically distributed  random variables, we have that 
 $$\frac{1}{N}\sum_{i=1}^N\left(\Lambda_i(\boldsymbol{\omega}_0, \boldsymbol{\omega}) - K_i(\boldsymbol{\omega}_0, \boldsymbol{\omega}) \right) \rightarrow 0$$
 a.s. $[P_{\boldsymbol{\omega}_0}]$. Thus, for each $\boldsymbol{\omega} \in \mathcal{C}(\boldsymbol{\omega}_0, \epsilon)$, with $P_{\boldsymbol{\omega}_0}$-probability  1, 
 $$\frac{1}{N}\sum_{i=1}^N\Lambda_i(\boldsymbol{\omega}_0, \boldsymbol{\omega}) \rightarrow \mathbb{E}(\overline{K_i(\boldsymbol{\omega}_0, \boldsymbol{\omega})}) < \epsilon < B,$$
 since $\boldsymbol{\omega} \in \mathcal{C}(\boldsymbol{\omega}_0, \epsilon)$. Therefore, we have that
 $$\int_{\mathcal{C}(\boldsymbol{\omega}_0, \epsilon)}\liminf_{N \rightarrow \infty}\text{exp}\left[\beta N -\sum_{i=1}^N \text{log}\left(\frac{f_i(\mathbf{Y}_i; \boldsymbol{\omega}_0)}{f_i(\mathbf{Y}_i; \boldsymbol{\omega})}\right)\right]\text{d}\boldsymbol{\Pi}(\boldsymbol{\omega}) \ge \int_{\mathcal{C}(\boldsymbol{\omega}_0, \epsilon)} \inf_{N \rightarrow \infty} \text{exp}\left\{N(\beta- \epsilon) \right\}\text{d}\boldsymbol{\Pi}(\boldsymbol{\omega}).$$
Since $\beta - \epsilon > 0$, and $\boldsymbol{\Pi}\left(\theta \in \mathcal{C}(\boldsymbol{\omega}_0, \epsilon)\right) > 0$ (lemma 3.1), we have that 
\begin{equation}
    e^{\beta N}\left(\int_{\boldsymbol{\Omega}}\prod_{i=1}^N \frac{f_i(\mathbf{Y}_i; \boldsymbol{\omega})}{f_i(\mathbf{Y}_i; \boldsymbol{\omega}_0)}\text{d}\boldsymbol{\Pi}(\boldsymbol{\omega}) \right) \rightarrow \infty
    \label{part_c}
\end{equation}
a.s. $[P_{\boldsymbol{\omega}_0}]$ for all $\beta > 0$. We will now show that there exist measurable mappings such that $P_{\boldsymbol{\omega}_0}^{N}\psi_N \rightarrow 0$ and $\text{sup}_{\boldsymbol{\omega} \in \mathcal{U}^c} P_{\boldsymbol{\omega}}^{N}(1-\psi_N) \rightarrow 0$. Consider weak neighborhoods $\mathcal{U}$ of $\boldsymbol{\omega}_0$ of the form
\begin{equation}
    \mathcal{U} = \left\{\boldsymbol{\omega}: \left|\int f_i \text{d}P_{\boldsymbol{\omega}} -\int f_i \text{d}P_{\boldsymbol{\omega}_0}  \right| < \epsilon_i, \;\; i = 1,2, \dots, r \right\},
    \label{weak_neighborhood}
\end{equation}
where $r \in \mathbb{N}$, $\epsilon_i > 0$, and $f_i$ are continuous functions such that $f_i:\boldsymbol{\mathcal{S}} \times \boldsymbol{\mathcal{Z}} \rightarrow [0,1]$. As shown in \citet{GhoshNonParametrics}, for any particular $f_i$ and $\epsilon_i > 0$,  $\left|\int f_i \text{d}P_{\boldsymbol{\omega}} -\int f_i \text{d}P_{\boldsymbol{\omega}_0}  \right| < \epsilon_i$ iff $\int f_i \text{d}P_{\boldsymbol{\omega}} -\int f_i \text{d}P_{\boldsymbol{\omega}_0} < \epsilon_i$ and $\int (1 -f_i) \text{d}P_{\boldsymbol{\omega}} -\int (1 -f_i) \text{d}P_{\boldsymbol{\omega}_0} < \epsilon$. Since $\tilde{f}_i:=(1 - f_i)$ is still a continuous function such that $\tilde{f}_i:\boldsymbol{\mathcal{S}} \times \boldsymbol{\mathcal{Z}} \rightarrow [0,1]$, we can rewrite Equation \ref{weak_neighborhood} as
\begin{equation}
    \mathcal{U} = \cap_{i=1}^{2r} \left\{\boldsymbol{\omega}: \int g_i \text{d}P_{\boldsymbol{\omega}} -\int g_i \text{d}P_{\boldsymbol{\omega}_0} < \epsilon_i \right\},
\end{equation}
where $g_i$ are continuous functions such that $g_i:\boldsymbol{\mathcal{S}} \times \boldsymbol{\mathcal{Z}} \rightarrow [0,1]$ and $\epsilon_i > 0$. Following \citet{ghosal2017fundamentals}, it can be shown by Hoeffding's inequality that using the test function $\tilde{\psi}$, defined as
\begin{equation}
    \tilde{\psi}_{iN}(\mathbf{Y}_1, \dots, \mathbf{Y}_N, \mathbf{z}_1, \dots, \mathbf{z}_N) := \mathbbm{1}\left\{\frac{1}{N}\sum_{j=1}^N g_i(\mathbf{Y}_j, \mathbf{z}_j) > \int g_i \text{d}P_{\boldsymbol{\omega}_0} + \frac{\epsilon_i}{2}\right\},
\end{equation}
leads to 
$$\int \tilde{\psi}_{iN}(\mathbf{Y}_1, \dots, \mathbf{Y}_N, \mathbf{z}_1, \dots, \mathbf{z}_N) \text{d}P_{\boldsymbol{\omega}_0} \le e^{-N\epsilon_i^2/2}$$ 
and 
$$\int \left(1 - \tilde{\psi}_{iN}(\mathbf{Y}_1, \dots, \mathbf{Y}_N, \mathbf{z}_1, \dots, \mathbf{z}_N)\right) \text{d}P_{\boldsymbol{\omega}} \le e^{-N\epsilon_i^2/2}$$
for any $\boldsymbol{\omega} \in \mathcal{U}^c$. Let $\boldsymbol{\psi}_n = \max_{i}\tilde{\psi}_{iN}$ be our test function and $\epsilon = \min_{i}\epsilon_i$. Using the fact that $\mathbb{E}(\max_{i}\tilde{\psi}_{iN}) \le \sum_{i}\mathbb{E}(\tilde{\psi}_{iN})$ and $\mathbb{E}(1 - \max_{i}\tilde{\psi}_{iN}) \le \mathbb{E}(1 - \tilde{\psi}_{iN})$, we have
\begin{equation}
    \int \psi_N(\mathbf{Y}_1, \dots, \mathbf{Y}_N, \mathbf{z}_1, \dots, \mathbf{z}_N) \text{d}P_{\boldsymbol{\omega}_0} \le (2r)e^{-N\epsilon^2/2}
    \label{test_bound1}
\end{equation}
and
\begin{equation}
    \int \left(1 - \psi_N(\mathbf{Y}_1, \dots, \mathbf{Y}_N, \mathbf{z}_1, \dots, \mathbf{z}_N)\right) \text{d}P_{\boldsymbol{\omega}} \le e^{-N\epsilon^2/2},
    \label{test_bound2}
\end{equation}
for any $\boldsymbol{\omega} \in \mathcal{U}^c$.
Using Markov's inequality on Equation \ref{test_bound1}, we have that 
\begin{align}
\nonumber
    P\left(\psi_N(\mathbf{Y}_1, \dots, \mathbf{Y}_N, \mathbf{z}_1, \dots, \mathbf{z}_N) \ge e^{-nC}\right) & \le \frac{\mathbb{E}\left(\psi_N(\mathbf{Y}_1, \dots, \mathbf{Y}_N, \mathbf{z}_1, \dots, \mathbf{z}_N)\right)}{e^{-NC}}\\
    \nonumber
    & \le (2r)e^{-N(\epsilon^2/2 - C)}
\end{align}
Thus letting $C < \epsilon^2 / 2$, we have that $\sum_{N=1}^\infty P\left(\psi_N(\mathbf{Y}_1, \dots, \mathbf{Y}_N, \mathbf{z}_1, \dots, \mathbf{z}_N) \ge e^{-NC}\right) < \infty$. Thus, by the Borel-Cantelli lemma, we know that 
$$P\left(\limsup_{N\rightarrow \infty}P\left(\psi_N(\mathbf{Y}_1, \dots, \mathbf{Y}_N, \mathbf{z}_1, \dots, \mathbf{z}_N) \ge e^{-NC} \right)\right) = 0$$
Thus we have that $\psi_N(\mathbf{Y}_1, \dots, \mathbf{Y}_N,\mathbf{z}_1,\dots, \mathbf{z}_N) \rightarrow 0$ a.s. $[P_{\boldsymbol{\omega}_0}]$ (Condition (a)). To prove condition (b), we will first start by taking the expectation with respect to $P_{\boldsymbol{\omega}_0}$:
\begin{align}
    \nonumber &
    \mathbb{E}_{P_{\boldsymbol{\omega}_0}^N}\left(e^{\beta N}\left( 1- \psi_N(\mathbf{Y}_1, \dots, \mathbf{Y}_N,\mathbf{z}_1,\dots, \mathbf{z}_N)\right)\int_{\mathcal{U}^c}\prod_{i=1}^N \frac{f_i(\mathbf{Y}_i; \boldsymbol{\omega})}{f_i(\mathbf{Y}_i; \boldsymbol{\omega}_0)}\text{d}\boldsymbol{\Pi}(\boldsymbol{\omega})\right)\\
    \nonumber
     = & \int_{\boldsymbol{\mathcal{S}}^N} \left(e^{\beta N}\left( 1- \psi_N(\mathbf{Y}_1, \dots, \mathbf{Y}_N,\mathbf{z}_1,\dots, \mathbf{z}_N)\right)\int_{\mathcal{U}^c}\prod_{i=1}^N \frac{f_i(\mathbf{Y}_i; \boldsymbol{\omega})}{f_i(\mathbf{Y}_i; \boldsymbol{\omega}_0)}\text{d}\boldsymbol{\Pi}(\boldsymbol{\omega})\right) \text{d}P_{\boldsymbol{\omega}_0}^N\\
     \nonumber
     = & \int_{\mathcal{U}^c}\left(\prod_{i=1}^N \int_{\mathcal{S}} e^{\beta N}\left( 1- \psi_N(\mathbf{Y}_1, \dots, \mathbf{Y}_N,\mathbf{z}_1,\dots, \mathbf{z}_N)\right)f_i(\mathbf{Y}_i; \boldsymbol{\omega})\text{d}\mathbf{Y}_i\right)\text{d}\boldsymbol{\Pi}(\boldsymbol{\omega})\\
     \nonumber
     = & e^{\beta N} \int_{\mathcal{U}^c}\mathbb{E}_{P_{\boldsymbol{\omega}}^N}\left( 1- \psi_N(\mathbf{Y}_1, \dots, \mathbf{Y}_N,\mathbf{z}_1,\dots, \mathbf{z}_N) \right)\text{d}\boldsymbol{\Pi}(\boldsymbol{\omega})\\
     \nonumber
     \le & e^{\beta_1N}e^{-N\epsilon^2/2},
\end{align} 
where the last inequality is from Equation \ref{test_bound2}. Thus by Markov's inequality and letting $\beta_1 < \epsilon^2/2$, we have that 
\begin{align}
    \nonumber
    & P\left( e^{\beta N}\left( 1- \psi_N(\mathbf{Y}_1, \dots, \mathbf{Y}_N,\mathbf{z}_1,\dots, \mathbf{z}_N)\right)\int_{\mathcal{U}^c}\prod_{i=1}^N \frac{f_i(\mathbf{Y}_i; \boldsymbol{\omega})}{f_i(\mathbf{Y}_i; \boldsymbol{\omega}_0)}\text{d}\boldsymbol{\Pi}(\boldsymbol{\omega}) \ge e^{-N((\epsilon^2/2 - \beta_1)/2)} \right)\\
    \nonumber
    \le & \frac{\mathbb{E}_{P_{\boldsymbol{\omega}_0}^N}\left(e^{\beta N}\left( 1- \psi_N(\mathbf{Y}_1, \dots, \mathbf{Y}_N,\mathbf{z}_1,\dots, \mathbf{z}_N)\right)\int_{\mathcal{U}^c}\prod_{i=1}^N \frac{f_i(\mathbf{Y}_i; \boldsymbol{\omega})}{f_i(\mathbf{Y}_i; \boldsymbol{\omega}_0)}\text{d}\boldsymbol{\Pi}(\boldsymbol{\omega})\right)}{e^{-N((\epsilon^2/2 - \beta_1)/2)}} \\
    \nonumber
    \le &  e^{-N((\epsilon^2/2 - \beta_1)/2)}
\end{align}
Letting $E_N$ be the event that $e^{\beta N}$ $\left( 1- \psi_N(\mathbf{Y}_1, \dots, \mathbf{Y}_N,\mathbf{z}_1,\dots, \mathbf{z}_N)\right)$
$\int_{\mathcal{U}^c}\prod_{i=1}^N \frac{f_i(\mathbf{Y}_i; \boldsymbol{\omega})}{f_i(\mathbf{Y}_i; \boldsymbol{\omega}_0)}\text{d}\boldsymbol{\Pi}(\boldsymbol{\omega})$
$\ge e^{-N((\epsilon^2/2 - \beta_1)/2)}$, we have that $\sum_{i=1}^\infty P(E_N) < \infty$. Thus, by the Borel-Cantelli lemma, we have that 
$$e^{\beta N}\left( 1- \psi_N(\mathbf{Y}_1, \dots, \mathbf{Y}_N,\mathbf{z}_1,\dots, \mathbf{z}_N)\right)\int_{\mathcal{U}^c}\prod_{i=1}^N \frac{f_i(\mathbf{Y}_i; \boldsymbol{\omega})}{f_i(\mathbf{Y}_i; \boldsymbol{\omega}_0)}\text{d}\boldsymbol{\Pi}(\boldsymbol{\omega}) \rightarrow 0$$ a.s. $[P_{\boldsymbol{\omega}_0}]$ for $0 <  \beta_1 < \epsilon^2/2$. Therefore, we have proved conditions (a), (b), and (c). Thus, letting $\beta$ in condition (c) be such that $\beta = \beta_1$, where $0 <  \beta_1 < \epsilon^2/2$, we can see that 
$\boldsymbol{\Pi}_N(\mathcal{U}^c| \mathbf{Y}_1, \dots, \mathbf{Y}_N) \rightarrow 0$ a.s. $[P_{\boldsymbol{\omega}_0}]$ for every weak neighborhood, $\mathcal{U}$ of $\boldsymbol{\omega}_0$.

\section{Case Studies}
\subsection{Simulation Study 1}
\label{Sim_Study_1_appendix}
In this simulation study, we examined how well we could recover the mean, covariance, and cross-covariance functions at different numbers of functional observations. For this simulation, we used 3 different numbers of functional observations ($N= 40, 80, 160$), and ran 50 MCMC chains for $500,000$ iterations on 50 different datasets. To help the chain converge, we used the Multiple Start Algorithm (Algorithm \ref{alg:MSA}) with $n\_try1 = 50$, $n\_try2 = 10$, $n\_MCMC1 = 2000$, and $n\_MCMC2 = 20000$. Due to our allocated computation budget, we did not use tempered transitions to help move around the space of parameters. The simulation took 134 hours to run using 5 CPU cores on an Apple M1 chip. To save memory, we only saved every $100$ iterations. We used eight functions to form the basis of the observed functions, such that the observed smooth functions lie in a space spanned by cubic b-spline basis functions with four equally spaced internal nodes ($P = 8$), and that 3 eigenfunctions can capture the entire covariance process ($M = 3$). For this simulation, we used the two-feature model ($K = 2$). We specified that $\sigma^2 = 0.001$, while the $\boldsymbol{\nu}$ parameters were drawn according to the following distributions:
$$\boldsymbol{\nu}_1 \sim \mathcal{N}\left((6, 4, \dots, -6, -8)', 4\mathbf{P} \right),$$
$$\boldsymbol{\nu}_2 \sim \mathcal{N}\left((-8, -6, \dots, 4, 6)', 4\mathbf{P} \right),$$
where $\mathbf{P}$ is the matrix corresponding with the first-order random walk penalty. We drew the $\boldsymbol{\Phi}$ parameters from the subspace orthogonal to the space spanned by the $\boldsymbol{\nu}$ parameters. Thus, let $\text{colsp}(\mathbf{B}^\perp) := \text{span}\{b^\perp_1, \dots,b^\perp_6\}\subset \mathbb{R}^8$ be the subspace orthogonal to the $\boldsymbol{\nu}$ parameters, which can be described as the span of $6$ vectors in $\mathbb{R}^8$. The $\boldsymbol{\Phi}$ parameters were drawn according to the following distributions:
$$\boldsymbol{\phi}_{km}  = \mathbf{q}_{km}\mathbf{B}^\perp \;\;\; k = 1, 2 \;\;\; m = 1,2,3,$$
where $\mathbf{q}_{k1} \sim \mathcal{N}(\mathbf{0}_{6}, 2.25\mathbf{I}_{6})$, $\mathbf{q}_{k2} \sim \mathcal{N}(\mathbf{0}_{6}, \mathbf{I}_{6})$, $\mathbf{q}_{k3} \sim \mathcal{N}(\mathbf{0}_{6}, 0.49\mathbf{I}_{6})$. Although this may not completely remove the effect of the non-identifiability mentioned in Section 2.2 of the main text, it should help minimize its impact on our recovery of the mean and covariance structures.

For the $\mathbf{z}_i$ and $\boldsymbol{\chi}_{im}$ parameters, we would draw 3 different sets of parameters (corresponding to the various number of functional observations). The parameters $\chi_{im}$ were drawn from a standard normal distribution. The $\mathbf{z}_i$ parameters were drawn from a mixture of Dirichlet distributions. Approximately 30\% of the $\mathbf{z}_i$ parameters were drawn from a Dirichlet distribution with $\alpha_1 = 10$ and $\alpha_2 = 1$. Another roughly 30\% of the $\mathbf{z}_i$ parameters were drawn from a Dirichlet distribution where $\alpha_1 = 1$ and $\alpha_2 = 10$. The rest of the $\mathbf{z}_i$ parameters were drawn from a Dirichlet distribution with $\alpha_1 = \alpha_2 = 1$. For each simulation, we used these parameters to simulate the observed functions of our model. 

\begin{figure}[htbp]
     \centering
     \subfigure[$C^{(1,1)}$]{\includegraphics[width=.49\textwidth]{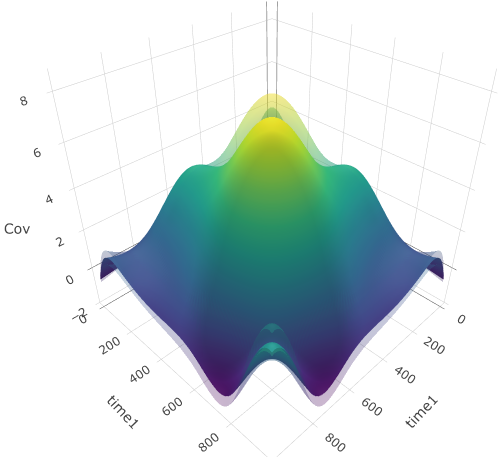}}
     \subfigure[$C^{(2,2)}$]{\includegraphics[width=.49\textwidth]{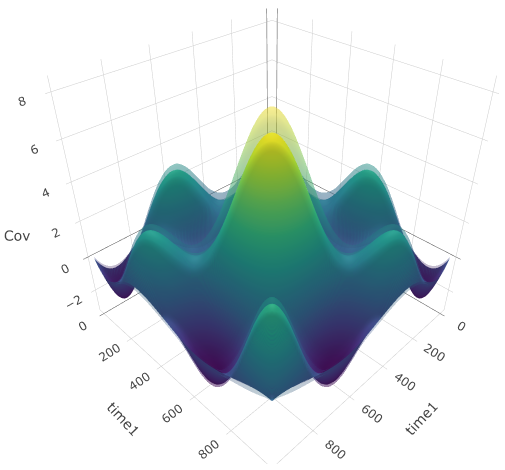}}
     \subfigure[$C^{(1,2)}$]{\includegraphics[width=.49\textwidth]{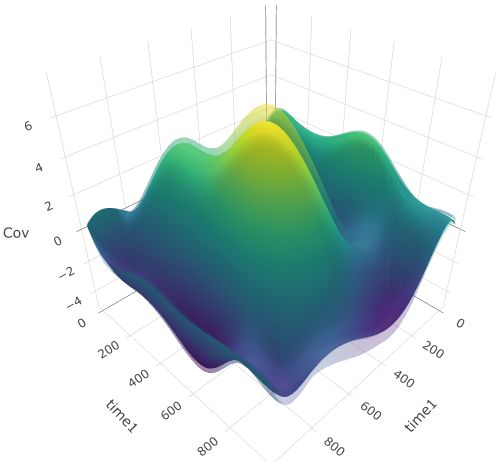}}
     \caption{Posterior median estimates of the covariance and cross-covariance functions (opaque) along with the true functions (transparent) for a simulated data set with 160 functional observations.}
     \label{fig:sim_1_cov}
\end{figure}

\begin{figure}[htbp]
    \centering
    \includegraphics[width = 0.99\textwidth]{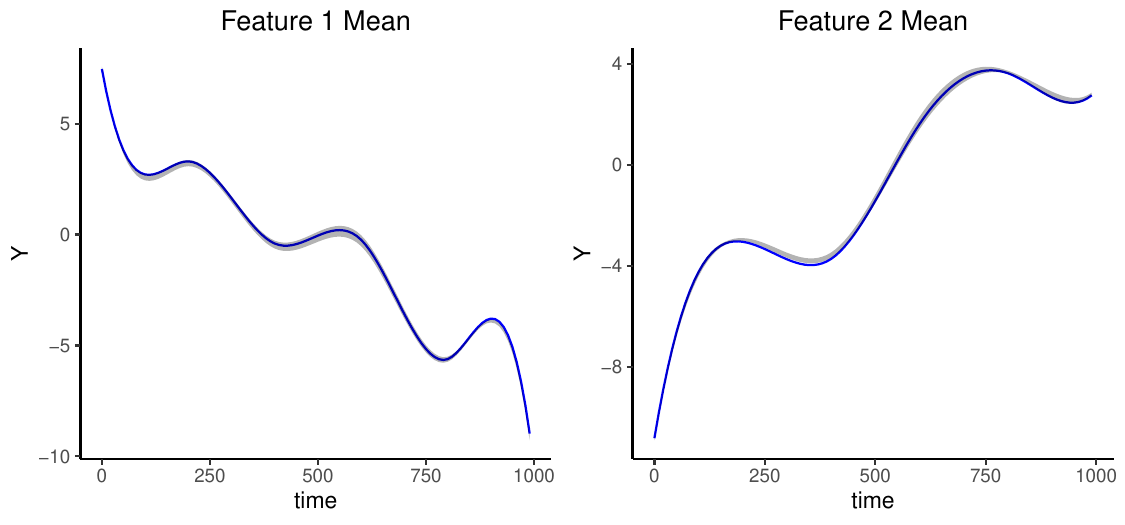}
    \caption{95\% credible interval of the posterior mean functions for the case when we have 160 functional observations.}
    \label{fig:sim_1_CI}
\end{figure}

Before getting the posterior median estimates of the functions of interest, we used the Membership Rescale Algorithm (Algorithm \ref{alg:MTA}) to help interpretability and identifiability. From figure \ref{fig:sim_1_cov}, we can see that we do a good job in recovering the covariance and cross-covariance functions. The estimated functions are slightly conservative as they tend to slightly underestimate the magnitude of the covariance functions. Figure \ref{fig:sim_1_CI} shows the median posterior mean recovered from one of the 50 datasets.

A supplementary simulation study was conducted in which we looked at the recovery of the mean, covariance, and cross-covariance functions when we have three functional features ($K = 3$). In this simulation study, we looked at how we recovered the functions of interest in 3 different sample sizes of the observed functions ($N = 50, 100, 200$). In this supplementary simulation study, we generated 10 different datasets for each of the three sample sizes. The parameters used to create the datasets were generated as follows. For this simulation study, $\sigma^2 = 0.001$, and the $\boldsymbol{\nu}$ parameters were drawn according to the following distributions:
$$\boldsymbol{\nu}_1 \sim \mathcal{N}\left((6, 4, \dots, -6, -8)', 2\mathbf{P} \right),$$
$$\boldsymbol{\nu}_2 \sim \mathcal{N}\left((-8, -6, \dots, 4, 6)', 2\mathbf{P} \right),$$
$$\boldsymbol{\nu}_3 \sim \mathcal{N}\left((4, 4, \dots, 4)', 2\mathbf{P} \right),$$
where $\mathbf{P}$ is the matrix corresponding with the first-order random walk penalty. Due to the non-identifiability described in Section 2.2 of the main text, we drew the $\boldsymbol{\Phi}$ parameters from the subspace orthogonal to the space spanned by the $\boldsymbol{\nu}$ parameters. Thus, let $\text{colsp}(\mathbf{B}^\perp) := \text{span}\{b^\perp_1, \dots,b^\perp_5\}\subset \mathbb{R}^8$ be the subspace orthogonal to the $\boldsymbol{\nu}$ parameters, which can be described as the span of $5$ vectors in $\mathbb{R}^8$. The $\boldsymbol{\Phi}$ parameters were drawn according to the following distributions:
$$\boldsymbol{\phi}_{km}  = \mathbf{q}_{km}\mathbf{B}^\perp \;\;\; k = 1, 2, 3 \;\;\; m = 1,2,3,$$
where $\mathbf{q}_{k1} \sim \mathcal{N}(\mathbf{0}_{6}, \mathbf{I}_{5})$, $\mathbf{q}_{k2} \sim 0.81\mathcal{N}(\mathbf{0}_{6}, \mathbf{I}_{5})$, $\mathbf{q}_{k3} \sim \mathcal{N}(\mathbf{0}_{6}, 0.49\mathbf{I}_{5})$. The parameters $\chi_{im}$ were drawn from a standard normal distribution. The $\mathbf{z}_i$ parameters were drawn from a mixture of Dirichlet distributions. Approximately 20\% of the $\mathbf{z}_i$ parameters were drawn from a Dirichlet distribution with $\alpha_1 = 100$, $\alpha_2 = 1$, and $\alpha_3 = 1$. Another roughly 20\% of the $\mathbf{z}_i$ parameters were drawn from a Dirichlet distribution where $\alpha_1 = 1$, $\alpha_2 = 100$, and $\alpha_3 = 1$. Another roughly 20\% of the $\mathbf{z}_i$ parameters were drawn from a Dirichlet distribution where $\alpha_1 = 1$, $\alpha_2 = 1$, and $\alpha_3 = 100$. Thus, the first 60\% of the allocation parameters were drawn so that most of the membership was in one of the three features, corresponding to the separability condition. The rest of the $\mathbf{z}_i$ parameters were drawn from a Dirichlet distribution with $\alpha_1 = \alpha_2 = \alpha_3 = 1$.

\begin{figure}
    \centering
    \includegraphics[width = 0.95\textwidth]{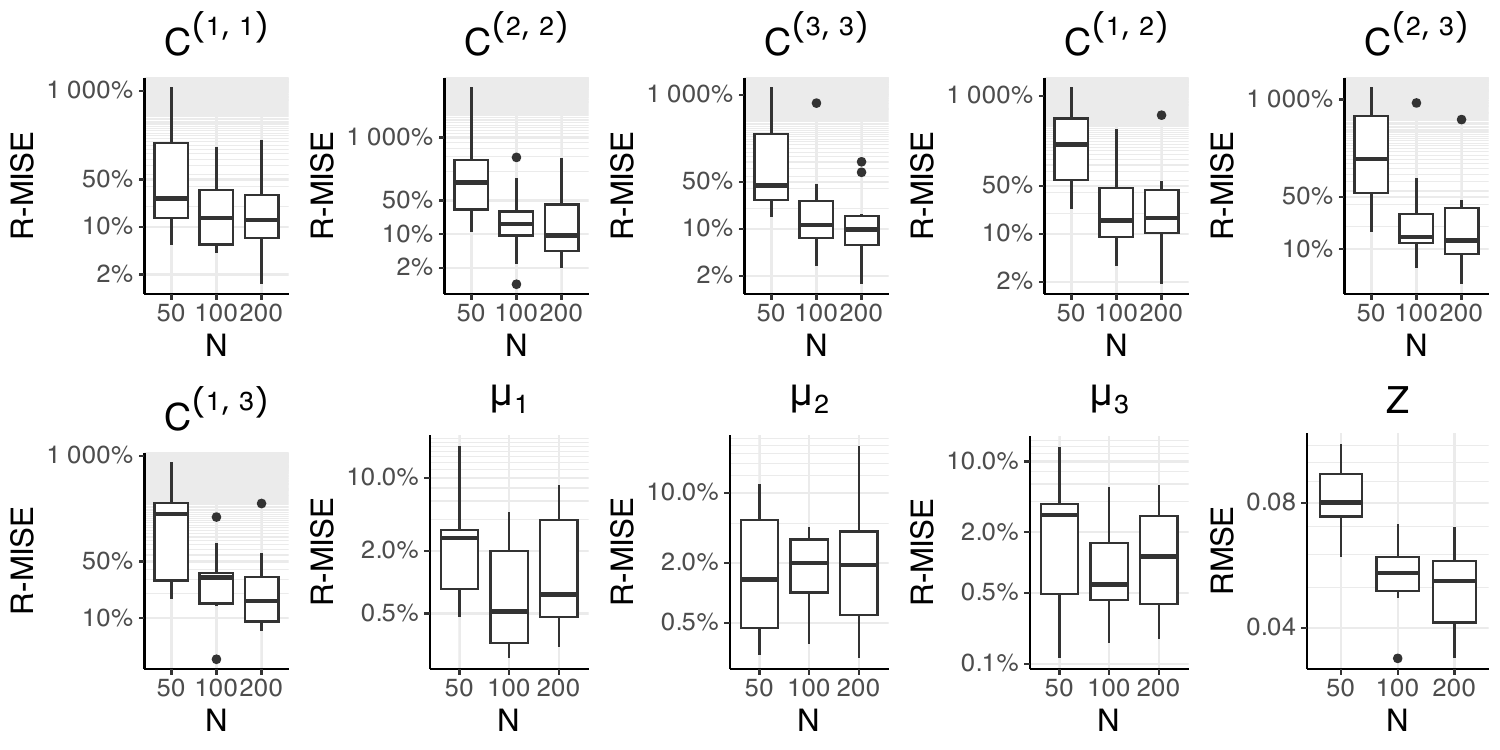}
    \caption{R-MISE values for the latent feature means and cross-covariances, as well as RMSE values for the allocation parameters, evaluated as sample size increases for the three feature model.}
    \label{fig:Sim1_appendix}
\end{figure}

The simulation took 56 hours to run using 5 CPU cores on an Apple M1 chip. The results of the simulation can be seen in Figure \ref{fig:Sim1_appendix}. Similarly to the two-feature case study, we have a relatively good recovery of the mean structure with as little as 50 functional observations. Similarly to the two-feature case study, we can see that the recovery of the covariance structure improves greatly as we add more functional observations. Compared to the relative error estimates obtained from the two-feature case study, we can see that the relative error is higher overall in the three-feature case study, suggesting that more functional observations are needed in order to obtain similar error rates. This is not surprising, since three-feature models have more parameters and represent more complicated generative processes. Overall, the results in this supplementary simulation study support the results in the two-feature simulation study and provide empirical evidence that given an identifiable model (i.e. assuming the separability condition) our estimates of the mean, covariance, cross-covariance, and allocation structures converge to the truth as the number of functional observations increases.

\subsection{Simulation Study 2}
\label{Sim_Study_2_appendix}
Picking the number of features can be a challenging task for many practitioners, especially when there is little scientific knowledge on the data. Practitioners often rely on information criteria to help choose the number of features. In this simulation, we simulate 10 different ``true'' data sets from a 3-feature model to see if the information criteria can help pick the correct number of features. This simulation study took 35 hours to complete using 5 CPU cores on an Apple M1 chip. For this simulation study, we considered testing the information criteria when $K = 2,3,4,$ and $5$ (where $K$ is the number of features in our model). For each $K$ and each data set, we run an MCMC chain for 100,000 iterations each. To help the chain converge, we use the Multiple Start Algorithm (Algorithm \ref{alg:MSA}) with $n\_try1 = 50$, $n\_try2 = 5$, $n\_MCMC1 = 2000$, and $n\_MCMC2 = 4000$. To save memory, we only saved once every 10 iterations. 

The BIC, proposed by \citet{schwarz1978estimating}, is defined as:
$$\text{BIC} = 2\text{log}P\left(\mathbf{Y}|\hat{\boldsymbol{\Theta}}\right) - d\text{log}(n)$$
where $d$ is the number of parameters and $\hat{\boldsymbol{\Theta}}$ are the maximum likelihood estimators (MLE) of our parameters. In the case of our proposed model, we have that
\begin{equation}
\text{BIC} = 2\text{log}P\left(\mathbf{Y}|\hat{\boldsymbol{\nu}}, \hat{\boldsymbol{\Phi}}, \hat{\sigma}^2, \hat{\mathbf{Z}}, \hat{\boldsymbol{\chi}}\right) - d\text{log}(\tilde{N})
    \label{BIC}
\end{equation}
where $\tilde{N} = \sum_{i}n_i$ (where $n_i$ is the number of observed time points observed for the $i^{th}$ function), and $d = (N + P)K + 2MKP + 4K + (N + K)M + 2 $.

Similarly, the AIC, proposed by \citet{akaike1974new}, can be written as
\begin{equation}
    \text{AIC} = -2\text{log}P\left(\mathbf{Y}|\hat{\boldsymbol{\nu}}, \hat{\boldsymbol{\Phi}}, \hat{\sigma}^2, \hat{\mathbf{Z}}, \hat{\boldsymbol{\chi}}\right) + 2d.
    \label{AIC}
\end{equation}

Following the work of \citet{roeder1997practical}, we will use the posterior mean instead of the MLE for our estimates of BIC and AIC. In particular, we will use the posterior mean of the mean function in Equation 11 of the main text for each functional observation, as well as the posterior mean of $\sigma^2$, to estimate the BIC and AIC.

The modified DIC, proposed by \citet{celeux2006deviance}, is advantageous to the original DIC proposed by \citet{spiegelhalter2002bayesian} when we have a posterior distribution with multiple modes and when identifiability may be a problem. The modified DIC (referred to as $\text{DIC}_3$ in \citet{celeux2006deviance}) is specified as
\begin{equation}
    \text{DIC} = -4 \mathbb{E}_{\boldsymbol{\Theta}}[\text{log} f(\mathbf{Y}|\boldsymbol{\Theta})|\mathbf{Y}] + 2 log \hat{f}(\mathbf{Y})
    \label{DIC}
\end{equation}
where $\hat{f}(y_{ij}) = \frac{1}{N_{MC}}\sum_{l=1}^{N_{MC}}P\left(y_{ij}|\boldsymbol{\nu}^{(l)}, \boldsymbol{\Phi}^{(l)}, \left(\sigma^2\right)^{(l)}, \mathbf{Z}^{(l)}\right)$, $\hat{f}(\mathbf{Y}) = \prod_{i=1}^{N}\prod_{j=1}^{n_i}\hat{f}(y_{ij})$, and $N_{MC}$ is the number of MCMC samples used for estimating $\hat{f}(y_{ij})$. We can approximate $\mathbb{E}_{\boldsymbol{\Theta}}[\text{log} f(\mathbf{Y}|\boldsymbol{\Theta})|\mathbf{Y}]$ by using the MCMC samples, such that
$$\mathbb{E}_{\boldsymbol{\Theta}}[\text{log} f(\mathbf{Y}|\boldsymbol{\Theta})|\mathbf{Y}] \approx \frac{1}{N_{MC}} \sum_{l=1}^{N_{MC}}\sum_{i=1}^{N}\sum_{j=1}^{n_i}\text{log}\left[P\left(y_{ij}|\boldsymbol{\nu}^{(l)},\boldsymbol{\Phi}^{(l)}, \left(\sigma^2\right)^{(l)}, \mathbf{Z}^{(l)}\right)\right].$$

For the 10 ``true'' datasets with 3 functional features ($K=3$) and 200 functional observations ($N = 200$, $n_i = 100$), we assumed that the observed smooth functions lie in a space spanned by cubic b-spline basis functions with 4 equally spaced internal nodes ($P = 8$), and that 3 eigenfunctions can capture the entire covariance process ($M = 3$). In this simulation, we assumed that $\sigma^2 = 0.001$, and randomly drew the $\boldsymbol{\nu}$ and $\boldsymbol{\Phi}$ parameters for each data-set according to the following distributions:
$$\boldsymbol{\nu}_1 \sim \mathcal{N}\left((6, 4, \dots, -6, -8)', 4\mathbf{P} \right),$$
$$\boldsymbol{\nu}_2 \sim \mathcal{N}\left((-8, -6, \dots, 4, 6)', 4\mathbf{P} \right),$$
$$\boldsymbol{\nu}_1 \sim \mathcal{N}\left(\mathbf{0}, 4\mathbf{P} \right),$$
$$\boldsymbol{\phi}_{k1} \sim \mathcal{N}\left(\mathbf{0}, \mathbf{I}_8 \right),$$
$$\boldsymbol{\phi}_{k2} \sim \mathcal{N}\left(\mathbf{0}, 0.5\mathbf{I}_8 \right),$$
$$\boldsymbol{\phi}_{k3} \sim \mathcal{N}\left(\mathbf{0}, 0.2\mathbf{I}_8 \right).$$

The $\chi_{im}$ parameters were drawn from a standard normal distribution, while the $\mathbf{Z}$ parameters were drawn from a mixture of Dirichlet distributions. 20\% of the $\mathbf{z}_i$ parameters were drawn from a Dirichlet distribution with $\alpha_1 = 10$, $\alpha_2 = 1$, and $\alpha_3 = 1$. Another 20\% of the $\mathbf{z}_i$ parameters were drawn from a Dirichlet distribution where $\alpha_1 = 1$, $\alpha_2 = 10$, and $\alpha_3 = 1$. Another 20\% of the $\mathbf{z}_i$ parameters were drawn from a Dirichlet distribution where $\alpha_1 = 1$, $\alpha_2 = 1$, and $\alpha_3 = 10$. The rest of the $\mathbf{z}_i$ parameters were drawn from a Dirichlet distribution with $\alpha_1 = \alpha_2 = 1$. Once all the parameters for the ``true'' data set were specified, the observed data points were generated according to the model. MCMC was then carried out with various values of $K$, but with the correct number of eigenfunctions, $M$, and the correct basis functions.

One may be tempted to put a prior on $K$, however extending the current model to allow for $K$ to have a prior on it is not a trivial task.  One popular way to do trans-dimensional sampling is through reversible-jump MCMC (RJMCMC). Unfortunately, RJMCMC requires a one-to-one and differentiable mapping between dimensions. While good mappings can be found for simple models, finding a good mapping for a model like like the functional mixed membership model is non-trivial. Without a good mapping, we will have poor mixing of the Markov Chain, where we would be unlikely to ever change directions.

Another option would be to use a hierarchical Dirichlet process \citep{teh2004sharing}. Dirichlet process mixture models have become somewhat popular, especially because inference can be performed using algorithms similar to those described in \citep{neal2000markov}. However, due to the continuous nature of the allocation parameters $Z_{ik}$ and the need to estimate the cross-covariance function between different features, the algorithms described in \citep{neal2000markov} could not be easily used to extend this framework to the infinite-dimensional features case. Furthermore, ensuring identifiability, through either the separability condition or the sufficiently scattered condition \citep{chen2022learning}, in a non-parametric model would also be challenging with changing dimensions.

Therefore, due to the poor mixing often found in RJMCMC and the non-trivial extension using a hierarchical Dirichlet process, we did not put a prior on $K$. While the user needs to currently specify the number of features in our model, we maintain that BIC and heuristics such as the elbow method can help users choose the number of features for their model. 

\subsection{A Case Study of EEG in ASD}
\label{EEG_case_study_appendix}
In this study, we grouped patients according to their T8 electrode signal. Each child in the study was instructed to look at a computer monitor that displayed bubbles for two minutes in a dark, sound-attenuated room \citep{dickinson2018peak}. While the children were watching the screen, EEG data were being recorded using a 128-channel HydroCel Geodesic Sensor Net. The data were then filtered to remove signals outside of the 0.1 - 100 Hz band, and then interpolated to match the international 10-20 system 25 channel montage. The data were also transformed into the frequency domain using a fast Fourier transform (FFT). In this study, we assume that the underlying smooth random functions will lie in a space spanned by cubic b-spline basis functions with 4 equally spaced internal nodes ($P = 8$) and that 3 eigenfunctions can sufficiently capture the covariance process ($M = 3$). We ran the MCMC chain for 500,000 iterations and used the Multiple Start Algorithm, described in Section C.2 of the web-based supporting materials, to obtain a good initial starting point.

We used a two-functional-feature model and found that patients in the first feature could be interpreted as $1/f$ noise, while the second feature could be interpreted as a distinct PAF.  We also fit a three-functional-feature mixed membership model, but found that the two-feature mixed membership model (AIC = -13091.31, BIC = 9369.66, DIC = -13783.5) appeared to be optimal compared to the three-feature mixed membership model (AIC = -12831.17, BIC = 8138.12, DIC= -13815.34). Figure \ref{fig:real_1_cov} shows the median of the posterior posterior distribution of the covariance and cross-covariance functions. We can see that the covariance function associated with feature 1 ($C^{(1,1)}$) has high covariance around 6 Hz, which is where we have the highest power of $1/f$ noise.

 Looking at the mean function for feature 2 (Figure 5 of the main text), we can see that the peak power occurs at around 9 Hz. However, for people who have a distinct PAF pattern, it is common for their peak power to occur anywhere between 9 Hz and 11 Hz. When looking at the covariance function associated with feature 2 ($C^{(2,2)}$), we can see that this is being modeled by the high variance at 9 Hz and at 11 Hz. We can also see that people who only have high Alpha power typically tend to have only one peak in the alpha band, which is also accounted for in our model by the negative covariance between 9 and 11 Hz. When looking at the cross-covariance function, we can see that there is high cross-covariance between 9 Hz in feature 1 and 6 Hz in feature 2, and negative cross-covariance between 11 Hz in feature 1 and 6 Hz in feature 2. This means that patients who are simultaneously in features 1 and 2 that have moderate $1/f$ noise are likely to have moderate alpha power around 9 Hz and are less likely to have a peak around 11 HZ. According to the scientific literature, this is likely to occur in younger TD individuals.
 
 From Figure 6 of the main text, we can see that on average children with ASD tended to belong to feature 1 more than feature 2. Thus, on average, ASD children tended to have a less distinct PAF when compared to TD children.

\begin{figure}
     \centering
     \includegraphics[width=.95\textwidth]{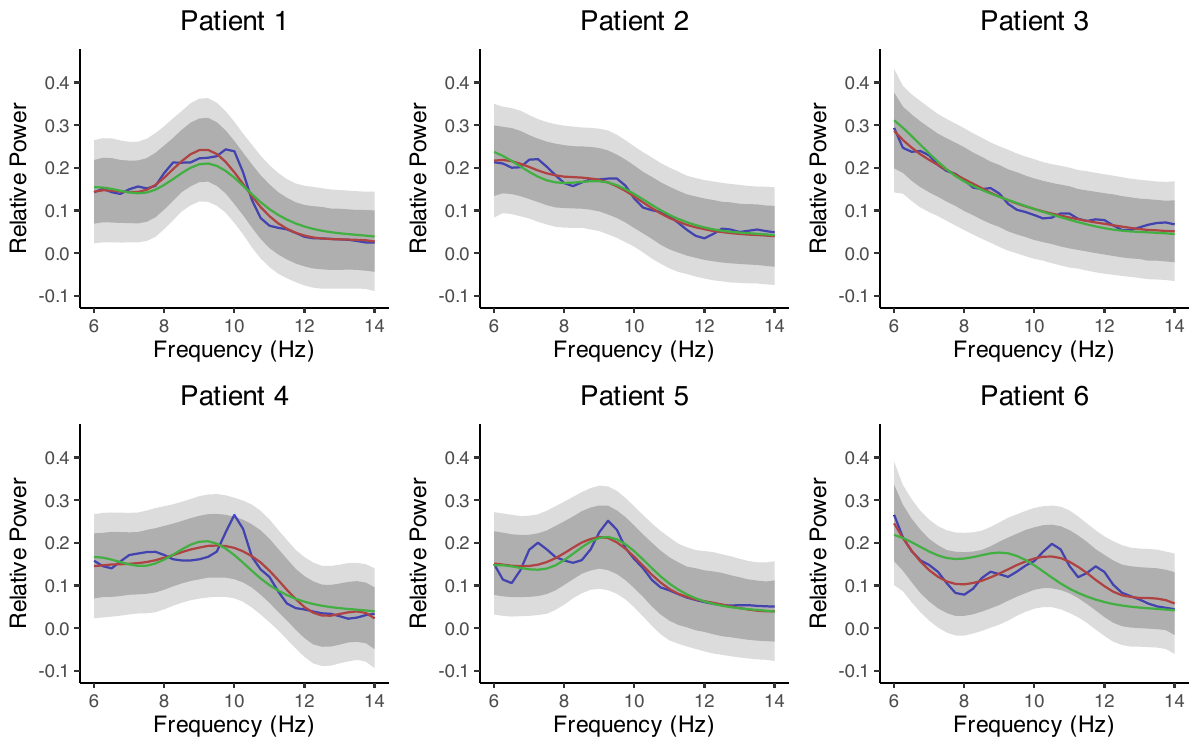}
     \caption{Posterior median and 95\% credible intervals (pointwise CI in dark gray and simultaneous CI in light gray) of the individual trajectories for the first 6 individuals.  The observed data is depicted as the blue curve, the median predicted trajectory is represented in red, and the median predicted trajectory using only the feature means (without the effects of covariance and cross-covariance functions) are represented in green.}
     \label{fig: posterior_pred}
 \end{figure}

\begin{figure}
     \centering
     \subfigure[$C^{(1,1)}$]{\includegraphics[width=.49\textwidth]{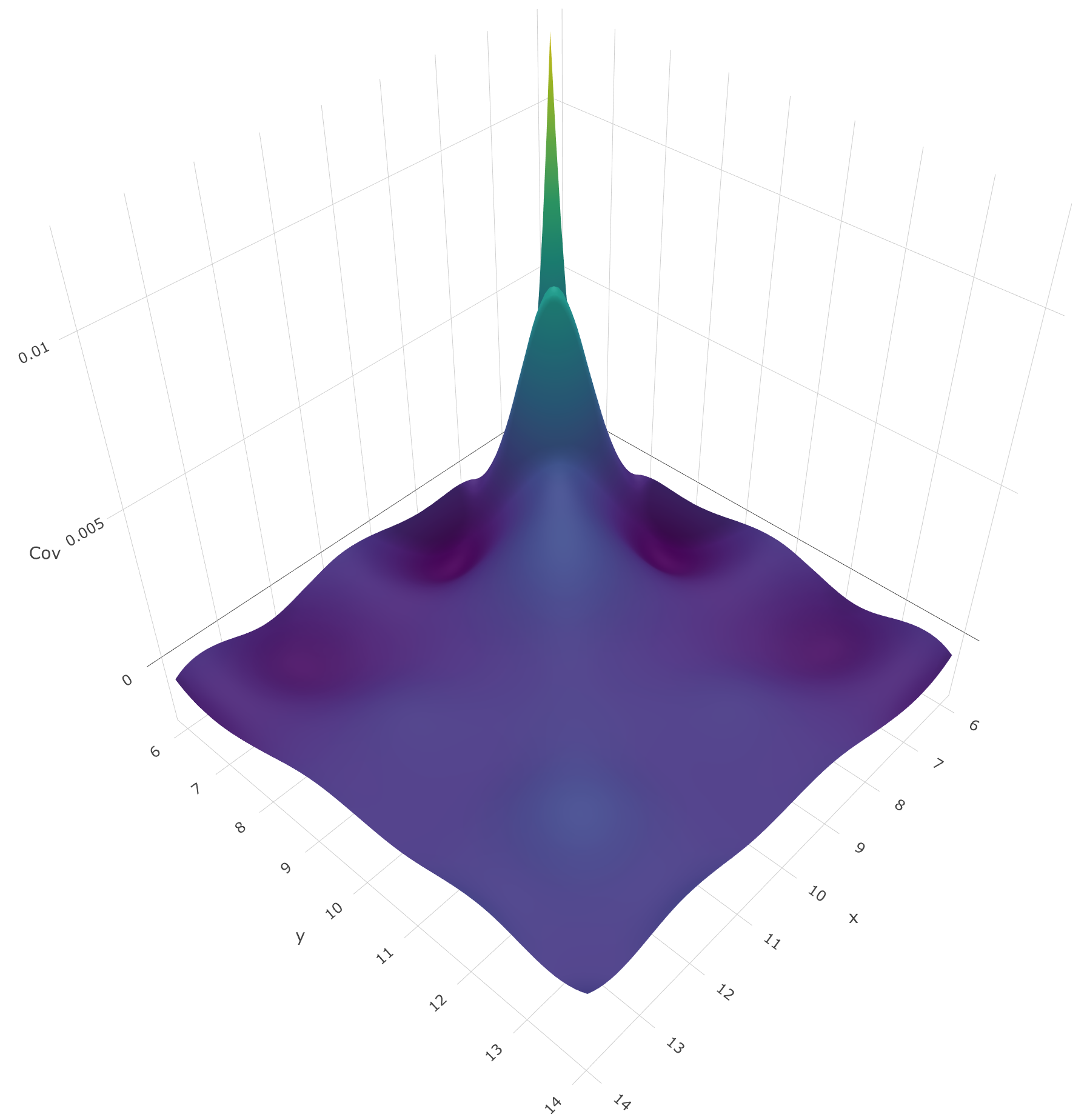}}
     \subfigure[$C^{(1,2)}$]{\includegraphics[width=.49\textwidth]{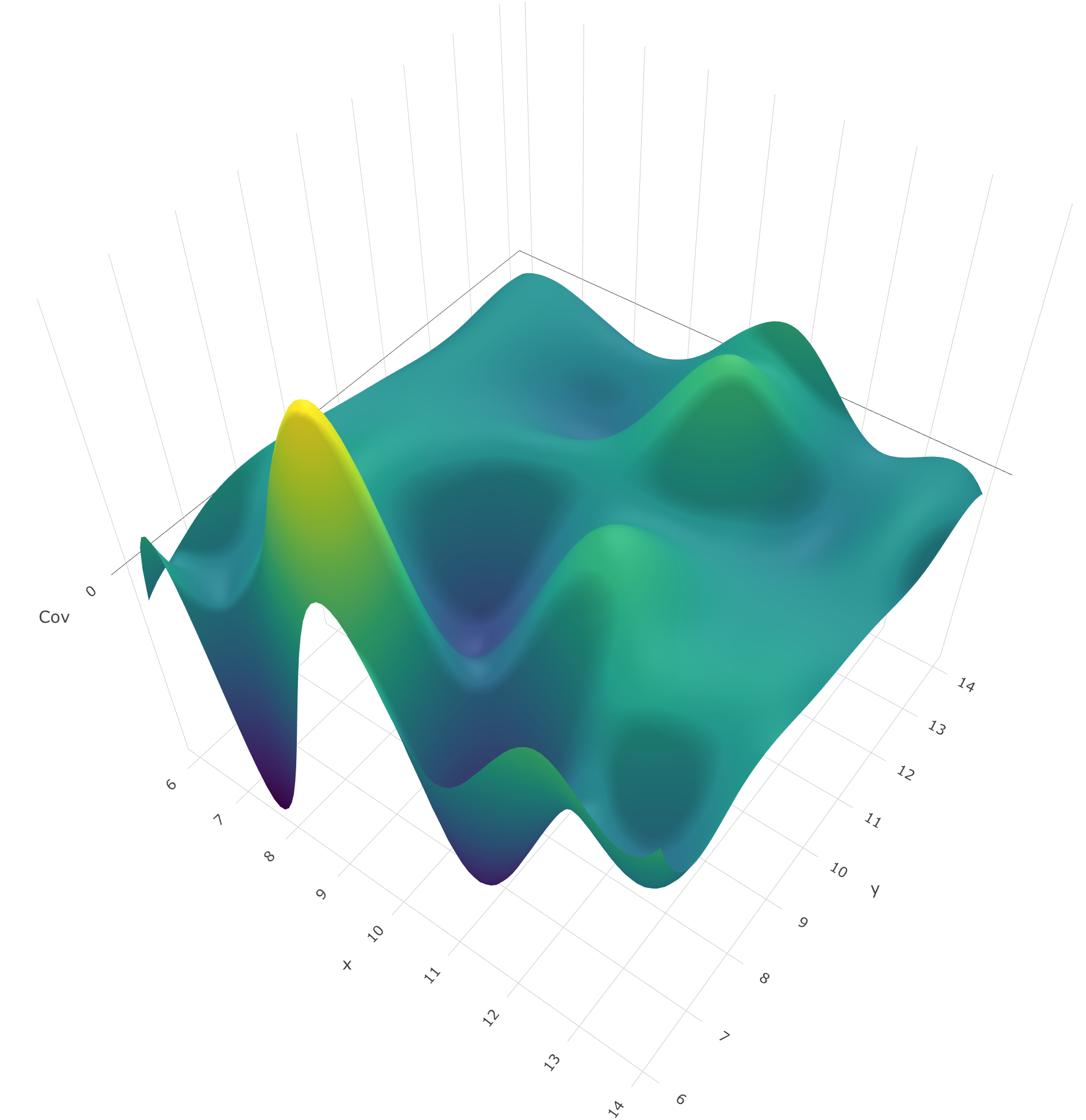}}
     \subfigure[$C^{(2,2)}$]{\includegraphics[width=.49\textwidth]{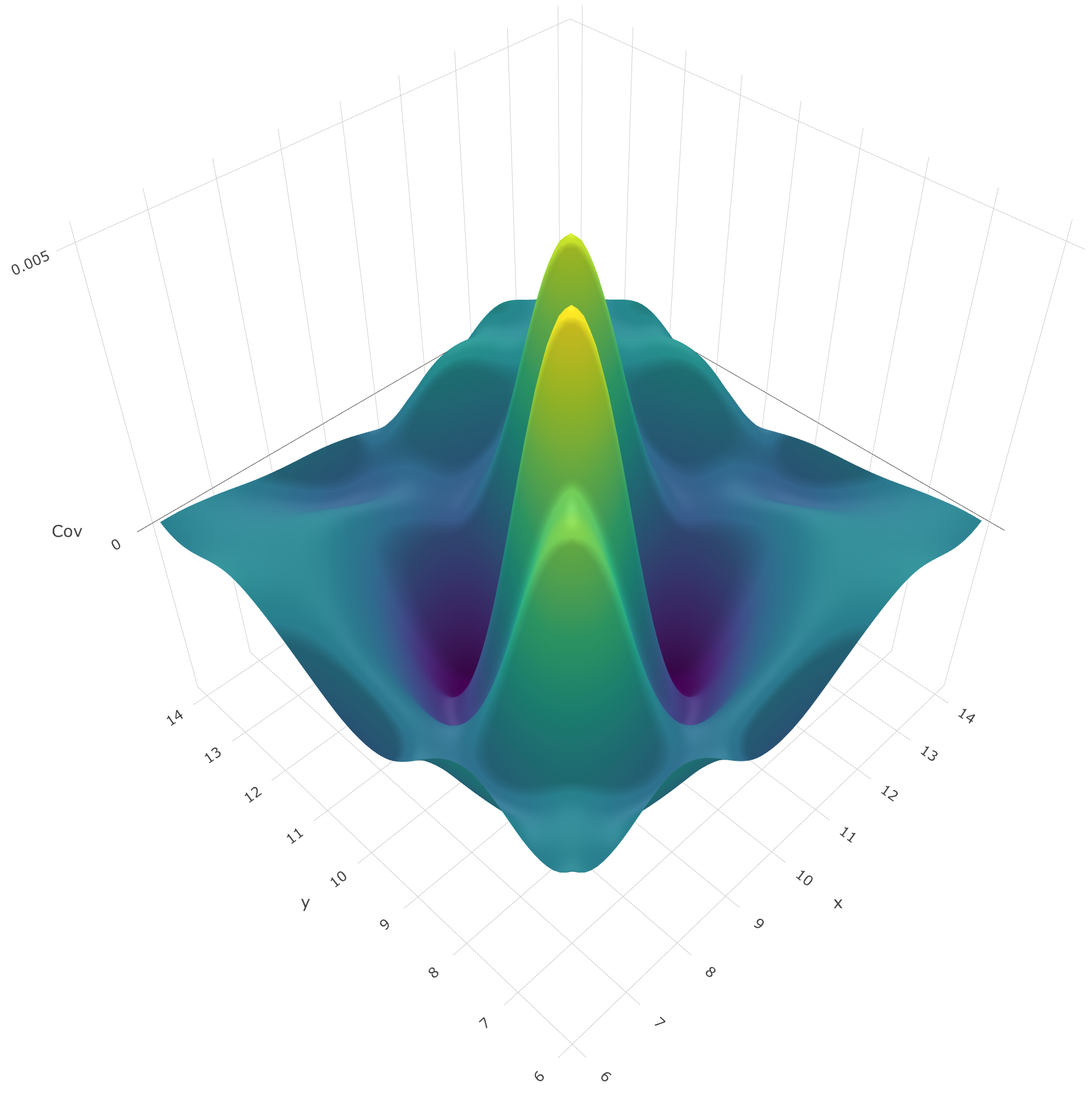}}
     \caption{Posterior estimates of the covariance and cross-covariance functions}
     \label{fig:real_1_cov}
\end{figure}

Figure \ref{fig: posterior_pred} shows a visualization of the posterior distribution of the individual trajectories compared to the observed sample paths. We can see that the observed paths are relatively noisy compared to the estimated trajectories. EEG signals can be fairly noisy due to contamination from biological artifacts (eye movements, muscle activity, pulse, ect.) and other external factors such as electrical interference \citep{fitzgibbon2007removal}. When considering EEG data, we often conceptualize the observed sample path as some underlying smooth process contaminated by noise. Even though the estimated paths are smoother, we can see that the proposed framework is able to model trajectories with shifted PAFs (i.e. patient 1 has a PAF around 9 Hz and patient 6 has a PAF slightly over 10 Hz), as well as model trajectories with various ratios of periodic signals to aperiodic signals (patients 2 and 3 have a high proportion of aperiodic signals, while patients 1 and 5 have a high proportion of periodic signals). In this case, the ability to model trajectories with shifted PAFs is possible because of the estimation of the covariance structure. This is particularly apparent in patient 6, where the mean predicts an alpha peak with a much lower PAF (represented in green), but the estimated PAF is shifted to the right by the eigenfunctions of the covariance operator (represented in red). As noted in the main text, it has been shown that the alpha power can shift as children age \citep{scheffler2019covariate}. Furthermore, \citet{haegens2014inter} noted that shifts in alpha oscillations can confound measurements of alpha power if not taken into account. Thus, the ability for us to model the covariance and cross-covariance functions is crucial in this application to prevent the PAF shifts from confounding measures of the alpha power.

Alternative analyzes were performed on the same data using a model-based functional clustering algorithm \citep{pya2021fdamocca} and using functional principal component analysis (FPCA) \citep{fdapace}. Due to the lack of packages to implement functional factor analysis models in R, we were unable to fit a functional factor analysis model to compare the results with our mixed membership model. However, due to the similarity between FPCA and functional factor analysis models, we should be able to get a general idea of what a functional factor analysis model would look like. 

We will start with a brief review of the functional forms of FPCA, factor analysis models, and our mixed membership models. Let $f: \mathcal{T} \rightarrow \mathbb{R}$ be a continuous square-integrable stochastic process, where $\mathcal{T}$ is a compact subset of $\mathbb{R}^d$. We can arrive at the functional form of FPCA by utilizing the Karhunen-Loève theorem, such that 
\begin{equation}
    f(t) - \mu(t) = \sum_{k=1}^\infty \xi_k\phi_k(t),
    \label{eq: FPCA}
\end{equation}
where $\xi_k := \int_{\mathcal{T}}\left(f(t) - \mu(t)\right)\phi_k(t)\text{d}t$ is the principal component associated with the $k^{th}$ eigenfunction $\phi_k(t)$. While the decomposition has an infinite expansion, in practice we only use a finite number of eigenfunctions to explain a majority of the variance. \citet{shang2014survey} gives a more comprehensive review of FPCA in their review paper. Factor analysis models for functional data can be thought of as a probabilistic representation of FPCA. Following \cite{hsing2015theoretical}, we can write a functional factor analysis model as:
\begin{equation}
    Y(t) - \mu(t) = \sum_{m= 1}^M Z_m \psi_m(t) + \epsilon(t),
    \label{eq: functional_factor}
\end{equation}
where $e(t)$ is an error term, $Z_m \in \mathbb{R}$ are known as the \textit{factors}, and $\psi_m(t)$ are known as the \textit{loadings}. In factor analysis, we assume the following:

\begin{assumption}
    $Z_m$ are mean zero, uncorrelated random variables with unit variance.
    \label{assumption: fun fact1}
\end{assumption}
\begin{assumption}
    $e \sim \mathcal{GP}(0, \mathcal{C})$.
    \label{assumption: fun fact2}
\end{assumption} 
\begin{assumption}
    $\psi_m$ are orthogonal functions
    \label{assumption: fun fact3}
\end{assumption}

The main difference between the two models is that the functional factor analysis model assumes one common mean, while the functional mixed membership model assumes feature-specific means. The covariance structure of the two models also differs in that the functional factor analysis models assume that the loadings are mutually independent (assumption \ref{assumption: fun fact1}), while the functional mixed membership model allows for dependence between features. The main differences between the two models can be seen in Table \ref{tab: Fun_Fact_Fun_Mixed}.

\begin{table}
    \begin{tabular}{|c|c|c|}
        \hline
         & Functional Factor Analysis Model & Functional Mixed Membership Model  \\
        \hline
        Mean & $\mu(t)$ & $\sum_{k=1}^K Z_k\mu^{(k)}$\\
        Covariance & $\sum_{m=1}^m\psi_m(t)\psi_m(t') + \mathcal{C}(t,t')$ & $\sum_{k=1}^K\sum_{k' =1}^K Z_{ik}Z_{ik'}C^{(k,k')}(t, t') + \sigma^2$ \\
        \hline
    \end{tabular}
    \label{tab: Fun_Fact_Fun_Mixed}
    \caption{Differences in mean and covariance structures of functional factor analysis models and functional mixed membership models.}
\end{table}

Alternatively, one may desire to use factors in a similar fashion as the allocation parameters in the mixed membership model. In that case, we can see that the mean structures between the two models are similar. In the case of the functional factor analysis model, the mean becomes $\mu(t) + \sum_{m=1}^M Z_m\psi_m(t)$, which is similar to the mean of the functional mixed membership model found in Table \ref{tab: Fun_Fact_Fun_Mixed}. The main difference between the two mean structures is the constraint in the functional mixed membership model that the allocation parameters have to lie in the unit simplex. While the mean structures are relatively similar, the covariance structures are completely different. The functional factor analysis model assumes that there is a common covariance, $\mathcal{C}(t,t')$, for all observations. The functional mixed membership model allows for the covariance to change depending on the allocation parameters.

\begin{figure}[htb]
    \centering
    \includegraphics[width=.9\textwidth]{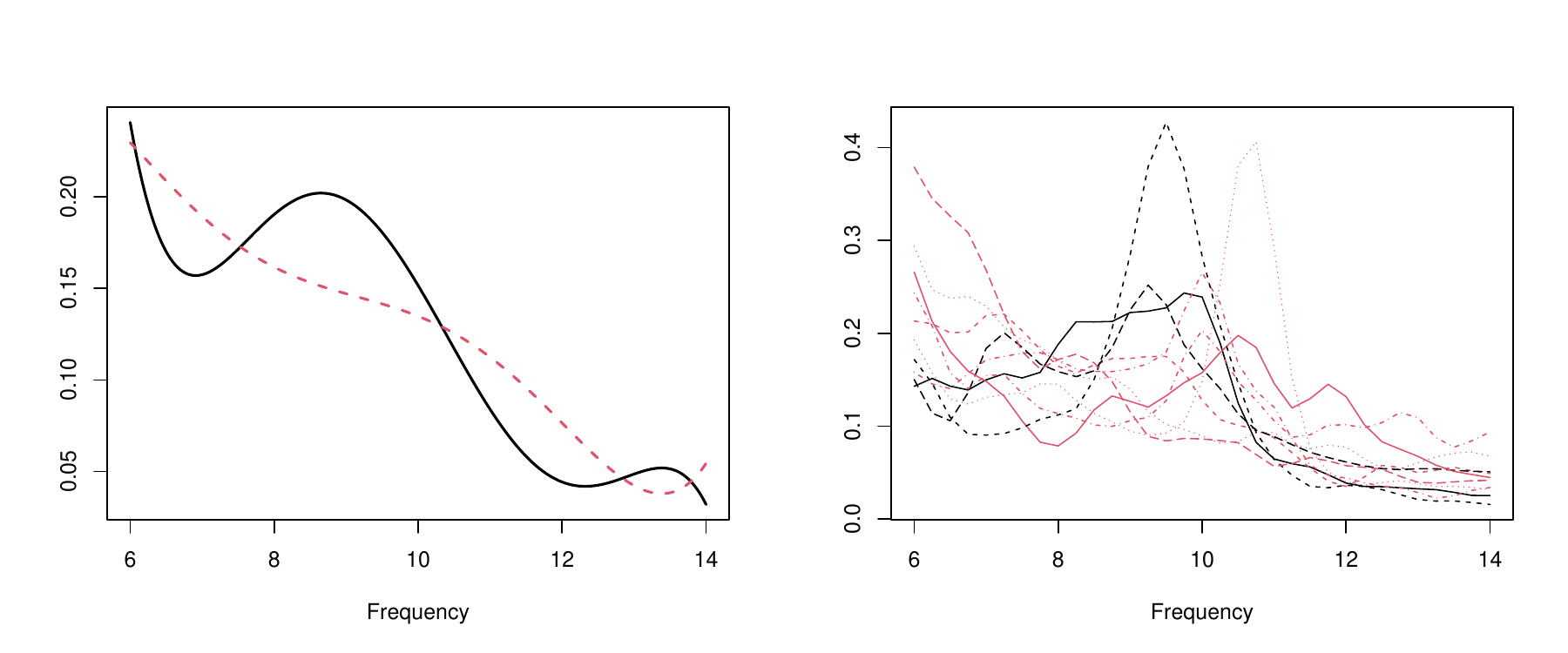}
    \caption{Model-based functional clustering with 2 clusters, fit on the data from the case study of EEG in ASD. The right subfigure illustrates the observed functions, colored based off of the estimated cluster membership.}
    \label{fig:clustering2}
\end{figure}

To illustrate the differences between our proposed model and alternative models, we fit a model-based clustering model and performed FPCA on the data from the EEG case study. Figure \ref{fig:clustering2} shows the results of fitting a model=based clustering model with 2 clusters. Compared to the results found in Section 4.3 of the main text, we can see that the cluster means seem like mixtures of the two features found in mixed membership model. The first cluster, denoted by the black functions in Figure \ref{fig:clustering2} can be interpreted as a distinct alpha peak with some $1/f$ trend, yet we can see that the PAF is relatively low with a frequency around 8 to 8.5 Hz. The second cluster, denoted by the red functions in Figure \ref{fig:clustering2} can be interpreted as mainly a $1/f$ trend, however we can see that there is relatively high power in the higher frequency range (11Hz to 12 Hz), which would also indicate a weaker alpha peak. Both of these clusters are not very easy to interpret, and do a poor result of explaining data heterogeneity. Namely, from the subfigure on the right of Figure \ref{fig:clustering2}, we can see that observed functions that have distinct alpha peaks with PAFs in the high frequency range are clustered in the second cluster (red group), and not the first cluster. In this case, a mixed membership model seems to be more appropriate because the mean of the features represents the mean of extreme observations, leading to interpretable features. Clustering tends to average out the periodic and aperiodic signals, leading to hard-to-interpret clusters.

\begin{figure}[H]
    \centering
    \includegraphics[width=.9\textwidth]{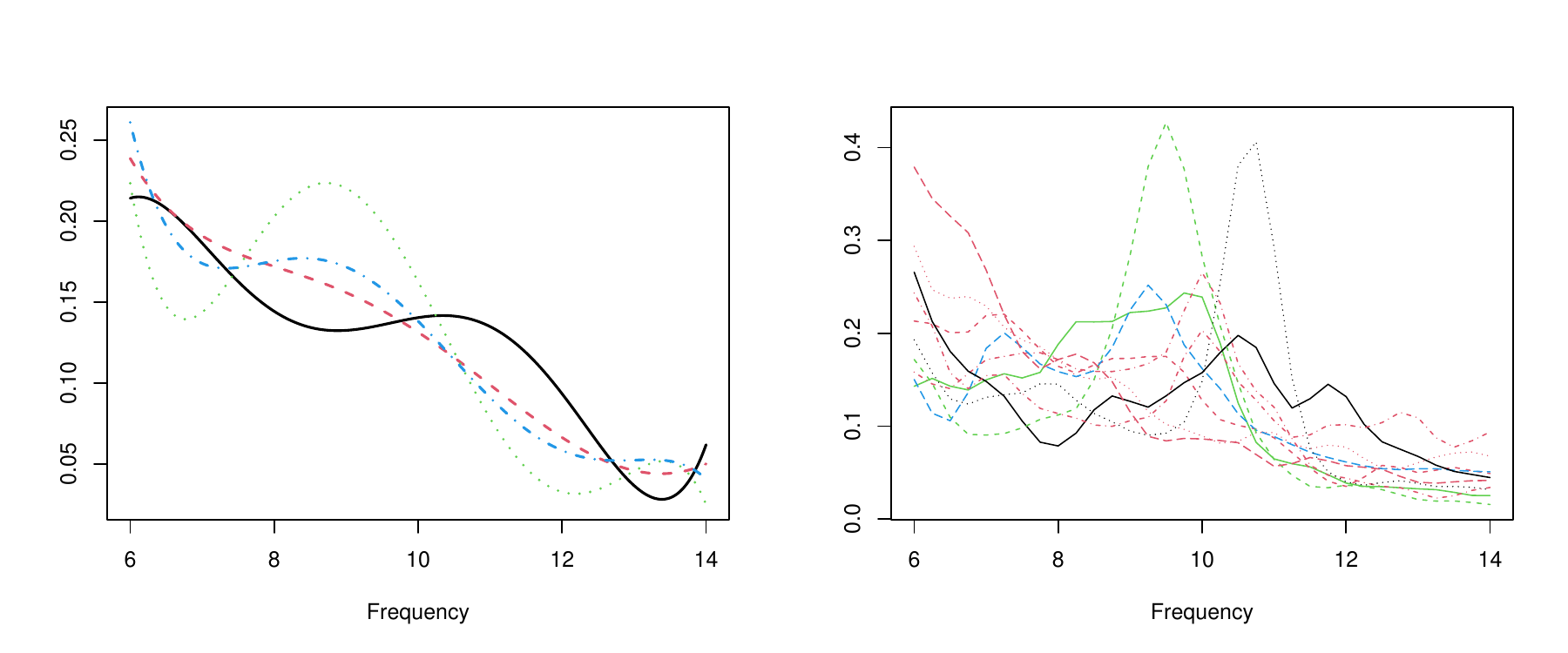}
    \caption{Model-based functional clustering with 4 clusters, fit on the data from the case study of EEG in ASD. The right subfigure illustrates the observed functions, colored based off of the estimated cluster membership.}
    \label{fig:clustering4}
\end{figure}

As stated in the main text, the added flexibility of mixed membership models allows you to potentially model data with fewer features/clusters than what you would need in a traditional clustering model. Therefore, we fit multiple clustering models \citep{pya2021fdamocca} with $K = 2, 3, 4,$ and $5$, and found that the model with 4 clusters was optimal using BIC.  From Figure \ref{fig:clustering4}, we can see that all 4 cluster means seem like an average of the 2 features found in the mixed membership model; however, none of the 4 clusters are easily interpretable. From a modeling standpoint, standard model-based clustering does not seem to do a good job of explaining the heterogeneity of the data. Instead, the mixed membership model gives us a more parsimonious model, with easily interpretable features that explain the heterogeneity of the data.

\begin{figure}[H]
    \centering
    \includegraphics[width=.9\textwidth]{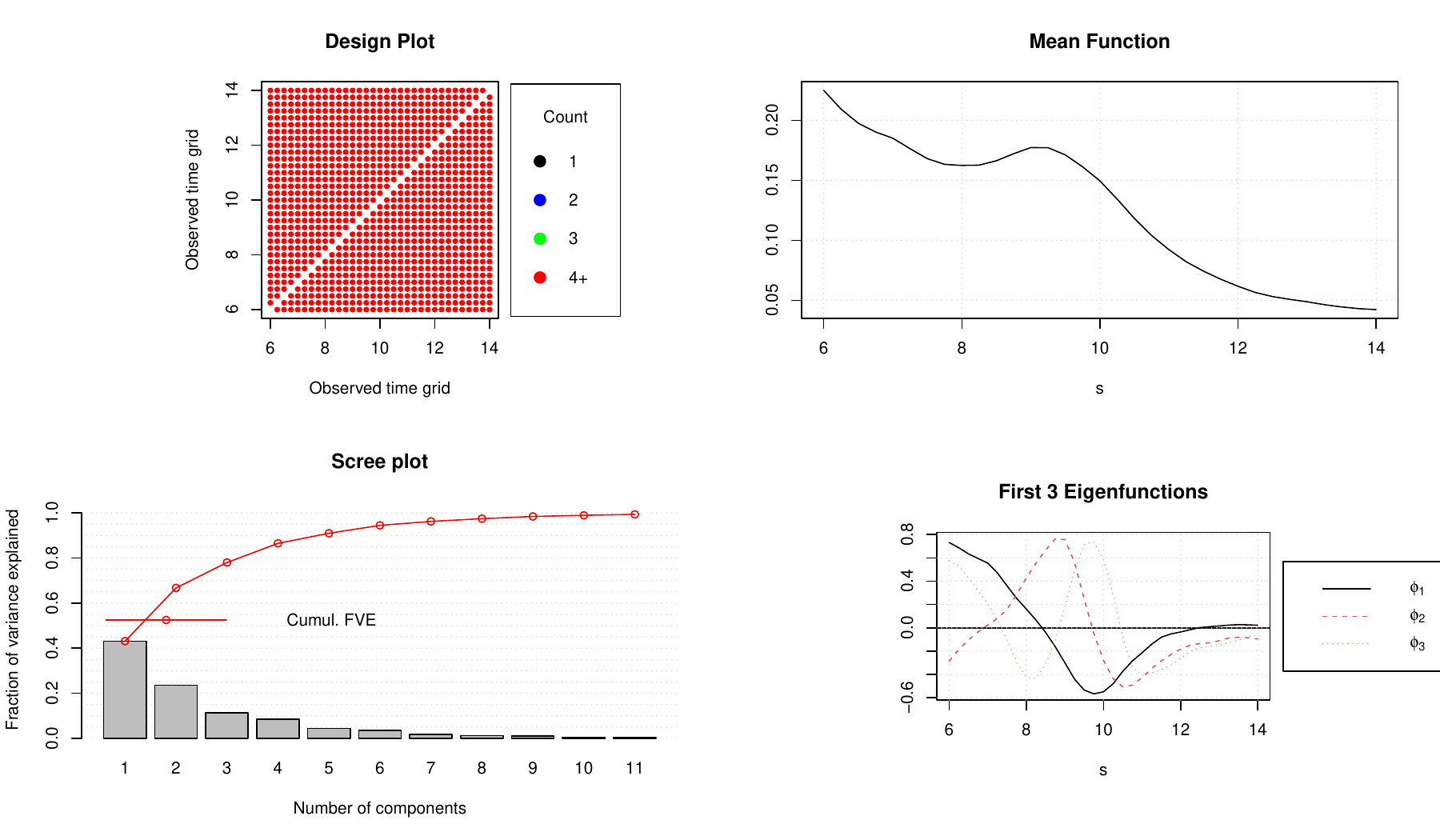}
    \caption{Summary of the FPCA results, including the mean function (top right), fraction of variance explained (bottom left), and top 3 eigenfunctions (bottom right).}
    \label{fig:FPCA}
\end{figure}

To compare the results of our functional mixed membership model to FPCA or functional factor analysis models, we performed FPCA on EEG case study data. The results of FPCA can be visualized in Figure \ref{fig:FPCA}. We can see that the mean function seems to be approximately an equal mixture between the first feature and the second feature found in Section 4.3 (roughly an equal mixture of the $1/f$ signal and the alpha peak). We can see that the first eigenfunction controls how prominent the alpha peak is (the alpha peak is more prominent if the first FPC is negative and less prominent if the first FPC is positive). The second eigenfunction also controls how prominent the alpha peak is; however, the PAF will have a lower frequency if the second FPC is positive (the alpha peak is more prominent if the second FPC is positive and less prominent if the second FPC is negative). We can see that the third eigenfunction accounts for a phase shift in the alpha peak. 

\begin{figure}[H]
    \centering
    \includegraphics[width=.9\textwidth]{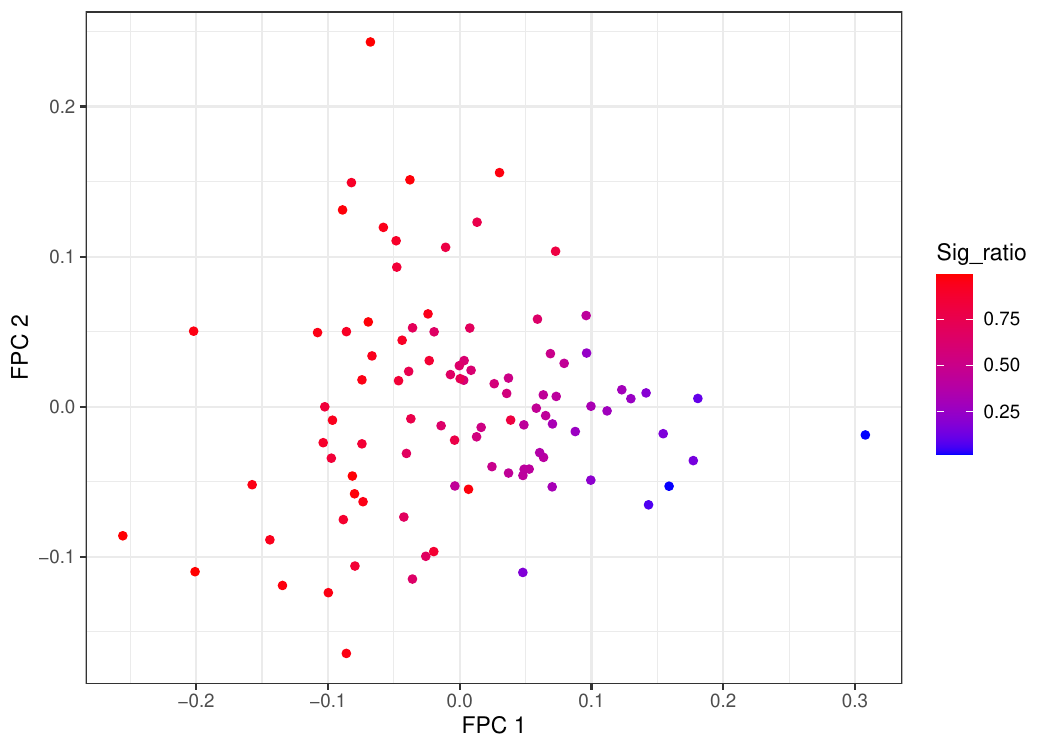}
    \caption{Plot of the first two functional principal components, colored by the allocations (feature 2) found from the functional mixed membership model (denoted sig\_ratio).}
    \label{fig:FPCA_Sig_Ratio}
\end{figure}

As stated earlier in this section, one may be tempted to use the FPCs as allocation parameters. Figure \ref{fig:FPCA_Sig_Ratio} shows the first two FPCs, colored by the allocation parameters found by the functional mixed membership model in Section 4.3. We can see that the first principal component has a relatively strong correlation with the allocation to feature 2 found in the functional mixed membership model. However, we can see that the allocations cannot be completely explained by the first FPC. For example, one subject has a mean allocation of 0.97 to the second feature (FPC1=0.030, FPC2=0.156) and another subject has a mean allocation of 0.18 to the second feature (FPC1=0.048, FPC2 = -0.110). Thus, we can see that while the FPCs are correlated to the estimated allocations, we cannot recover the allocations directly through the first FPC or even the second FPC. Moreover, estimates of the feature-specific covariance surfaces, like the ones found in Figure \ref{fig:real_1_cov}, are unobtainable using functional feature allocation models.

\subsection{Analysis of Multi-Channel EEG Data}
\label{HD_EEG_Case_Study_Appendix}
The proposed modeling framework is suitable for the analysis of functional data evaluated on $\mathcal{T}\subset R^d$. Therefore, we extend our analysis in the main manuscript to include EEG data measured on the entire cortex. Specifically, we will use a model with $2$ latent functional features ($K=2$) where $\mathcal{T}\subset \mathbb{R}^3$. Two of the three indices denote the spatial position on the scalp, while the third index contains information on the frequency observed. Similarly, the value of the function at some point $t \in \mathcal{T}$ represents the spectral power of the observed signal. For computational purposes, we project the true three-dimensional coordinates of the electrodes to a two-dimensional bird's eye view of electrodes using the `eegkit' package developed by \citet{eegkit}. In this section, we used two eigenfunctions to capture the covariance process ($M = 2$). We used a tensor product of B-splines to create a basis for our space of functions. For each dimension, we used quadratic B-splines, with $3$ internal nodes for each spatial index and $2$ internal nodes for the frequency index ($P = 180$). Since we are using functional data analysis techniques to model the EEG data, we assume smoothness over the spatial and frequency domains. Since the EEG data has poor spatial resolution \citep{grinvald2004vsdi} and we have relatively sparse sampling in the spatial domain ($25$ channels), the smoothness assumption can be thought of as a type of regularization over the domain of our function. Due to computational limitations, we ran the Multiple Start Algorithm (algorithm \ref{alg:MSA}) with $n\_try1 = 6$, $n\_try2 = 1$, $n\_MCMC1 = 3000$, and $n\_MCMC2 = 4000$. We then ran the chain for 19,000 iterations, saving only every 10 iterations.

\begin{figure}
     \centering
     \includegraphics[width=.99\textwidth, , height=5.5cm]{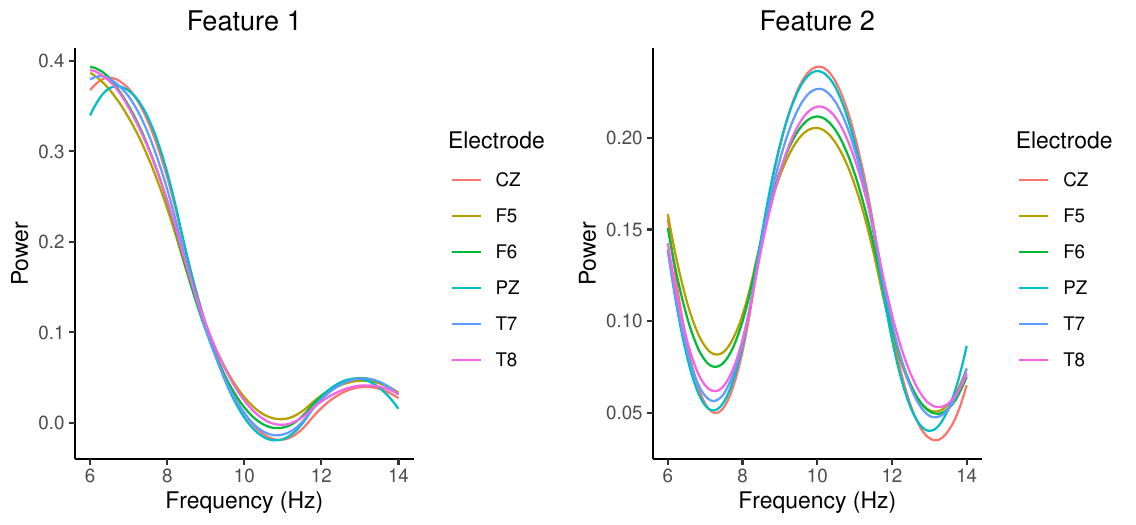}
     \caption{Posterior estimates of the means of the two functional features viewed at specific electrodes.}
     \label{fig:real_mult_mean}
\end{figure}

\begin{figure}
    \centering
     \subfigure[$6$ Hz]{\includegraphics[width=.49\textwidth]{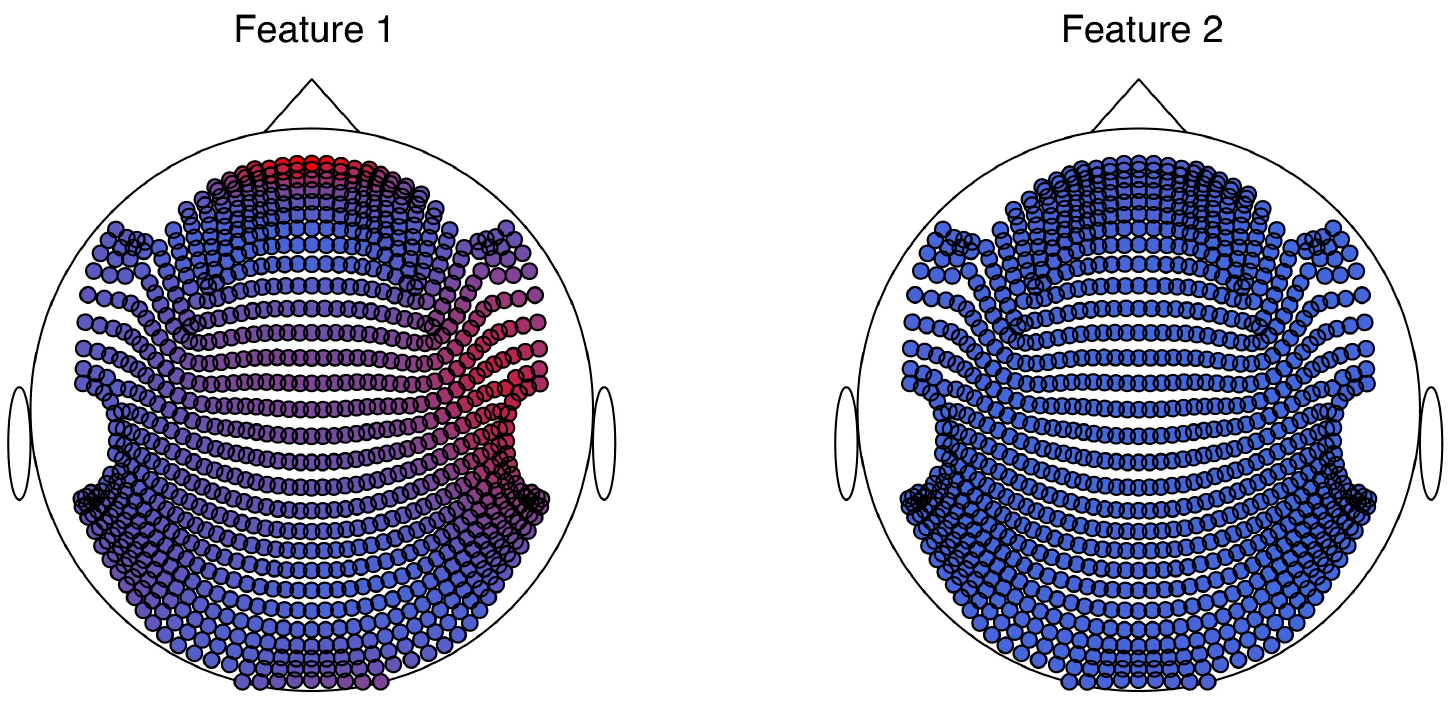}}
     \subfigure[$10$ Hz]{\includegraphics[width=.49\textwidth]{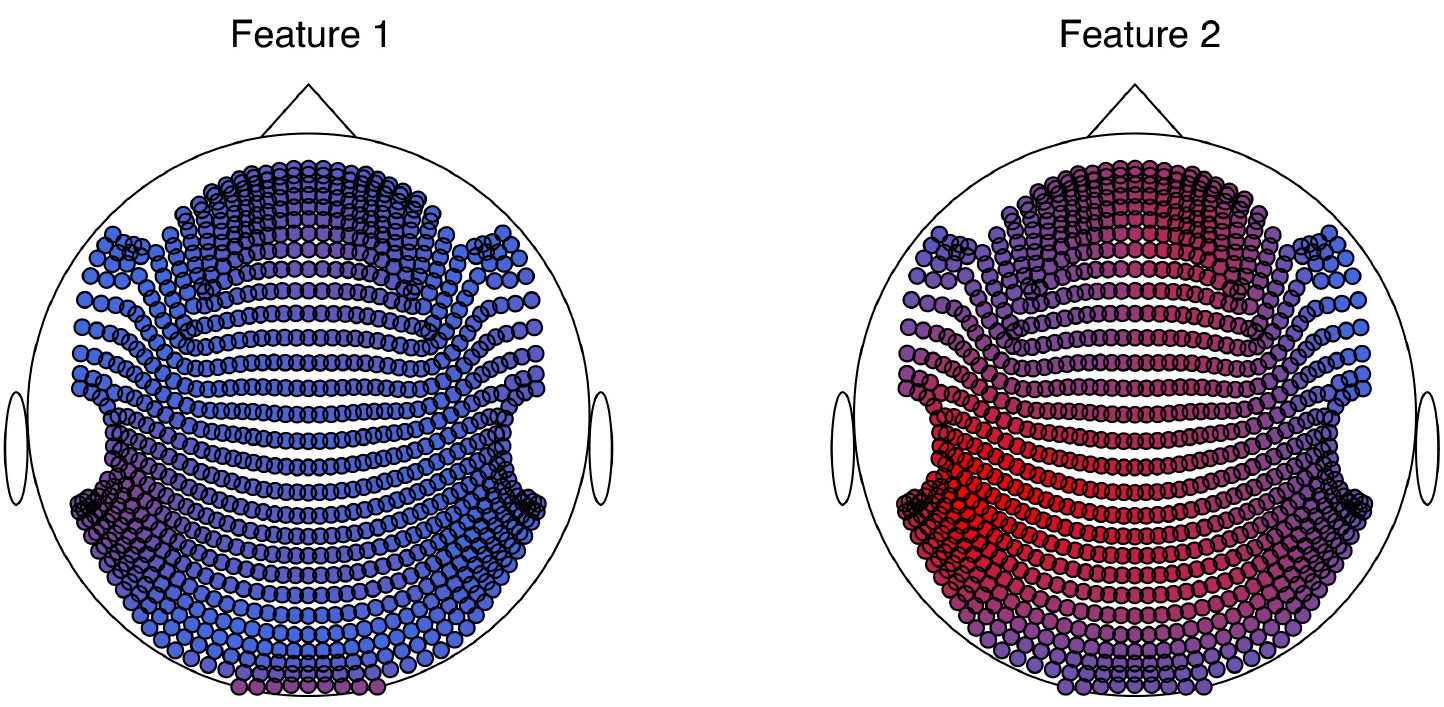}}
    \caption{Variance of the electrodes at 6 and 10 Hz for each functional feature. The relative magnitude of the variance of each electrode is indicated by the color of the electrode (red is relatively high variance, while blue is relatively low variance).}
    \label{fig:cov_Multivariate}
\end{figure}

Figure \ref{fig:real_mult_mean} reports posterior mean estimates for the feature means over a sample of electrodes.  Our findings are similar to our results on electrode T8, analyzed in the main manuscript; one latent feature corresponding to $1/f$ noise, and the other exhibiting well-defined PAF across electrodes. Figure \ref{fig:cov_Multivariate}, reports the electrode-specific variance at frequencies of 6 Hz and 10 Hz, respectively corresponding to the highest relative power in the first latent feature and the average peak alpha frequency in the second latent feature. 

The right temporal region around electrode T8 is found to exhibit a high level of heterogeneity (high relative variance) at 6 Hz, within latent feature 1 (poorly defined PAF), which relates to the findings of \citet{scheffler2019covariate}, who identified patterns of variation in the right temporal region as the highest contributor to log-odds of ASD versus TD discrimination. On the contrary, feature 2 (well-defined PAF) exhibits high levels of heterogeneity (relative variance) throughout the cortex at frequency 10 Hz, corresponding to the location of the PAF in feature 2.  Overall, the results of our analysis on the entire set of electrodes agree with our findings for electrode T8 in the main manuscript.

\begin{figure}[htbp]
    \centering
    \includegraphics[width=.99\textwidth]{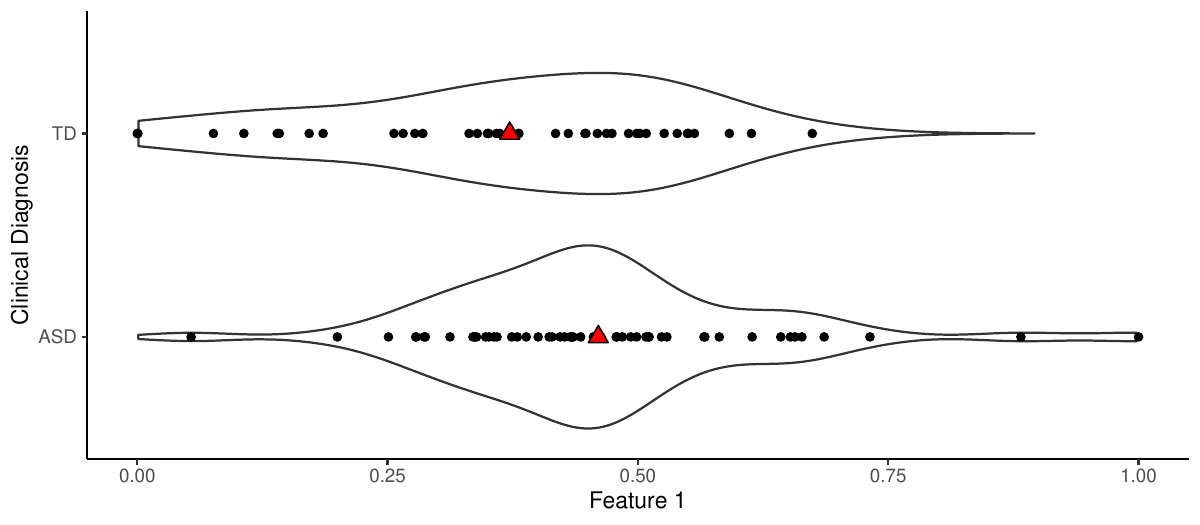}
    \caption{Posterior estimates of the median membership to the first functional feature.}
    \label{fig:mult_Z}
\end{figure}

Figure \ref{fig:mult_Z} shows the posterior median estimates of the membership allocations for each individual. We can see from both the mean functions and membership allocations that these results seem to match the univariate results in Section 4.3 of the main text.


\section{Computation}
\subsection{Posterior Distributions and Computation}
\label{Posterior_appendix}
In this section, we will discuss the computational strategy used to perform Bayesian inference. In cases where the posterior distribution is a known distribution, a Gibbs update will be performed. We will let $\boldsymbol{\Theta}$ be the collection of all parameters, and $\boldsymbol{\Theta}_{-\zeta}$ be the collection of all parameters, excluding the $\zeta$ parameter. We will first start with the $\boldsymbol{\phi}_{km}$ parameters, for $j = 1,\dots, K$ and $m = 1, \dots, M$. Let $\mathbf{D}_{jm} = \tilde{\tau}_{mj}^{-1} diag\left(\gamma_{j1m}^{-1}, \dots, \gamma_{jPm}^{-1}\right)$. By letting
$$\begin{aligned}
\mathbf{m}_{jm} = & \frac{1}{\sigma^2} \sum_{i=1}^N \sum_{l = 1}^{n_i}\left(B(t_{il})\chi_{im} \left(y_i(t_{il})Z_{ij} -  Z_{ij}^2 \boldsymbol{\nu}_{j}'B(t_{il}) - Z_{ij}^2\sum_{n \ne m}\left[\chi_{in} \boldsymbol{\phi}_{jn}' B(t_{il})\right] -\right. \right.\\
& \left. \left. \sum_{k \ne j} Z_{ij}Z_{ik}\left[\boldsymbol{\nu}_{k}' B(t_{il}) + \sum_{n=1}^M \chi_{in} \boldsymbol{\phi}_{kn}'B(t_{il}) \right] \right) \right),
\end{aligned}$$
and
$$\mathbf{M}_{jm}^{-1} = \frac{1}{\sigma^2}\sum_{i=1}^N \sum_{l = 1 }^{n_i} \left(Z_{ij}^2\chi_{im}^2B(t_{il})B'(t_{il})\right) + \mathbf{D}_{jm}^{-1},$$
 we have that 
$$\boldsymbol{\phi}_{jm} | \boldsymbol{\Theta}_{-\boldsymbol{\phi}_{jm} }, \mathbf{Y}_1, \dots, \textbf{Y}_N \sim \mathcal{N}(\mathbf{M}_{jm}\mathbf{m}_{jm}, \mathbf{M}_{jm}).$$

The posterior distribution of $\delta_{1k}$, for $k = 1, \dots, K$, is 
$$\begin{aligned}
\delta_{1k} | \boldsymbol{\Theta}_{-\delta_{1k}}, \mathbf{Y}_1, \dots, \textbf{Y}_N \sim & \Gamma\left(a_{1k} + (PM/2), 1 + \frac{1}{2} \sum_{r=1}^P \gamma_{k,r,1}\phi_{k,r,1}^2  \right.  \\
& \left. + \frac{1}{2}\sum_{m=2}^M \sum_{r=1}^P \gamma_{k,r,m}\phi_{k,r,m}^2\left( \prod_{j=2}^m \delta_{jk} \right)\right).
\end{aligned}.$$
The posterior distribution for $\delta_{ik}$, for $i = 2, \dots, M$ and $k = 1, \dots, K$, is 
$$\begin{aligned}
\delta_{ik} | \boldsymbol{\Theta}_{-\delta_{ik}}, \mathbf{Y}_1, \dots, \textbf{Y}_N  \sim & \Gamma\Bigg(a_{2k} + (P(M - i + 1)/2), 1   \\
& \left. +\frac{1}{2}\sum_{m = i}^M \sum_{r=1}^P \gamma_{k,r,m}\phi_{k,r,m}^2\left( \prod_{j=1; j \ne i}^m \delta_{j} \right)\right).
\end{aligned}$$
The posterior distribution for $a_{1k}$ is not a commonly known distribution; however, we have that
$$P(a_{1k}|\boldsymbol{\Theta}_{-a_{1k}}, \mathbf{Y}_1, \dots, \textbf{Y}_N) \propto \frac{1}{\Gamma(a_{1k})}\delta_{1k}^{a_{1k} -1} a_{1k}^{\alpha_{1} -1} exp \left\{-a_{1k}\beta_{1} \right\}.$$
Since this is not a known kernel of a distribution, we will have to use Metropolis-Hastings algorithm. Consider the proposal distribution $Q(a_{1k}'| a_{1k}) = \mathcal{N}\left(a_{1k}, \epsilon_1\beta_{1}^{-1}, 0, + \infty\right)$ (Truncated Normal) for some small $\epsilon_1 > 0$. Thus, the probability of accepting any step is
$$A(a_{1k}',a_{1k}) = \min \left\{1, \frac{P\left(a_{1k}'| \boldsymbol{\Theta}_{-a_{1k}'}, \mathbf{Y}_1, \dots, \textbf{Y}_N \right)}{P\left(a_{1k}| \boldsymbol{\Theta}_{-a_{1k}}, \mathbf{Y}_1, \dots, \textbf{Y}_N \right)} \frac{Q\left(a_{1k}|a_{1k}'\right)}{Q\left(a_{1k}'|a_{1k}\right)}\right\}.$$

Similarly for $a_{2k}$, we have
$$P(a_{2k} | \boldsymbol{\Theta}_{-a_{2k}}, \mathbf{Y}_1, \dots, \textbf{Y}_N) \propto \frac{1}{\Gamma(a_{2k})^{M-1}}\left(\prod_{i=2}^M\delta_{ik}^{a_{2k} -1}\right) a_{2k}^{\alpha_{2k} -1} exp \left\{-a_{2k}\beta_{2} \right\}.$$
We will use a similar proposal distribution, such that $Q(a_{2k}'| a_{2k}) = \mathcal{N}\left(a_{2k}, \epsilon_2\beta_{2}^{-1}, 0, + \infty\right)$ for some small $\epsilon_2 > 0$. Thus, the probability of accepting any step is
$$A(a_{2k}',a_{2k}) = \min \left\{1, \frac{P\left(a_{2k}'| \boldsymbol{\Theta}_{-a_{2k}'}, \mathbf{Y}_1, \dots, \mathbf{Y}_N\right)}{P\left(a_{2k}| \boldsymbol{\Theta}_{-a_{2k}}, \mathbf{Y}_1, \dots, \mathbf{Y}_N\right)} \frac{Q\left(a_{2k}|a_{2k}'\right)}{Q\left(a_{2k}'|a_{2k}\right)}\right\}.$$
For the parameters $\gamma_{j,r,m}$, for $j = 1, \dots K$, $r = 1, \dots, P$, and $m = 1, \dots, M$, we have 
$$\gamma_{j,r,m}| \boldsymbol{\Theta}_{-\gamma_{j,r,m}}, \mathbf{Y}_1, \dots, \mathbf{Y}_N \sim \Gamma\left(\frac{\nu_\gamma + 1}{2},\frac{\phi_{j,r,m}^2\tilde{\tau}_{mj} + \nu_\gamma}{2} \right).$$
The posterior distribution for the $\mathbf{z}_i$ parameters are not a commonly known distribution, so we will have to use the Metropolis-Hastings algorithm. We know that
$$\begin{aligned}
p(\mathbf{z}_i| \boldsymbol{\Theta}_{-\mathbf{z}_i}, \mathbf{Y}_1, \dots, \mathbf{Y}_N) & \propto \prod_{k=1}^K Z_{ik}^{\alpha_3\pi_k - 1}\\
& \times \prod_{l=1}^{n_i} exp\left\{-\frac{1}{2\sigma^2}\left(y_i(t_{il}) -  \sum_{k=1}^K Z_{ik}\left(\boldsymbol{\nu}_k'B(t_{il}) + \sum_{n=1}^M\chi_{in}\boldsymbol{\phi}_{kn}'B(t_{il})\right)\right)^2\right\}.
\end{aligned}$$
We will use $Q(\mathbf{z}_i'| \mathbf{z}_i) = Dir(a_{\mathbf{z}} \mathbf{z}_i)$ for some large $a_{\mathbf{z}} \in \mathbb{R}^+$ as the proposal distribution. Thus, the probability of accepting a proposed step is 
$$A(\mathbf{z}_i', \mathbf{z}_i) = \min \left\{1, \frac{P\left(\mathbf{z}_i'| \boldsymbol{\Theta}_{-\mathbf{z}_i'}, \mathbf{Y}_1, \dots, \mathbf{Y}_N\right)}{P\left(\mathbf{z}_i| \boldsymbol{\Theta}_{-\mathbf{z}_i}, \mathbf{Y}_1, \dots, \mathbf{Y}_N\right)} \frac{Q\left(\mathbf{z}_i|\mathbf{z}_i'\right)}{Q\left(\mathbf{z}_i'|\mathbf{z}_i\right)}\right\}.$$

Similarly, a Gibbs update is not available for an update of the $\boldsymbol{\pi}$ parameters. We have
$$\begin{aligned}
p(\boldsymbol{\pi}|\boldsymbol{\Theta}_{-\boldsymbol{\pi}}, \mathbf{Y}_1,\dots, \mathbf{Y}_N) & \propto \prod_{k=1}^K \pi_k^{c_k - 1} \\
& \times \prod_{i=1}^N\frac{1}{B(\alpha_3\boldsymbol{\pi})}\prod_{k=1}^K Z_{ik}^{\alpha_3\pi_k - 1}.
\end{aligned}$$
Letting the proposal distribution be such that $Q(\boldsymbol{\pi}'| \boldsymbol{\pi}) = Dir(a_{\boldsymbol{\pi}} \boldsymbol{\pi})$, for some large $a_{\boldsymbol{\pi}} \in \mathbb{R}^+$, we find that our probability of accepting any proposal is
$$A(\boldsymbol{\pi}', \boldsymbol{\pi}) = \min \left\{1, \frac{P\left(\boldsymbol{\pi}'| \boldsymbol{\Theta}_{-\boldsymbol{\pi}'}, \mathbf{Y}_1, \dots, \mathbf{Y}_N\right)}{P\left(\boldsymbol{\pi}| \boldsymbol{\Theta}_{-\boldsymbol{\pi}}, \mathbf{Y}_1, \dots, \mathbf{Y}_N\right)} \frac{Q\left(\boldsymbol{\pi}|\boldsymbol{\pi}'\right)}{Q\left(\boldsymbol{\pi}'|\boldsymbol{\pi}\right)}\right\}.$$
The posterior distribution of $\alpha_3$ is also not a commonly known distribution, so we will use the Metropolis-Hastings algorithm to sample from the posterior distribution. We have
$$\begin{aligned}
p(\alpha_3|\boldsymbol{\Theta}_{-\alpha_3}, \mathbf{Y}_1, \dots, \mathbf{Y}_N) & \propto e^{-b\alpha_3} \\
& \times \prod_{i=1}^N\frac{1}{B(\alpha_3\boldsymbol{\pi})}\prod_{k=1}^K Z_{ik}^{\alpha_3\pi_k - 1}.
\end{aligned}$$
Using a proposal distribution such that $Q(\alpha_3'|\alpha_3) = \mathcal{N}(\alpha_3, \sigma^2_{\alpha_3}, 0, +\infty)$ (Truncated Normal), we are left with the probability of accepting a proposed state as
$$A(\alpha_3',\alpha_3) = \min \left\{1, \frac{P\left(\alpha_3'| \boldsymbol{\Theta}_{-\alpha_3'}, \mathbf{Y}_1, \dots, \mathbf{Y}_N\right)}{P\left(\alpha_3| \boldsymbol{\Theta}_{-\alpha_3}, \mathbf{Y}_1, \dots, \mathbf{Y}_N\right)} \frac{Q\left(\alpha_3|\alpha_3'\right)}{Q\left(\alpha_3'|\alpha_3\right)}\right\}.$$
Let $\mathbf{P}$ be the following tridiagonal matrix:
$$\mathbf{P}= \begin{bmatrix}
1 & -1 & 0 &  & \\
-1 & 2 & -1 &  &  \\
 & \ddots & \ddots & \ddots&  \\
 &  & -1 & 2 & -1  \\
 &  & 0 & -1 & 1 \\
\end{bmatrix}.$$
Thus, letting
$$\mathbf{B}_j = \left( \tau_j\mathbf{P} + \frac{1}{\sigma^2} \sum_{i =1}^N \sum_{l=1}^{n_i}Z_{ij}^2B(t_{il})B'(t_{il})  \right)^{-1}$$
and
$$\mathbf{b}_{j} = \frac{1}{\sigma^2}\sum_{i=1}^N\sum_{l=1}^{n_i}Z_{ij}B(t_{il})\left(y_i(t_{il}) - \left(\sum_{k\ne j}Z_{ik}\boldsymbol{\nu}'_{k}B(t_{il})\right)  - \left(\sum_{k=1}^K \sum_{m=1}^M Z_{ik}\chi_{im}\boldsymbol{\phi}_{km}'B(t_{il}) \right)\right),$$
we have that 
$$\boldsymbol{\nu}_j| \boldsymbol{\Theta}_{-\boldsymbol{\nu}_j}, \mathbf{Y}_1, \dots, \mathbf{Y}_N \sim \mathcal{N}(\mathbf{B}_j\mathbf{b}_j, \mathbf{B}_j),$$
for $j = 1, \dots, K$. Thus, we can perform a Gibbs update to update our $\boldsymbol{\nu}$ parameters. The $\tau_l$ parameters for $l = 1, \dots K$, can also be updated by using a Gibbs update since the posterior distribution is:
$$\tau_l| \boldsymbol{\Theta}_{-\tau_l}, \mathbf{Y}_1, \dots, \mathbf{Y}_N \sim \Gamma\left(\alpha + P/2, \beta + \frac{1}{2}\boldsymbol{\nu}'_l\mathbf{P}\boldsymbol{\nu}_l\right).$$
The parameter $\sigma^2$ can be updated by using a Gibbs update. If we let 
$$\beta_{\sigma} =\frac{1}{2}\sum_{i=1}^N\sum_{l=1}^{n_i}\left(y_i(t_{il}) -  \sum_{k=1}^K Z_{ik}\left(\boldsymbol{\nu}_k'B(t_{il}) + \sum_{n=1}^M\chi_{in}\boldsymbol{\phi}_{kn}'B(t_{il})\right)\right)^2,$$
then we have
$$\sigma^2| \boldsymbol{\Theta}_{-\sigma^2}, \mathbf{Y}_1, \dots, \mathbf{Y}_N  \sim  IG\left(\alpha_0 + \frac{\sum_{i=1}^N n_i}{2} , \beta_0 +\beta_{\sigma}\right),$$
where $n_i$ are the number of time points observed for the $i^{th}$ observed function. Lastly, we can update the parameters $\chi_{im}$ for $i = 1, \dots, N$ and $m = 1, \dots, M$, using a Gibbs update. If we let 
$$\mathbf{w}_{im} = \frac{1}{\sigma^2}\left(\sum_{l=1}^{n_i} \left(\sum_{k = 1}^K Z_{ik}\boldsymbol{\phi}_{km}'B(t_{il})\right)\left(y_i(t_{il}) - \sum_{k = 1}^K Z_{ik}\left(\boldsymbol{\nu}_k'B(t_{il})  + \sum_{n\ne m}\chi_{in}\boldsymbol{\phi}_{kn}'B(t_{il})\right)\right)\right)$$
and 
$$\mathbf{W}_{im}^{-1} = 1 + \frac{1}{\sigma^2} \sum_{l=1}^{n_i}\left(\sum_{k = 1}^K Z_{ik}\boldsymbol{\phi}_{km}'B(t_{il})\right)^2,$$
then we have that 
$$\chi_{im}| \boldsymbol{\zeta}_{-\chi_{im}}, \mathbf{Y}_1, \dots, \mathbf{Y}_N \sim \mathcal{N}(\mathbf{W}_{im}\mathbf{w}_{im}, \mathbf{W}_{im}).$$

In our model, we relax the assumption that the $\boldsymbol{\Phi}$ parameters are orthogonal. Although we relaxed the assumption, we proved that many of the desirable properties still hold. However, if users do not want to relax this assumption, \citet{kowal2017bayesian} describes a framework that allows us to sample when orthogonality constraints are imposed. In our model, orthogonality is defined by the inner product in Equation 6 in the main text. Therefore, for $p$ such that $1 \le p \le KP$, we must have that
$$\left\langle\boldsymbol{\Phi}_p, \boldsymbol{\Phi}_j \right\rangle_{\boldsymbol{\mathcal{H}}}  = 0 \;\; \forall j \ne p.$$
By rearranging terms, we can see that we have
\begin{align}
    \left\langle\boldsymbol{\Phi}_p, \boldsymbol{\Phi}_j \right\rangle_{\boldsymbol{\mathcal{H}}} & =\sum_{k=1}^K\int_{\mathcal{T}}\boldsymbol{\phi}_{kp}'\mathbf{B}(t)\boldsymbol{\phi}_{kj}'\mathbf{B}(t)\text{d}t \nonumber \\
    & = \sum_{k \ne i} \int_{\mathcal{T}}\boldsymbol{\phi}_{kp}'\mathbf{B}(t)\boldsymbol{\phi}_{kj}'\mathbf{B}(t)\text{d}t  + \boldsymbol{\phi}_{ip}' \int_{\mathcal{T}} \mathbf{B}(t) \mathbf{B}(t)' \text{d}t\boldsymbol{\phi}_{ij}, \nonumber
\end{align}
where $\int_{\mathcal{T}} \mathbf{B}(t) \mathbf{B}(t)' \text{d}t$ is the element-wise integration of the $P \times P$ matrix. Letting 
$$\mathbf{L}_{-ip} = \begin{bmatrix}
\int_{\mathcal{T}} \mathbf{B}(t) \mathbf{B}(t)' \text{d}t\boldsymbol{\phi}_{i1}\\
\vdots \\
\int_{\mathcal{T}} \mathbf{B}(t) \mathbf{B}(t)' \text{d}t\boldsymbol{\phi}_{i(p-1)} \\
\int_{\mathcal{T}} \mathbf{B}(t) \mathbf{B}(t)' \text{d}t\boldsymbol{\phi}_{i(p+1)} \\
\vdots \\
\int_{\mathcal{T}} \mathbf{B}(t) \mathbf{B}(t)' \text{d}t\boldsymbol{\phi}_{i(KP)} \\
\end{bmatrix} \;\text{ and }\; \mathbf{c}_{-ip} =\begin{bmatrix}
\sum_{k \ne i} \int_{\mathcal{T}}\boldsymbol{\phi}_{kp}'\mathbf{B}(t)\boldsymbol{\phi}_{k1}'\mathbf{B}(t)\text{d}t\\
\vdots \\
\sum_{k \ne i} \int_{\mathcal{T}}\boldsymbol{\phi}_{kp}'\mathbf{B}(t)\boldsymbol{\phi}_{k(p-1)}'\mathbf{B}(t)\text{d}t\\
\sum_{k \ne i} \int_{\mathcal{T}}\boldsymbol{\phi}_{kp}'\mathbf{B}(t)\boldsymbol{\phi}_{k(p+1)}'\mathbf{B}(t)\text{d}t\\
\vdots\\
\sum_{k \ne i} \int_{\mathcal{T}}\boldsymbol{\phi}_{kp}'\mathbf{B}(t)\boldsymbol{\phi}_{k(KP)}'\mathbf{B}(t)\text{d}t\\
\end{bmatrix},$$
we can write our orthogonality constraint for $\boldsymbol{\phi}_{ip}$ given the other $\boldsymbol{\phi}$ parameters as
$$\boldsymbol{\phi}_{ip}' \mathbf{L}_{-ip} = -\mathbf{c}_{-ip}.$$
 Thus using the results in \citet{kowal2017bayesian}, we have that $\boldsymbol{\phi}_{ip} \sim \mathcal{N}(\tilde{\mathbf{M}}_{ip}\mathbf{m}_{ip}, \tilde{\mathbf{M}}_{ip})$, where $$\tilde{\mathbf{M}}_{ip} = \mathbf{M}_{ip} - \mathbf{M}_{ip}\mathbf{L}_{-ip}\left(\mathbf{L}_{-ip}' \mathbf{M}_{ip} \mathbf{L}_{-ip} \right)^{-1} \left(\mathbf{L}_{-ip}'\mathbf{M}_{ip} + \mathbf{c}_{-ip} \right).$$ Like in \citet{kowal2017bayesian}, $\mathbf{M}_{ip}$ and $\mathbf{m}_{ip}$ are such that when we relax the orthogonal constraints, we have $\boldsymbol{\phi}_{ip} \sim \mathcal{N}(\mathbf{M}_{ip}\mathbf{m}_{ip}, \mathbf{M}_{ip})$ (defined in Section \ref{Posterior_appendix}). Thus, one can use the modified Gibbs update to ensure orthogonality. However, by using this alternative update, the mixing of the Markov chain will likely suffer.

\subsection{Multiple Start Algorithm}
\label{MSA_appendix}
One of the main computational challenges that we encounter in this model is the multi-modal posterior distribution. Often times, the MCMC chain can get stuck in a mode, and it can have trouble moving through areas of low posterior density. One way to traverse through areas of low posterior density is to use tempered transitions. However, tempered transitions are computationally intensive and the hyperparameters can be somewhat difficult to tune. Thus, one of the best ways to converge to the correct mode is to have a good starting point. The Multiple Start Algorithm, found in algorithm \ref{alg:MSA}, is a way to pick an optimal starting point. To get optimal performance our of this algorithm, we recommend that the initial data is standardized before running this algorithm.

\begin{algorithm}
\caption{Multiple Start Algorithm}\label{alg:MSA}
\begin{algorithmic}
\Require $\text{n\_try1},\text{n\_try2}, Y, \text{time}, K, n\_MCMC1, n\_MCMC2, \dots$
\State $P \gets \texttt{BFPMM\_Nu\_Z(Y, \text{time}, K, n\_MCMC1, \dots)}$ \Comment{Returns the likelihood and estimates for $\boldsymbol{\nu}$ and $\mathbf{Z}$}
\State $\text{max\_likelihood} \gets P[\text{``likelihood''}]$
\State $i \gets 1$
\While{$i \le \text{n\_try1}$}
\State $P_i \gets \texttt{BFPMM\_Nu\_Z(Y, \text{time}, K, n\_MCMC1, \dots)}$
\If{max\_likelihood $< P_i[\text{``likelihood''}]$}
    \State \text{max\_likelihood} $\gets P[\text{``likelihood''}]$
    \State $P \gets P_i$
\EndIf
\State $i \gets i + 1$
\EndWhile
\State $\theta \gets \texttt{BFPMM\_Theta(P, Y, \text{time}, K, n\_MCMC2, \dots)}$ \Comment{Returns estimates for the rest of the parameters}
\State $\text{max\_likelihood} \gets \theta[\text{``likelihood''}]$
\State $i \gets 1$
\While{$i \le \text{n\_try2}$}
\State $\theta_i \gets \texttt{BFPMM\_Theta(P, Y, \text{time}, K, n\_MCMC2, \dots)}$
\If{max\_likelihood $< \theta_i[\text{``likelihood''}]$}
    \State \text{max\_likelihood} $\gets \theta_i[\text{``likelihood''}]$
    \State $\theta \gets \theta_i$
\EndIf
\State $i \gets i + 1$
\EndWhile
\State \Return $(\theta, P)$ \Comment{Returns estimates for all model parameters}
\end{algorithmic}
\end{algorithm}

The function calls two other functions, \texttt{BFPMM\_Nu\_Z(Y, \text{time}, K, n\_MCMC1, \dots)} and \texttt{BFPMM\_Theta(P, Y, \text{time}, K, n\_MCMC2, \dots)}. The first function, \texttt{BFPMM\_Nu\_Z(Y, \text{time}, K, n\_MCMC1, \dots)}, starts with random parameter values for $\boldsymbol{\nu}$, $\mathbf{Z}$, $\sigma^2$, and other hyperparameters relating to these parameters. We then run an MCMC chain with the values of the $\chi$ and $\boldsymbol{\phi}$ variables fixed as $0$ (or as the matrix $\mathbf{O}$). The function returns the mean likelihood for the last $20\%$ of the MCMC chain as well as the entire MCMC chain. The variable \texttt{n\_MCMC1} is assumed to be picked such that the chain converges in the first $80\%$ of the MCMC iterations. Since the starting points are random, the MCMC chains are likely to explore different modes. Once we have a good initial starting point, we estimate the $\chi$, $\boldsymbol{\phi}$ , and other parameters that have not already been estimated using the function \texttt{BFPMM\_Theta(P, Y, \text{time}, K, n\_MCMC2, \dots)}. In this function, we run an MCMC chain while fixing the values of $\boldsymbol{\nu}$ and $\mathbf{Z}$ to their optimal values found previously. We will use the outputs of algorithm \ref{alg:MSA} as a starting point for our final MCMC chain.
\subsection{Tempered Transitions}
\label{Tempered_Transition_appendix}
Tempered transitions are used to help traverse areas of low posterior probability density when running MCMC chains. In problems that have multimodal posterior distributions, traditional methods often have difficulty moving from one mode to another, which can cause the chain to not explore the entire state-space and therefore not converge to the true posterior distribution. Thus, by using tempered transitions, we are potentially able to traverse the state-space to explore multiple modes. In simulations, we found that the tuning parameters can be difficult to tune to get acceptable acceptance probabilities; however, in this section we will outline a way to use tempered transitions with our model. 

We will follow the works of \citet{behrens2012tuning} and \citet{pritchard2000inference} and only temper the likelihood. The target distribution that we want to temper is usually assumed to be written as
$$p(x) \propto \pi(x)exp\left(-\beta_h h(x)\right),$$
where $\beta_h$ controls how much the distribution is tempered. We will assume $1 = \beta_0 < \dots < \beta_h < \dots < \beta_{N_t}$. The hyperparameters $N_t$ and $\beta_{N_t}$ are specified by the user and will depend on the complexity of the model. For more complex models, we are likely to need a larger $N_t$. We will also assume that the parameters $\beta_h$ follow a geometric scheme. We can rewrite our likelihood to fit the above form:
$$\begin{aligned}
p_h(y_i(t)|\boldsymbol{\Theta}) & \propto exp\left\{- \beta_h\left(\frac{1}{2}log(\sigma^2) + \frac{1}{2\sigma^2} \left(y_i(t) -  \sum_{k=1}^K Z_{ik}\left(\boldsymbol{\nu}_k'B(t) + \sum_{n=1}^M\chi_{in}\boldsymbol{\phi}_{kn}'B(t)\right)\right)^2\right)\right\} \\
& =\left(\sigma^2\right)^{-\beta_h / 2}exp\left\{-\frac{\beta_h}{2\sigma^2}\left(y_i(t) -  \sum_{k=1}^K Z_{ik}\left(\boldsymbol{\nu}_k'B(t) + \sum_{n=1}^M\chi_{in}\boldsymbol{\phi}_{kn}'B(t)\right)\right)^2\right\}.
\end{aligned}$$
Let $\boldsymbol{\Theta}_h$ be the set of parameters generated from the model using the tempered likelihood associated with $\beta_h$. The tempered transition algorithm can be summarized by the following steps:
\begin{enumerate}
    \item Start with initial state $\boldsymbol{\Theta}_0$.
    \item Transition from $\boldsymbol{\Theta}_0$ to $\boldsymbol{\Theta}_1$ using the tempered likelihood associated with $\beta_1$.
    \item Continue in this way until we transition from $\boldsymbol{\Theta}_{N_t - 1}$ to $\boldsymbol{\Theta}_{N_t}$ using the tempered likelihood associated with $\beta_{N_t}$.
    \item Transition from $\boldsymbol{\Theta}_{N_t}$ to $\boldsymbol{\Theta}_{N_t +1}$ using the tempered likelihood associated with $\beta_{N_t}$.
    \item Continue in this manner until we transition from $\boldsymbol{\Theta}_{2N_t -1}$ to $\boldsymbol{\Theta}_{2N_t}$ using $\beta_1$.
    \item Accept transition from $\boldsymbol{\Theta}_0$ to $\boldsymbol{\Theta}_{2N_t}$ with probability 
    $$\min \left\{1, \prod_{h=0}^{N_t - 1} \frac{\prod_{i=1}^N\prod_{l=1}^{n_i} p_{h+1}(y_i(t_{il})|\boldsymbol{\Theta}_h)}{\prod_{i=1}^N\prod_{l=1}^{n_i} p_{h}(y_i(t_{il})|\boldsymbol{\Theta}_h)} \prod_{h=N_t + 1}^{2N_t} \frac{\prod_{i=1}^N\prod_{l=1}^{n_i} p_{h}(y_i(t_{il})|\boldsymbol{\Theta}_h)}{\prod_{i=1}^N\prod_{l=1}^{n_i} p_{h+1}(y_i(t_{il})|\boldsymbol{\Theta}_h)}\right\}.$$
\end{enumerate}
Since we only temper the likelihood, we can use many of updates in Section \ref{Posterior_appendix}. However, we will have to modify how we update the parameters $\boldsymbol{\nu}$, $\boldsymbol{\phi}$, $\sigma^2$, $\mathbf{Z}$, and $\chi$. By letting
$$\begin{aligned}
(\mathbf{m}_{jm})_h = & \frac{\beta_h}{(\sigma^2)_h} \sum_{i=1}^N \sum_{l = 1}^{n_i}\Bigg(B(t_{il})(\chi_{im})_h \Bigg(y_i(t_{il})(Z_{ij})_h -  (Z_{ij})_h^2 (\boldsymbol{\nu}_{j})_h'B(t_{il})\\
& - (Z_{ij})_h^2\sum_{n \ne m}\left[(\chi_{in})_h (\boldsymbol{\phi}_{jn})_h' B(t_{il})\right] \\
& \left. \left.-\sum_{k \ne j} (Z_{ij})_h(Z_{ik})_h\left[(\boldsymbol{\nu}_{k})_h' B(t_{il}) + \sum_{n=1}^M (\chi_{in})_h (\boldsymbol{\phi}_{kn})_h'B(t_{il}) \right] \right) \right),
\end{aligned}$$
and
$$(\mathbf{M}_{jm})_h^{-1} = \frac{\beta_h}{(\sigma^2)_h}\sum_{i=1}^N \sum_{l = 1 }^{n_i} \left((Z_{ij})_h^2(\chi_{im})_h^2B(t_{il})B'(t_{il})\right) + (\mathbf{D}_{jm})_h^{-1},$$
 we have that 
$$(\boldsymbol{\phi}_{jm})_h  | \boldsymbol{\Theta}_{-(\boldsymbol{\phi}_{jm})_{h}}, \mathbf{Y}_1, \dots, \mathbf{Y}_N\sim \mathcal{N}((\mathbf{M}_{jm})_h(\mathbf{m}_{jm})_h, (\mathbf{M}_{jm})_h).$$
The posterior distribution for $(\mathbf{z}_i)_h$ is still not a commonly known distribution, so we will still have to use the Metropolis-Hastings algorithm. The new posterior distribution when using tempered transitions changes into
$$\begin{aligned}
p((\mathbf{z}_i)_h| (\boldsymbol{\Theta}_{-(\mathbf{z}_i)_h})_h, \mathbf{Y}_1, \dots, \mathbf{Y}_N) & \propto \prod_{k=1}^K (Z_{ik})_h^{(\alpha_3)_h(\pi_k)_h - 1}\\
& \times \prod_{l=1}^{n_i} exp\left\{-\frac{\beta_h}{2(\sigma^2)_h}\Bigg( y_i(t_{il})  \right. \\
& \left. \left. - \sum_{k=1}^K (Z_{ik})_h\left((\boldsymbol{\nu}_k)_h'B(t_{il}) +  \sum_{n=1}^M(\chi_{in})_h(\boldsymbol{\phi}_{kn})_h'B(t_{il})\right)\right)^2\right\}.
\end{aligned}$$
We will use $Q((\mathbf{z}_i)_h'| (\mathbf{z}_i)_{h-1}) = Dir(a_{\mathbf{z}} (\mathbf{z}_i)_{h-1})$ for some large $a_{\mathbf{z}} \in \mathbb{R}^+$ as the proposal distribution. Thus the probability of accepting a proposed step is 
$$A((\mathbf{z}_i)_{h}', (\mathbf{z}_i)_{h-1}) = \min \left\{1, \frac{P\left((\mathbf{z}_i)_{h}'| (\boldsymbol{\Theta}_{-(\mathbf{z}_i)_{h}'})_h, \mathbf{Y}_1, \dots, \mathbf{Y}_N\right)}{P\left((\mathbf{z}_i)_{h-1}| \boldsymbol{\Theta}_{-(\mathbf{z}_i)_{h-1}}, \mathbf{Y}_1, \dots, \mathbf{Y}_N\right)} \frac{Q\left((\mathbf{z}_i)_{h-1}|(\mathbf{z}_i)_{h}'\right)}{Q\left((\mathbf{z}_i)_{h}'|(\mathbf{z}_i)_{h-1}\right)}\right\}.$$
Letting
$$(\mathbf{B}_j)_h = \left( (\tau_j)_h\mathbf{P} + \frac{\beta_h}{(\sigma^2)_h} \sum_{i =1}^N \sum_{l=1}^{n_i}(Z_{ij})_h^2B(t_{il})B'(t_{il})  \right)^{-1}$$
and
$$\begin{aligned}
(\mathbf{b}_{j})_h = & \frac{\beta_h}{(\sigma^2)_h}\sum_{i=1}^N\sum_{l=1}^{n_i}(Z_{ij})_hB(t_{il})\left(y_i(t_{il}) - \left(\sum_{k\ne j}(Z_{ik})_h(\boldsymbol{\nu}'_{k})_hB(t_{il})\right) \right. \\
& \left. - \left(\sum_{k=1}^K \sum_{n=1}^M(Z_{ik})_h(\chi_{in})_h(\boldsymbol{\phi}_{kn})_h'B(t_{il}) \right)\right),
\end{aligned}$$
we have that 
$$(\boldsymbol{\nu}_j)_h| \boldsymbol{\Theta}_{-(\boldsymbol{\nu}_j)_h}, \mathbf{Y}_1, \dots, \mathbf{Y}_N \sim \mathcal{N}((\mathbf{B}_j)_h(\mathbf{b}_j)_h, (\mathbf{B}_j)_h).$$

The parameter $(\sigma^2)_h$ can be updated using a Gibbs update. If we let 
$$(\beta_{\sigma})_h =\frac{\beta_h}{2}\sum_{i=1}^N\sum_{l=1}^{n_i}\left(y_i(t_{il}) -  \sum_{k=1}^K (Z_{ik})_h\left((\boldsymbol{\nu}_k)_h'B(t_{il}) + \sum_{n=1}^M(\chi_{in})_h(\boldsymbol{\phi}_{kn})_h'B(t_{il})\right)\right)^2,$$
then we have
$$(\sigma^2)_h| \boldsymbol{\Theta}_{-(\sigma^2)_h}, \mathbf{Y}_1, \dots, \mathbf{Y}_N  \sim  IG\left(\alpha_0 + \frac{\beta_h\sum_{i=1}^N n_i}{2} , \beta_0 +(\beta_{\sigma})_h\right).$$
Lastly, letting let 
$$\begin{aligned}
(\mathbf{w}_{im})_h = &\frac{\beta_h}{(\sigma^2)_h}\Bigg(\sum_{l=1}^{n_i} \left(\sum_{k = 1}^K (Z_{ik})_h(\boldsymbol{\phi}_{km})_h'B(t_{il})\right)\Bigg(y_i(t_{il})  \\ 
& \left. \left. - \sum_{k = 1}^K (Z_{ik})_h\left((\boldsymbol{\nu}_k)_h'B(t_{il})  + \sum_{n\ne m}(\chi_{in})_h(\boldsymbol{\phi}_{kn})_h'B(t_{il})\right)\right)\right)
\end{aligned}$$
and 
$$(\mathbf{W}^{-1}_{im})_h = 1 + \frac{\beta_h}{(\sigma^2)_h} \sum_{l=1}^{n_i}\left(\sum_{k = 1}^K (Z_{ik})_h(\boldsymbol{\phi}_{km})_h'B(t_{il})\right)^2,$$
then we have that 
$$(\chi_{im})_h| \boldsymbol{\zeta}_{-(\chi_{im})_h}, \mathbf{Y}_1, \dots, \mathbf{Y}_N \sim \mathcal{N}((\mathbf{W}_{im})_h(\mathbf{w}_{im})_h, (\mathbf{W}_{im})_h).$$
One of the biggest drawbacks to using tempered transition is the computational cost of just one iteration. It is common for $N_t$ to be in the thousands, especially when dealing with a complex model, so each tempered transition will take thousands of times longer than an untempered transition. Thus, we recommend using a mixture of tempered transitions and untempered transitions to speed up computation. From Proposition 1 in \citet{roberts2007coupling}, we know that an independent mixture of tempered transitions and untempered transitions will still preserve our stationary distribution of our Markov chain.

\subsection{Membership Rescale Algorithm}
\label{MRA_appendix}
As discussed in Section 2.2 of the main text, our model suffers from the rescaling problem. Specifically,  
consider a model with 2 features, and let $\boldsymbol{\Theta}_0$ be the set of ``true'' parameters. Let $Z_{i1}^* = 0.5(Z_{i1})_0$ and $Z_{i2}^* =  (Z_{i2})_0 + 0.5(Z_{i1})_0$ (transformation preserves the constraint that $Z_{i1}^* + Z_{i2}^* = 1$). If we let $\boldsymbol{\nu}_1^* = 2(\boldsymbol{\nu}_1)_0 - (\boldsymbol{\nu}_2)_0$, $\boldsymbol{\nu}_2^* = (\boldsymbol{\nu}_2)_0$, $\boldsymbol{\phi}_{1m}^* = 2(\boldsymbol{\phi}_{1m})_0 - (\boldsymbol{\phi}_{2m})_0$,  $\boldsymbol{\phi}_{2m}^* = (\boldsymbol{\phi}_{2m})_0 $, $\chi_{im}^* = (\chi_{im})_0$, and $(\sigma^2)^* = \sigma^2_0$, we have $P(Y_i(t)|\boldsymbol{\Theta}_0) = P(Y_i(t)|\boldsymbol{\Theta}^*)$ (Equation 11 in the main text). To help interpretability, we will apply a linear transformation to the $\mathbf{Z}$ matrix to ensure that we use as much of the unit simplex as possible. In the case where $K = 2$, this will correspond to rescaling the observations so that at least one observation is entirely in each feature. This specific assumption that at least one observation belongs entirely to each feature is known as the \textit{seperability} condition \citep{papadimitriou1998latent, mcsherry2001spectral, azar2001spectral, chen2022learning}. Thus, to ensure identifiability, algorithm \ref{alg:MTA} can be used when we only have two features. In the case of a two-feature model, the seperability condition is a very weak assumption; however, as we move to models with more features, it can be a relatively strong assumption. Weaker geometric assumptions such as the \textit{sufficiently scattered} condition \citep{huang2016anchor, jang2019minimum, chen2022learning}. In cases when we have more than two features, we can use the idea of the sufficiently scattered condition to help with identifiability. From \citet{chen2022learning}, we see that an allocation matrix $\mathbf{Z}$ is sufficiently scattered if:
\begin{enumerate}
    \item cone$(\mathbf{Z}')^* \subseteq \mathcal{K}$
    \item cone$(\mathbf{Z}')^* \cap bd\mathcal{K} \subseteq \{\lambda \mathbf{e}_f, f = 1, \dots, k, \lambda \ge 0 \}$
\end{enumerate}
where $\mathcal{K} := \left\{\mathbf{x}\in \mathbb{R}^K | \|\mathbf{x}\|_2 \le \mathbf{x}'\mathbf{1}_K\right\}$, $bd\mathcal{K}  := \left\{\mathbf{x}\in \mathbb{R}^K | \|\mathbf{x}\|_2 = \mathbf{x}'\mathbf{1}_K\right\}$, cone$(\mathbf{Z}')^* := \left\{\mathbf{x} \in \mathbb{R}^K | \mathbf{x}\mathbf{Z}' \ge 0 \right\}$, and $\mathbf{e}_f$ is a vector with the $i^{th}$ element equal to 1 and zero elsewhere. The first condition can be interpreted as the allocation parameters should form a convex polytope that contains the dual cone $\mathcal{K}^*$. Thus, we have
$$\text{Conv}(\mathbf{Z}')\subseteq \mathcal{K}^*,$$
where $\mathcal{K}^* := \left\{\mathbf{x}\in \mathbb{R}^K |  \mathbf{x}'\mathbf{1}_K \ge \sqrt{k -1}\|\mathbf{x}\|_2\right\}$ and $\text{Conv}(\mathbf{Z}'):= \{\mathbf{x} \in \mathbb{R}^K | \mathbf{x} = \mathbf{Z}'\lambda, \lambda \in \Delta^N \}$, where $\Delta^k$ denotes the N-dimensional simplex. Ensuring that these two conditions are met is not trivial in our setting. Therefore, we will focus on trying to promote allocation structures such that the first condition is satisfied. Similarly to the case of two functional features, we aim to find a linear transformation such that the convex polytope of our transformed allocation parameters covers the most area. Thus letting $\mathbf{T} \in \mathbb{R}^K \times \mathbb{R}^K$ be our transformation matrix, we aim to solve the following optimization problem:
\begin{align}
    \nonumber \max_{\mathbf{T}} & \;\; |\text{Conv}(\mathbf{T}\mathbf{Z}')|\\
    \nonumber s.t. &  \;\; \mathbf{z}_i\mathbf{T} \in \mathcal{C}\;\; \forall \;i,
\end{align}
where $|\text{Conv}(\mathbf{T}\mathbf{Z}')|$ denotes the volume of the convex polytope constructed by the allocation parameters. Although this will not ensure that the first condition is met, it will promote an allocation structure that uses the entire simplex. Although this algorithm does not ensure identifiability in the case where we have more than two functional features, it does make inference on the model more interpretable. Once the memberships are rescaled, we can perform posterior inference on the mean function and covariance functions using the rescaled parameters. In practice, maximizing the area of a convex polytope can be difficult as formulated. Thus, in the case when we have a three-feature model, we will first project the data onto a two-dimensional space (since we have the condition that $Z_{i1} + Z_{i2} + Z_{i3} = 1$, we do not lose any information). From there, we find the minimum minimum enclosing triangle that encloses the convex polytope created by the allocation parameters in the two-dimensional space \citep{parvu2016implementation}. We then use the vertices of the triangle to transform the data to maximize the area of the convex polytope. Implementation details can be found in the code for the three-feature supplemental simulation study (\url{https://github.com/ndmarco/BFMMM_Functional_Sims/tree/main/Sim_Study_1}).

\begin{algorithm}
\caption{Membership Rescale Algorithm}\label{alg:MTA}
\begin{algorithmic}
\Require $\mathbf{Z}, \boldsymbol{\nu}, \boldsymbol{\Phi}, M$
\State $T \gets \texttt{matrix}(0, 2, 2)$ \Comment{Initialize inverse transformation matrix (2 x 2)}
\State $i \gets 1$
\While{$i \le 2$}
\State $\text{max\_ind} \gets \texttt{max\_ind}(\mathbf{Z}[,i])$ \Comment{Find index of max entry in $i^{th}$ column}
\State $T[i,] \gets (\mathbf{Z}[\text{max\_ind},])$
\State $i \gets i + 1$
\EndWhile
\State $\mathbf{Z}\_t \gets \mathbf{Z} * \texttt{inv}(T)$ \Comment{Transform the $\mathbf{Z}$ parameters}
\State $\boldsymbol{\nu}\_t \gets T * \boldsymbol{\nu}$ \Comment{Transform the $\boldsymbol{\nu}$ parameters}
\State $i \gets 1$
\While{$i \le M$}
\State $\boldsymbol{\Phi}\_t[,,i] \gets T * \boldsymbol{\Phi}[,,i]$ \Comment{Transform the $\boldsymbol{\Phi}$ parameters}
\State $i \gets i + 1$
\EndWhile
\State \Return $(\mathbf{Z}\_t, \boldsymbol{\nu}\_t, \boldsymbol{\Phi}\_t)$
\end{algorithmic}
\end{algorithm}

\section{Simulation-Based Posterior Inference}
Statistical inference is based on Markov chain Monte Carlo samples from the posterior distribution. To achieve this, we used the Metropolis-within-Gibbs algorithm. By introducing the latent variables $\chi_{im}$, many of the posterior distributions related to the covariance process were easily sampled through Gibbs updates. More details on the sampling scheme can be found in Section C.1 of the web-based supporting materials. The sampling scheme is relatively simple and was implemented using the RcppArmadillo package created by \citet{eddelbuettel2014rcpparmadillo} to speed up computation. 

While the na\"{i}ve sampling scheme is relatively simple, ensuring good exploration of the posterior target can be challenging due to the potentially multimodal nature of the posterior distribution. Specifically, some sensitivity of the results to the starting values of the chain can be observed for some data. Section C.2 of the web-based supporting materials outlines an algorithm for the selection of informed starting values. Furthermore, to mitigate sensitivity to chain initialization, we also implemented a tempered transition scheme, which improves the mixing of the Markov chain by allowing for transitions between modal configurations of the target.  Implementation details for the proposed tempered transition scheme are reported in Section C.3 of the web-based supporting materials. 

Given Monte Carlo samples from the posterior distribution of all parameters of interest, posterior inference is implemented descriptively; either directly on the Monte Carlo samples for parameters of interest, such as the mixed membership proportions $\mathbf{z}_i$, or indirectly through the evaluation of relevant functions of the parameters of interest, e.g.
the mean and cross-covariance functions of the latent features. 

In this setting, to calculate the simultaneous credible intervals, we will use the simultaneous credible intervals proposed by \citet{crainiceanu2007spatially}. Let $\mathbf{g}_n$ be simulated realizations using the MCMC samples of the function of interest, and let $\{t_1, \dots, t_R\}$ be a fine grid of time points in $\mathcal{T}$. Let $\mathbb{E}(\mathbf{g}(t_i))$ be the expected value of the function evaluated at time point $t_i \in \mathcal{T}$, and $\text{SD}(\mathbf{g}(t_i))$ be the standard deviation of the function evaluated at time point $t_i \in \mathcal{T}$. Let $M_\alpha$ be the $(1- \alpha)$ quantile of 
$\max_{1 \le i \le R} \left|\frac{\mathbf{g}_n(t_i) - \mathbb{E}(\mathbf{g}(t_i))}{\text{SD}(\mathbf{g}(t_i))} \right|,$
for $1 \le n \le N_{MC}$, where $N_{MC}$ are the number of MCMC samples of the converged MCMC chain. Thus, the simultaneous credible intervals can be constructed as
$$\mathcal{I}(t_i) = \left[\mathbb{E}(\mathbf{g}(t_i)) - M_\alpha \text{SD}(\mathbf{g}(t_i)), \mathbb{E}(\mathbf{g}(t_i)) + M_\alpha \text{SD}(\mathbf{g}(t_i))\right].$$
Thus we estimate simultaneous credible intervals for all mean functions, $\mu^{(k)}$, and similarly generalize this procedure to define simultaneous credible intervals for the cross-covariance functions, $C^{(k,k')}$. Figure 5 in the main text, illustrates the difference between a simultaneous credible interval and a pointwise credible interval for one of the EEG case studies.

\section{Basis Functions}
In the proposed model, we assume that the stochastic processes, $f^{(k)}$, take values in $\boldsymbol{\mathcal{S}}$ (a space spanned by a pre-specified collection of $P$ basis functions), which is a $P$-dimensional subspace of $L^2(\mathcal{T})$. While this may seem restrictive, we note that the assumptions that functions can be approximated by a finite set of functions (or similar assumptions) are commonly made in the functional data analysis literature. 

Functional regression is one of the most comprehensive topics studied in functional data analysis \citep{reiss2017methods, morris2015functional}, consisting of function-on-scalar regression, scalar-on-function regression, and function-on-function regression. In any of the three types of functional regression, the functional coefficient $\boldsymbol{\beta}(t)$ is represented using a finite expansion of (1) pre-specified basis functions or (2) data-driven basis functions \citep{reiss2017methods}. When considering data-driven basis, we typically use the first few eigenfunctions of the covariance operator, which are often obtained by conducting functional principal component analysis (FPCA). Although this seems to not rely on a pre-specified set of basis functions, most FPCA methods use a basis function expansion to estimate the FPCs \citep{shang2014survey}.

In the non-parametric functional literature \citep{ferraty2006nonparametric, petersen2019frechet} kernel assumptions are commonly made, which typically assume local smoothness. While seemingly different from a finite basis expansion, \cite{donoho1989projection} illustrates the duality between projection based approaches and kernel based approaches. Moreover, \citet{de1978practical} has shown that for any smooth function, the distance between the true function and the approximation constructed using B-splines with uniform knots goes to zero as the number of knots goes to infinity. Specifically, if we have a set of B-splines of degree $\mathcal{D}$ with $\mathcal{K}$ uniform knots, then the bound of the distance between the smooth function and the B-spline approximation is proportional to $\left(\frac{|\mathcal{T}|}{\mathcal{K}}\right)^{\mathcal{D}}$, where $|\mathcal{T}|$ is the length of the domain of our function of interest. Thus, we can see that B-splines do a particularly good job in approximating smooth functions. We note that other types of basis systems can be used, as long as the basis functions are uniformly continuous. If the data modeled are periodic in nature, a wavelet basis or Fourier basis may be a more suitable basis compared to B-splines.

While finite-dimensional basis expansions are common and can be relatively flexible, having two few basis functions can lead to a lack of flexibility and a model that is too restrictive. However, a large $P$ leads to a heavier computational burden, as the number of parameters for the mean and covariance functions grow linearly with $P$. In practice, the number of basis functions used are often proportional to the number of time points observed. Prior to fitting a model, a practitioner can approximate the raw functional data using a proposed set of basis functions prior to fitting the functional mixed membership model proposed in the manuscript. Poor approximations would indicate that more basis functions may be needed. On the other hand, $P$ should not be so large that the computational burden of fitting the model is greater than the computational budget available.

\section{Effects of the Cross-Covariance Function}

\begin{figure}[H]
   \centering
    \includegraphics[width=.9\textwidth]{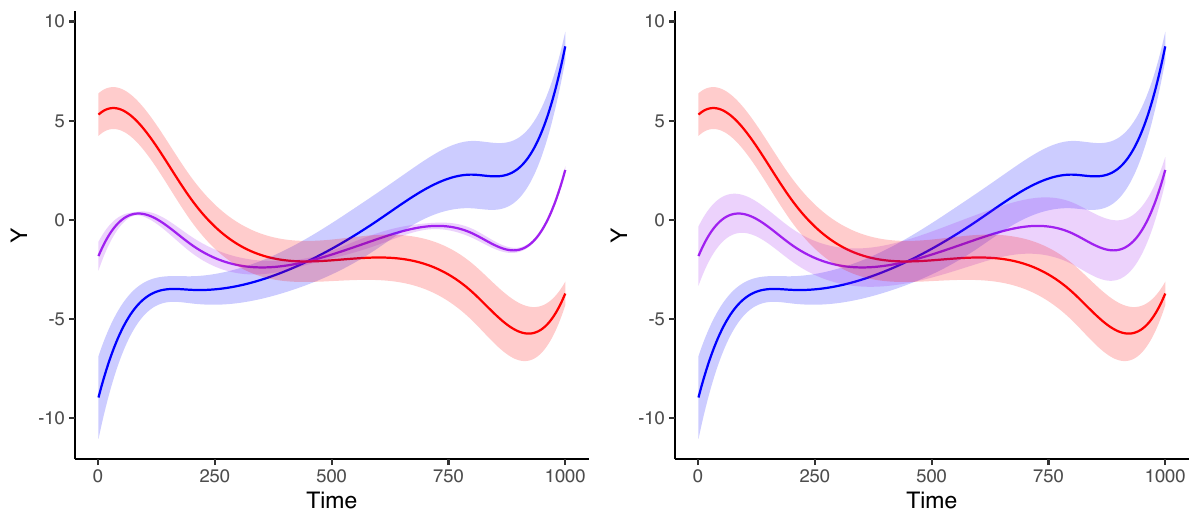}
   \caption{Visualizations of data generated from the proposed model with various cross-covariance functions. The red and blue functions represent the distribution of functions belonging in only one functional feature, while the purple function represents the distribution of functions belonging equally to both functional features.}
   \label{fig:sim_data}
\end{figure}

In a mixed membership model, we often want to infer the distribution of observations that belong partially to multiple functional features. Figure \ref{fig:sim_data} shows two examples of data generated from a two feature mixed membership model, with no measurement error ($\sigma^2 = 0$). 
The blue function represents the mean of observations that belong entirely to one functional feature, whereas the red function represents the mean of observations that belong to the other functional feature. The function in purple depicts the mean function of observations that belong equally to both functional features. 
The two examples were generated with the same mean and covariance functions, however they have different cross-covariance functions.
The graph on the left allows us to visualize what observed functions would look like when data are generated from the proposed mixed membership model with negative cross-covariance between the two functional features. In contrast, the graph on the right illustrates the case when there is positive cross-covariance between the two functional features. In this case, a negative cross-covariance function refers to the eigenvalues of the symmetric part of the cross-covariance operator being negative. 
We can see that the cross-covariance functions are crucial to ensuring a flexible model, as it controls the variation of observations that belong to more than one functional feature. 

In the ASD case study, discussed in Section 4.3 of the main text and Section \ref{EEG_case_study_appendix}, we could see that assuming that the two features were dependent allowed us to model some of the complex scenarios seen in EEG recordings such as a shifting PAF. This is particularly evident in Figures \ref{fig: posterior_pred} and \ref{fig:real_1_cov}. A detailed discussion on the effects of the covariance function can be found in Section \ref{EEG_case_study_appendix}.

\bibliographystyle{rss}
\bibliography{BFPMM}